# A plastics hierarchy of fates: sustainable choices for a circular future

**Kristoffer Kortsen[a], Siobhan Kilbride[a], Stephen R. Lowe[a], Adam Peirce[a] & Michael P. Shaver[a,*]**

[a] Department of Materials, The University of Manchester, Manchester, M1 7DN UK
[*] Corresponding author. E-mail address: michael.shaver@manchester.ac.uk

## Graphical abstract

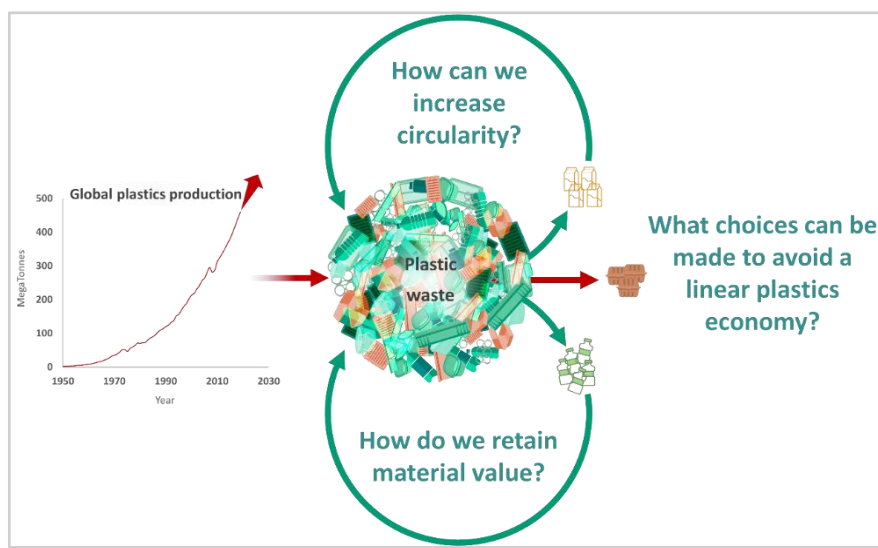

## Abstract

Plastics are ubiquitous in modern society, but the linear model of produce, use, and dispose results in massive amounts of resource consumption and pollution. Landfill and incineration of plastic waste is endemic, with no consensus on a clear path towards a more sustainable model. Progress is often hampered by a lack of clarity on what choices will enable a more sustainable circular plastics economy. Our 'Plastics Hierarchy of Fates' tool addresses this by bringing together scattered information on the end-of-life fates of plastics in a more accessible format. This tool will support manufacturing, processing, and policy decisions in the push for a more sustainable future. Potential sorting and recycling decisions for plastics waste are discussed in the hierarchy and the consequences of different decisions are highlighted. The hierarchy is meant to inform potential outcomes but can also be used to help shape future interventions.

## The Hierarchy

The hierarchy is created as an interactive information portal, with fates and decisions layered over each other to show interdependencies. The following pages are a compiled PDF of all information segments in the 'Plastics Hierarchy of Fates' tool, which can be accessed via:

The Plastics Hierarchy of Fates - https://lucid.app/documents/embedded/0ad93c05-0179-40bb-9468-63e25fc4dfae

# Welcome to our hierarchy of end-of-life fates!

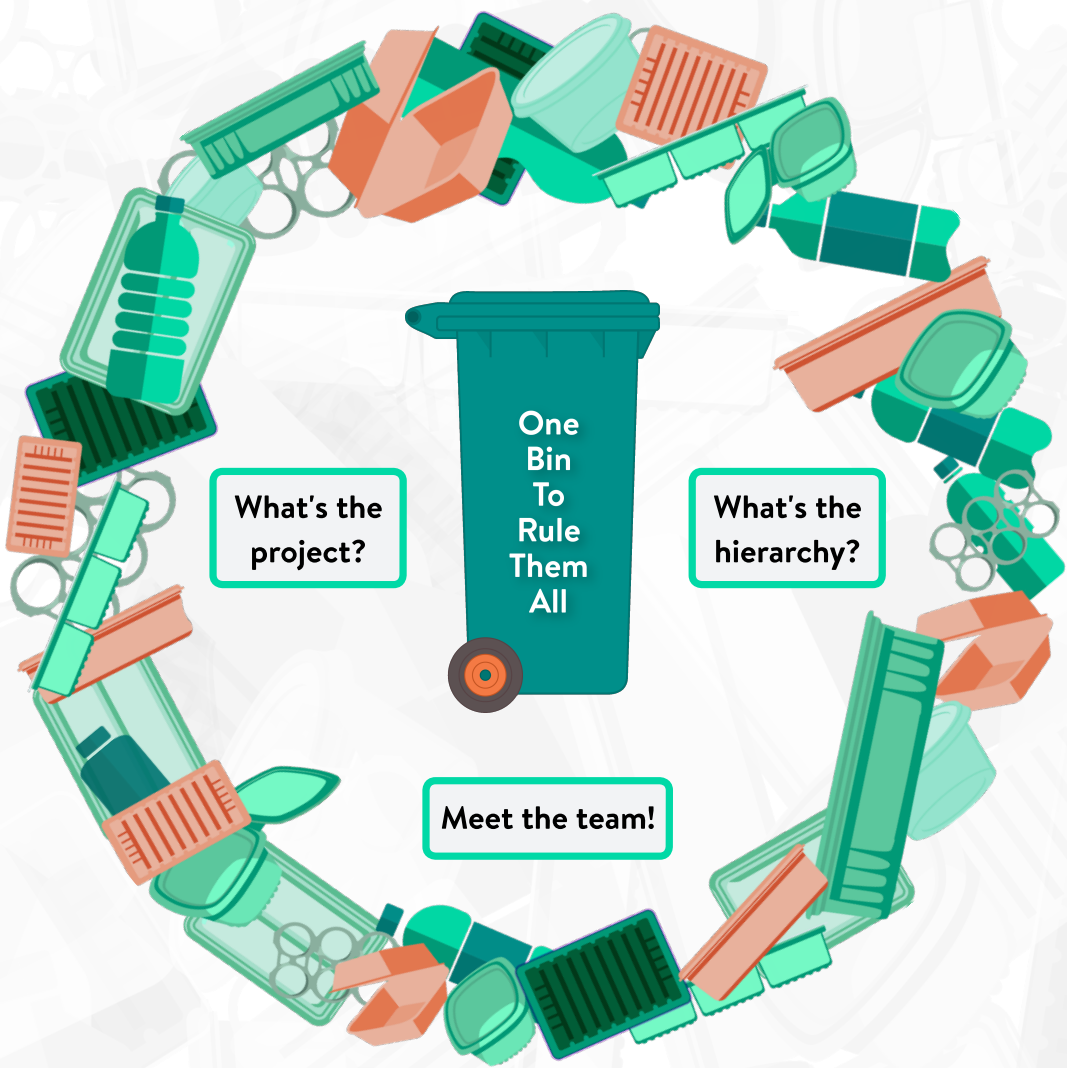

One Bin To Rule Them All

What's the project?

What's the hierarchy?

Meet the team!

## Hierarchy guidelines

Clicking on an **Arrow block** takes you through the various pages of the hierarchy.

All **Blocks** with a green border can be clicked to reveal more details and information.

Click on any *reference block* to reveal a hyperlink directly to the source of information.

Some context around words used in the hierarchy is accessed by clicking on the **Jargon Buster** icon.

End-of-life outcomes for plastic waste are colour-coded using a traffic light system:

- Best recycling choice for the material
- Good material outcome, but with limitations
- Material outcome that needs to be avoided

**Click here to explore the hierarchy**

# Welcome to our hierarchy of end-of-life fates!

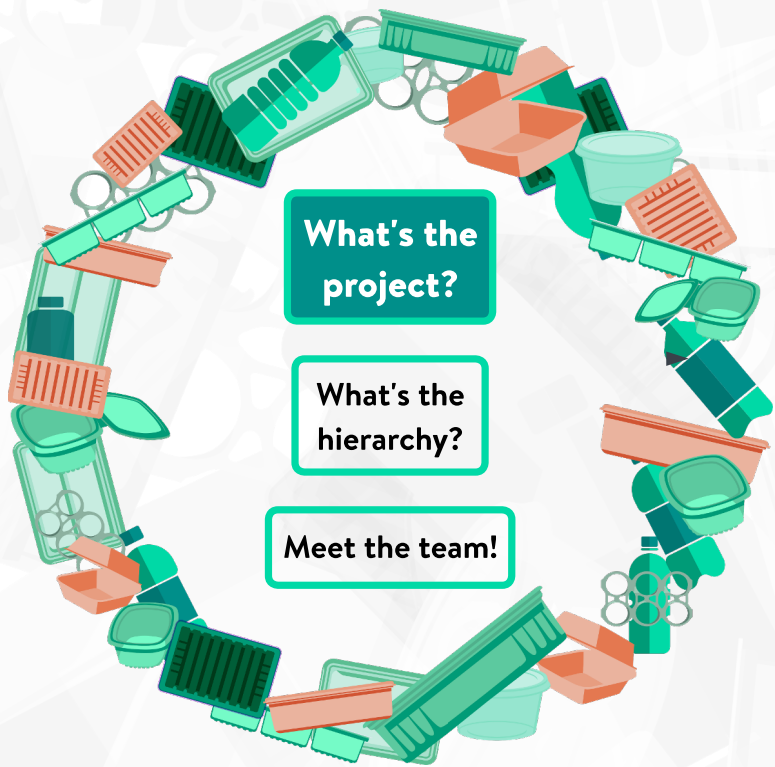

What's the project?

What's the hierarchy?

Meet the team!

**'One Bin to Rule Them All'**

A UKRI funded project developing **best practice** for future UK **household plastic recycling** policy. We're an interdisciplinary team of researchers, working in material science, social science and economics supported by over 30 policy and industry stakeholders.

**Materials science** is fundamental to recycling and this **hierarchy of end-of-life fates** has been developed as part of the 'One Bin To Rule Them All' project to **highlight the potential of** better household waste **recycling.**

## Hierarchy guidelines

Clicking on an [Arrow block] takes you through the various pages of the hierarchy.

All [Blocks] with a green border can be clicked to reveal more details and information.

Click on any [reference block] to reveal a hyperlink directly to the source of information.

Some context around words used in the hierarchy is accessed by clicking on the [Jargon Buster] icon.

End-of-life outcomes for plastic waste are colour-coded using a traffic light system:

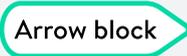 Best recycling choice for the material

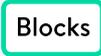 Good material outcome, but with limitations

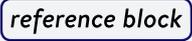 Material outcome that needs to be avoided

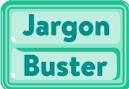

Click here to explore the hierarchy

```
V0.9 - public beta
Email any comments to Kris
```

# Welcome to our hierarchy of end-of-life fates!

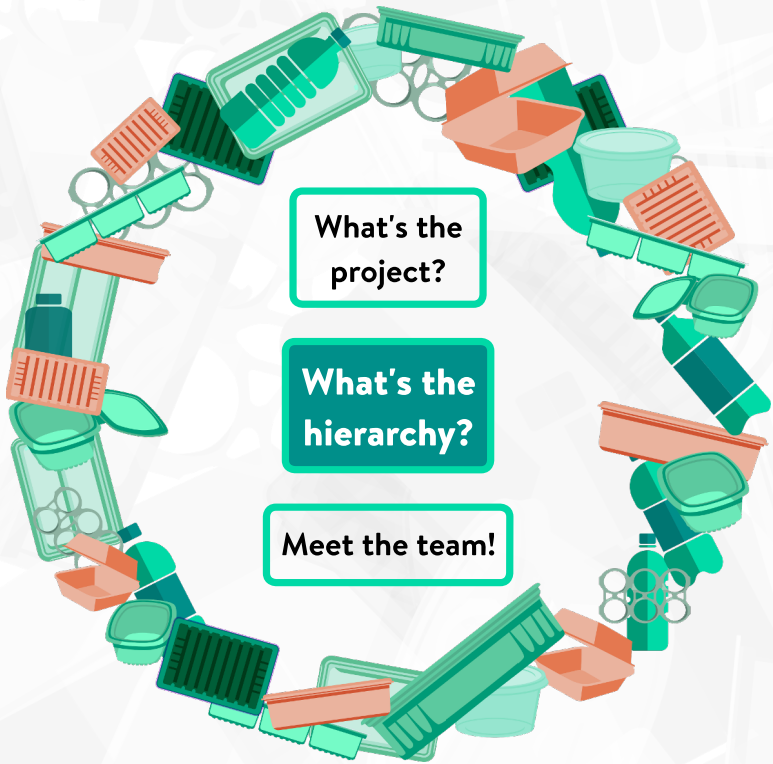

What's the project?

What's the hierarchy?

Meet the team!

This is an **interactive tool** designed to show how a bin **for dry mixed recycling** (DMR) waste can be processed. Currently scattered information on household waste recycling is brought together and different waste streams are analysed.

Starting from 'One Bin To Rule Them All', the hierarchy shows possible sorting and recycling paths, with an in-depth focus on plastics. The hierarchy is intended to be an **evolving overview of household waste recycling**, informing decisions across the supply chain and policy. It aims at fostering sustainability.

## Hierarchy guidelines

Clicking on an [ Arrow block ⟩ takes you through the various pages of the hierarchy.

All [ Blocks ] with a green border can be clicked to reveal more details and information.

Click on any [ reference block ] to reveal a hyperlink directly to the source of information.

Some context around words used in the hierarchy is accessed by clicking on the [Jargon Buster] icon.

End-of-life outcomes for plastic waste are colour-coded using a traffic light system:

- 🟩 Best recycling choice for the material
- ⬛ Good material outcome, but with limitations
- 🟧 Material outcome that needs to be avoided

[ Click here to explore the hierarchy ⟩

```
V0.9 - public beta
Email any comments to Kris
```

# Welcome to our hierarchy of end-of-life fates!

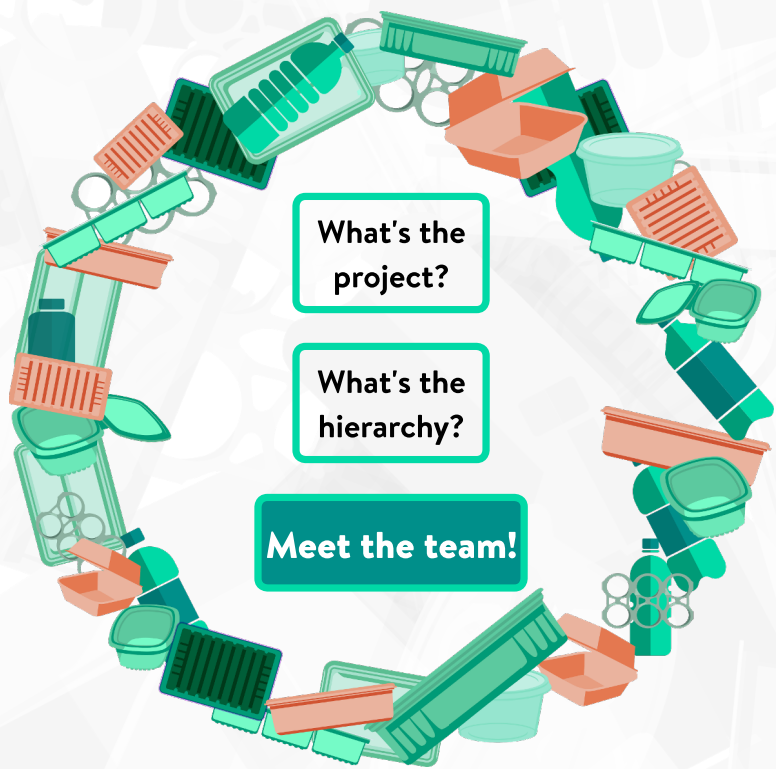

What's the project?

What's the hierarchy?

Meet the team!

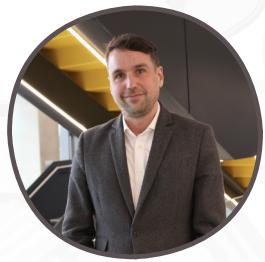

Prof. Mike Shaver

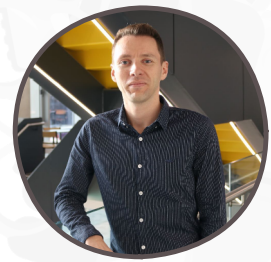

Dr Kristoffer Kortsen

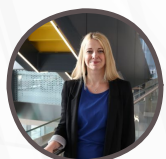

Dr Helen Holmes

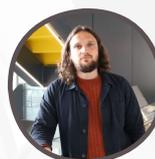

Dr Torik Holmes

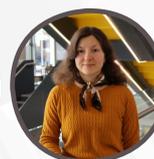

Dr Maria Sharmina

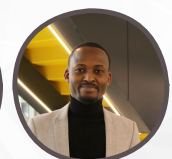

Dr Adeyemi Adelekan

## Hierarchy guidelines

Clicking on an ⟨ Arrow block ⟩ takes you through the various pages of the hierarchy.

All ⟨ Blocks ⟩ with a green border can be clicked to reveal more details and information.

Click on any ⟨ reference block ⟩ to reveal a hyperlink directly to the source of information.

Some context around words used in the hierarchy is accessed by clicking on the ⟨ Jargon Buster ⟩ icon.

End-of-life outcomes for plastic waste are colour-coded using a traffic light system:

- 🟩 Best recycling choice for the material
- ⬛ Good material outcome, but with limitations
- 🟧 Material outcome that needs to be avoided

**Click here to explore the hierarchy** ⟩

```
V0.9 - public beta
Email any comments to Kris
```

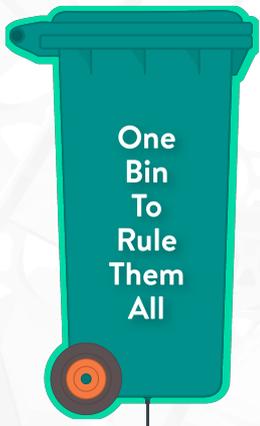

# One Bin To Rule Them All

- Reuse schemes
- Paper and card
- Metal packaging
- Glass
- Aseptic cartons
- Textiles
- Medical waste
- WEEE
- Biodegradable Plastics
- Flexible Plastics
- Rigid Plastics

Jargon Buster

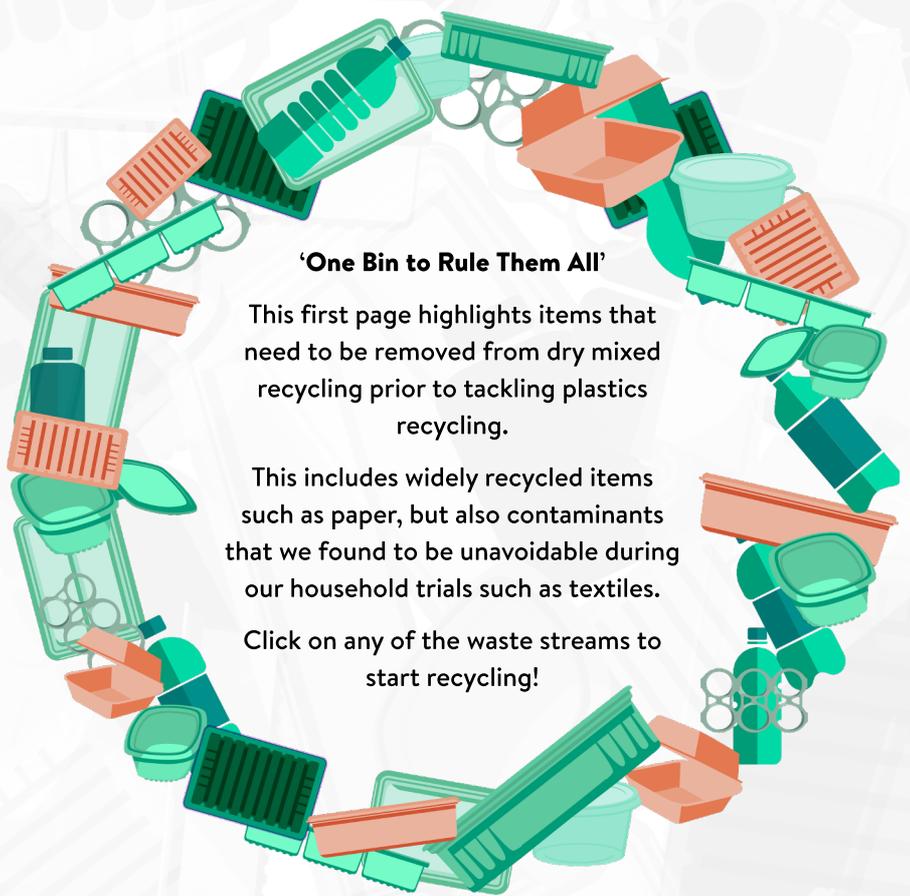

**'One Bin to Rule Them All'**

This first page highlights items that need to be removed from dry mixed recycling prior to tackling plastics recycling.

This includes widely recycled items such as paper, but also contaminants that we found to be unavoidable during our household trials such as textiles.

Click on any of the waste streams to start recycling!

```
V0.9 - public beta
Email any comments to Kris
```

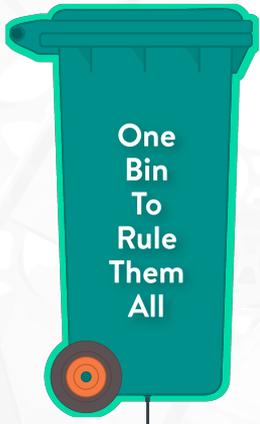
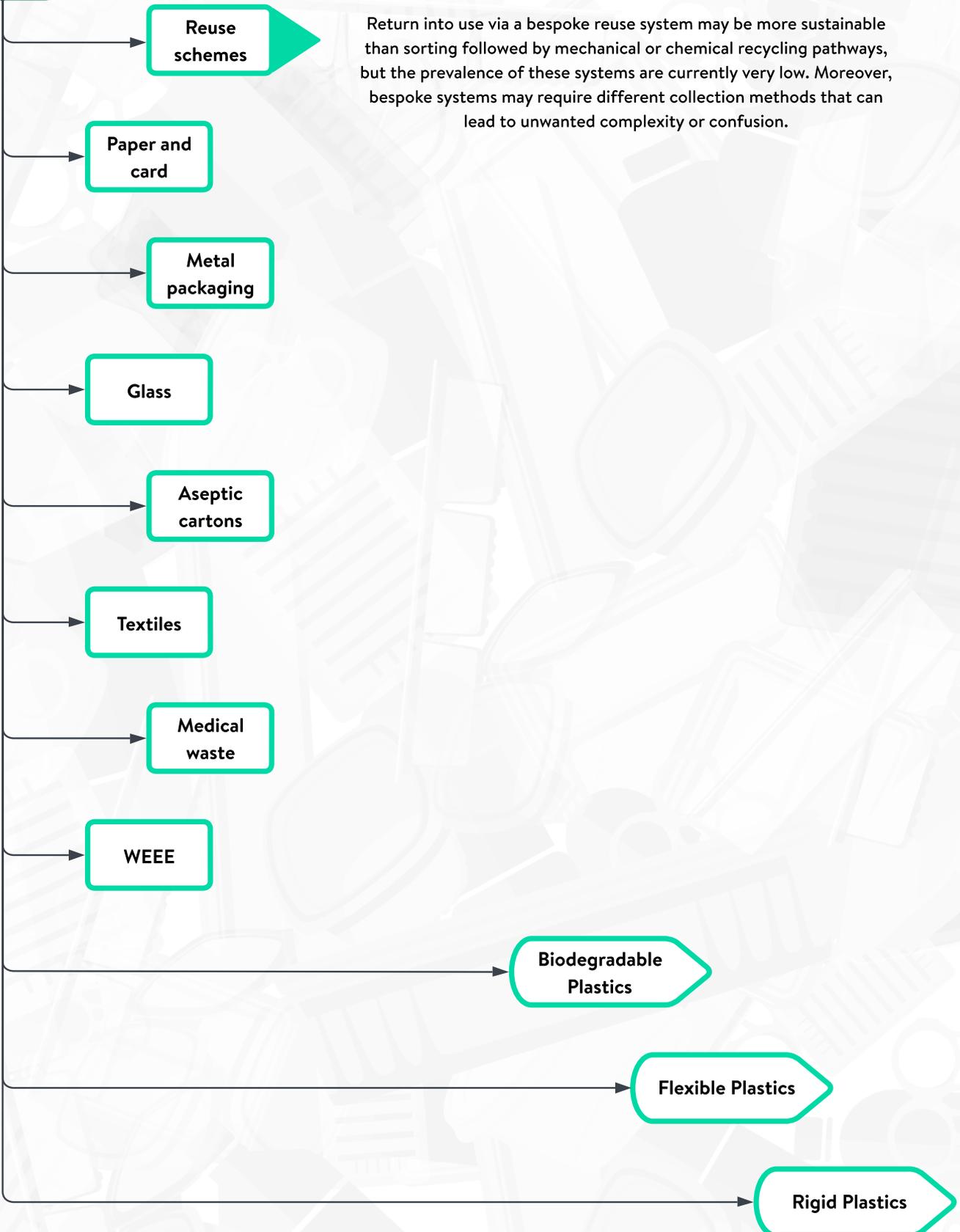

BackJargon Buster

One Bin To Rule Them All

- Reuse schemes
- Paper and card
- Metal packaging
- Glass
- Aseptic cartons
- Textiles
- Medical waste
- WEEE
- Biodegradable Plastics
- Flexible Plastics
- Rigid Plastics

Materials with dedicated reuse systems, such as tagged plastics, should be recovered prior to further sorting.

Return into use via a bespoke reuse system may be more sustainable than sorting followed by mechanical or chemical recycling pathways, but the prevalence of these systems are currently very low. Moreover, bespoke systems may require different collection methods that can lead to unwanted complexity or confusion.

V0.9 - public beta
Email any comments to Kris

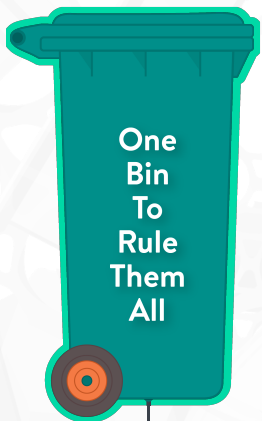

Jargon Buster

- One Bin To Rule Them All
  - Reuse schemes
  - **Paper and card**
  - Metal packaging
  - Glass
  - Aseptic cartons
  - Textiles
  - Medical waste
  - WEEE
  - Biodegradable Plastics
  - Flexible Plastics
  - Rigid Plastics

Paper and card recycling is widely conducted to high standards, with a wealth of knowledge available on the subject.[1] A recent EPRC report shows that the recycling rate of paper is above 80%, making it the most efficiently recycled packaging material type.[2]

Despite the success of the paper and card recycling industry, it is important to keep pushing for further improvements. 4evergreen, a cross-value chain alliance initiated by Cepi, highlights design guidelines, evaluation protocols, and improved collection targets to improve recycling by 2030.

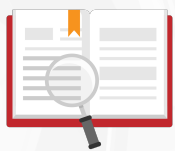

1. Bajpai, P., Recycling and De-inking of Recovered Paper, Elsevier, 2013.

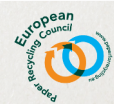

2. Monitoring Report 2021: Eropean Declaration on Paper Recycling 2021-2030, EPRC and CEPI, 2021.

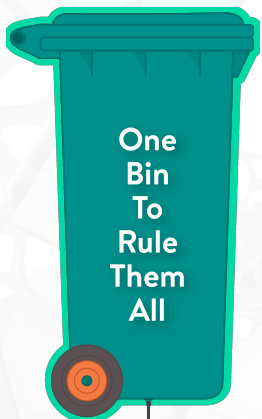

Jargon Buster

One Bin To Rule Them All

- Reuse schemes
- Paper and card
- Metal packaging
  - Ferrous metals
  - Non-ferrous metals
- Glass
- Aseptic cartons
- Textiles
- Medical waste
- WEEE
- Biodegradable Plastics
- Flexible Plastics
- Rigid Plastics

Ferrous (e.g. steel) and non-ferrous metals (e.g. aluminium) are recyclable after separation. Ferrous metals are retrieved with a magnet, while non-ferrous metals can be recovered using an eddy current separator.

Ferrous scrap metals are melted down and reprocessed into virgin like materials, affording a product of equally high quality using less energy and depletable resources.[1]

Non-ferrous metals are a diverse mix of high value elements such as aluminium, copper, tungsten, gold, etc. The different metals are separated through a range of techniques by metal processors, and the recovered metals are used as virgin replacements.[1]

While metal recycling is a profitable and often sustainable business, the degree of recycling taking place is often unclear. A broad assessment of recycling rates on a metal-by-metal basis has highlighted areas for potential improvement.[2]

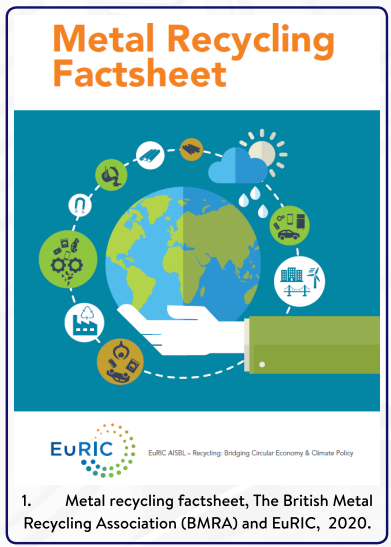

1. Metal recycling factsheet, The British Metal Recycling Association (BMRA) and EuRIC, 2020.

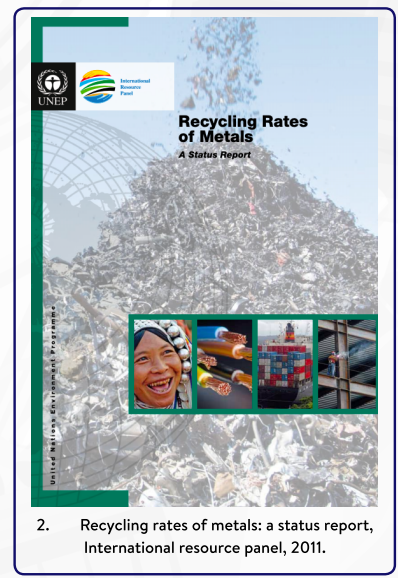

2. Recycling rates of metals: a status report, International resource panel, 2011.

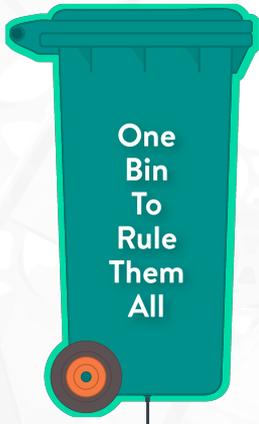

Jargon Buster

One Bin To Rule Them All

- Reuse schemes
- Paper and card
- Metal packaging
- **Glass**
- Aseptic cartons
- Textiles
- Medical waste
- WEEE
- Biodegradable Plastics
- Flexible Plastics
- Rigid Plastics

Just above 70% of glass is currently recycled in the UK, with a target collection rate of 90% by 2030. The large-scale glass recycling infrastructure that already exists in the UK provides high quality closed-loop recycling, saving raw materials, and avoiding carbon emissions.[1]

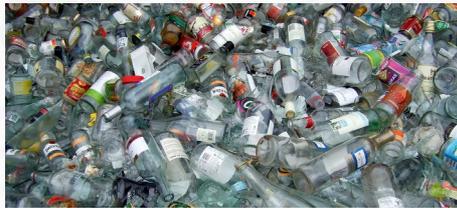

1. How is glass recycled?, WRAP.

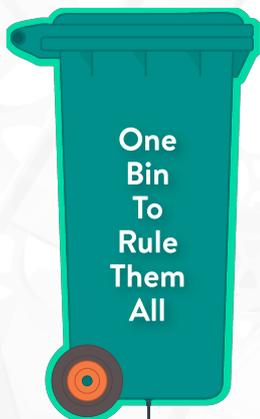
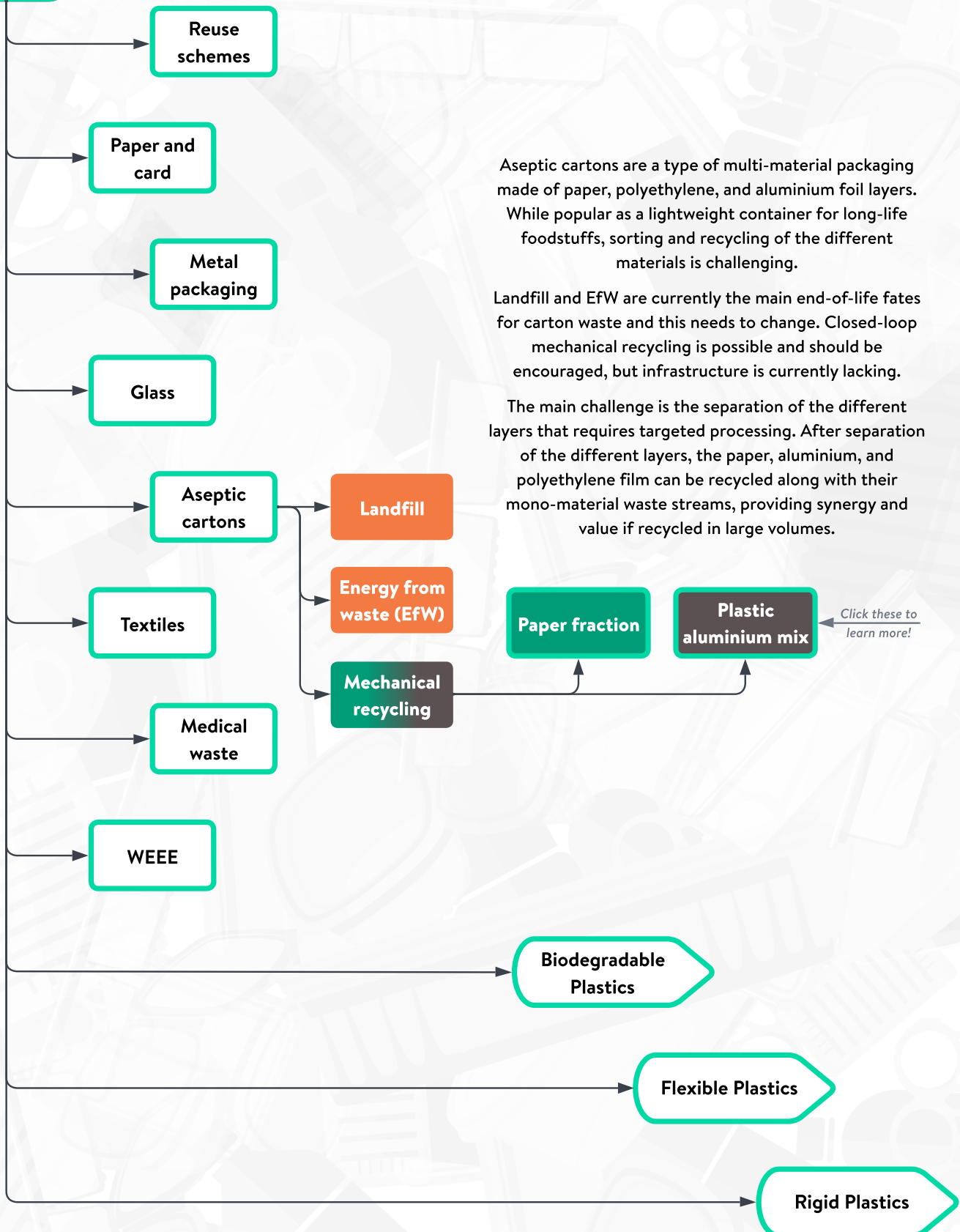

Back | Jargon Buster

**One Bin To Rule Them All**

- Reuse schemes
- Paper and card
- Metal packaging
- Glass
- Aseptic cartons → Landfill
- Aseptic cartons → Energy from waste (EfW)
- Aseptic cartons → Mechanical recycling → Paper fraction
- Aseptic cartons → Mechanical recycling → Plastic aluminium mix
- Textiles
- Medical waste
- WEEE
- Biodegradable Plastics
- Flexible Plastics
- Rigid Plastics

Aseptic cartons are a type of multi-material packaging made of paper, polyethylene, and aluminium foil layers. While popular as a lightweight container for long-life foodstuffs, sorting and recycling of the different materials is challenging.

Landfill and EfW are currently the main end-of-life fates for carton waste and this needs to change. Closed-loop mechanical recycling is possible and should be encouraged, but infrastructure is currently lacking.

The main challenge is the separation of the different layers that requires targeted processing. After separation of the different layers, the paper, aluminium, and polyethylene film can be recycled along with their mono-material waste streams, providing synergy and value if recycled in large volumes.

*Click these to learn more!*

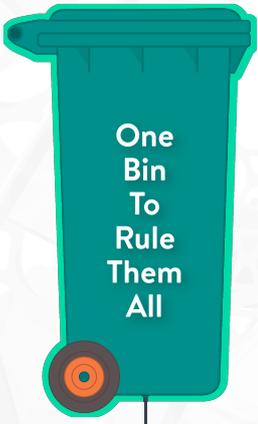

One Bin To Rule Them All

Jargon Buster

- Reuse schemes
- Paper and card
- Metal packaging
- Glass
- Aseptic cartons
  - Landfill
  - Energy from waste (EfW)
  - Mechanical recycling
    - Paper fraction
    - Plastic aluminium mix
- Textiles
- Medical waste
- WEEE
- Biodegradable Plastics
- Flexible Plastics
- Rigid Plastics

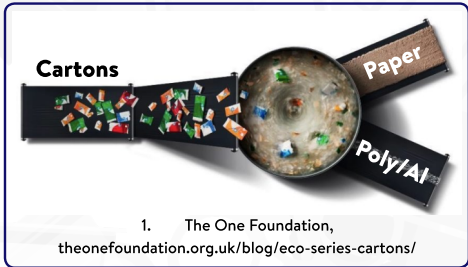

1. The One Foundation, theonefoundation.org.uk/blog/eco-series-cartons/

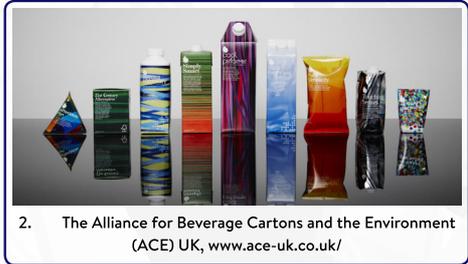

2. The Alliance for Beverage Cartons and the Environment (ACE) UK, www.ace-uk.co.uk/

Paper makes up 75% of the weight of a typical carton and it can be separated from the plastic and aluminium fractions through hydropulping.[1]

There is infrastructure for this in parts of the UK, but expansion is needed to recycle all aseptic cartons currently used.[2] Paper is separated with >95% efficiency through hydropulping and can be merged back into regular paper recycling streams.

"Tetra Pak® packages with FSC® label" by Tetra Pak is licensed under CC BY-SA 2.0.

[Back]

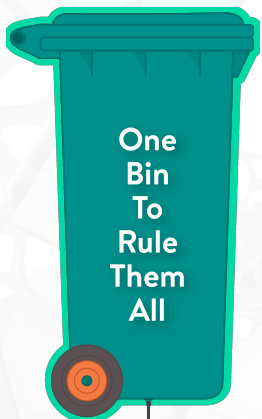

One Bin To Rule Them All

- Reuse schemes
- Paper and card
- Metal packaging
- Glass
- Aseptic cartons
  - Landfill
  - Energy from waste (EfW)
  - Mechanical recycling
    - Paper fraction
    - Plastic aluminium mix
- Textiles
- Medical waste
- WEEE
- Biodegradable Plastics
- Flexible Plastics
- Rigid Plastics

[Jargon Buster]

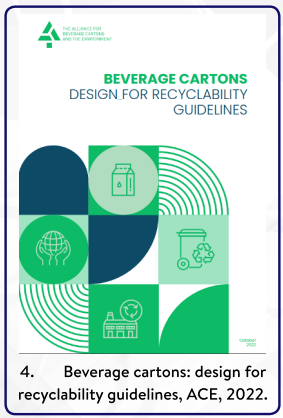

4. Beverage cartons: design for recyclability guidelines, ACE, 2022.

1. Şahin, G. G.; Karaboyacı, M., Process and machinery design for the recycling of tetra pak components. Journal of Cleaner Production, 2021.

2. Georgiopoulou, I.; Pappa, G. D.; Vouyiouka, S. N.; Magoulas, K., Recycling of post-consumer multilayer Tetra Pak packaging with the Selective Dissolution-Precipitation process. Resources, Conservation and Recycling, 2021.

3. Varžinskas, V.; Staniškis, J. K.; Knašytė, M., Decision-making support system based on LCA for aseptic packaging recycling. Waste Management & Research, 2012.

A mixed PE/aluminium fraction is recovered after hydropulping of aseptic cartons. This PE/aluminium mixture is currently exported to the EU for recycling by microwave enabled pyrolysis through the Enval process (patent US20080099325A1), where the plastic is converted to fuels and the aluminium is recycled.

Separation and recycling of both fractions has been demonstrated on lab scale, with life cycle analyses (LCAs) showing a positive impact, but this is currently not commercially feasible.[1,2,3]

Guidelines for recyclable carton design published by ACE, show desired carton composition for optimal recyclability, but also what recycling methods and capacity exists across Europe.[4]

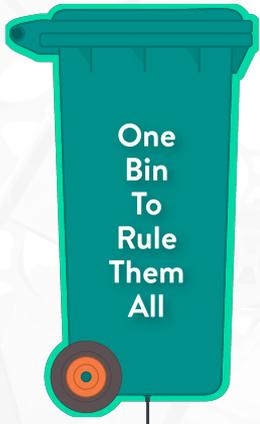

Jargon Buster

One Bin To Rule Them All

- Reuse schemes
- Paper and card
- Metal packaging
- Glass
- Aseptic cartons
- **Textiles**
- Medical waste
- WEEE
- Biodegradable Plastics
- Flexible Plastics
- Rigid Plastics

Non-reusable polyester and polycotton blended textiles can be decontaminated, separated, and processed by a range of techniques, affording raw materials for reuse in textile and other applications.

Companies such as Worn Again Technologies and BlockTexx use solvent-based separation methods to recover PET and cellulose. These materials can then be used to make new textiles in a mechanical recycling process.

Businesses such as Ambercycle, Ioncell, Infinitedfiber, and Renewcell turn used cellulose based textiles into high quality fibres through closed-loop chemical recycling processes.

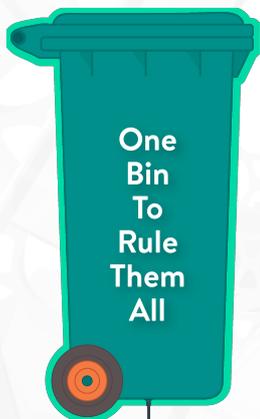

One Bin To Rule Them All

Jargon Buster

- Reuse schemes
- Paper and card
- Metal packaging
- Glass
- Aseptic cartons
- Textiles
- Medical waste
- WEEE
- Biodegradable Plastics
- Flexible Plastics
- Rigid Plastics

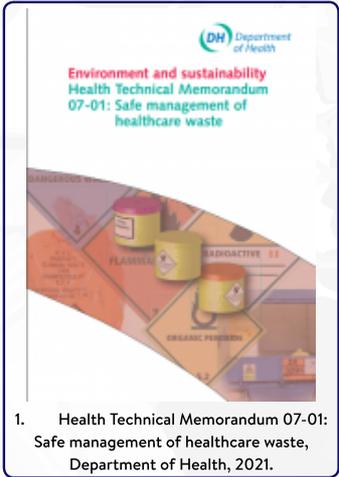

1. Health Technical Memorandum 07-01: Safe management of healthcare waste, Department of Health, 2021.

While this waste should not be included with dry mixed recycling, our trials have shown that it will be present. Separation of medical waste is not possible with current sorting techniques, so specific attention is needed to enable this in the future (e.g. fluorescent tagging).

The potential biohazardous contamination of medical waste necessitates the separate treatment of this waste stream. Non-selective chemical recycling has the potential to provide the best end-of-life fate, though energy recovery systems such as pyrolysis are the most likely choice due to the restricted handling of medical waste.[1]

Appropriate handling of household medical waste should feed back into the existing NHS healthcare waste management process, where chemical recycling and energy recovery systems are being explored on a larger scale.[1]

- Chemical recycling
- Energy from waste (EfW)
- Landfill

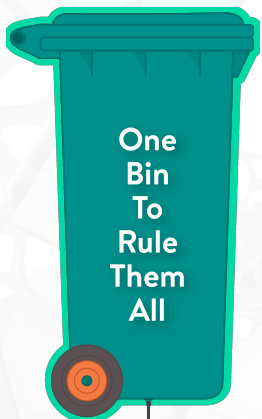

**One Bin To Rule Them All**

- Reuse schemes
- Paper and card
- Metal packaging
- Glass
- Aseptic cartons
- Textiles
- Medical waste
- **WEEE**
- Biodegradable Plastics
- Flexible Plastics
- Rigid Plastics

**Jargon Buster**

Waste Electrical and Electronic Equipment (WEEE) is of high value due to the presence of many rare metals, but bespoke end-of-life treatment is required. UK governmental guidance by HSE gives a clear overview of the challenges and intricacies of WEEE processing and recycling, highlighting the diversity of materials and the hazards of some substances and components. A WEEE treatment guide by WRAP discusses best practice of collection, processing, and treatment by specialist facilities.

Due to safety concerns, batteries and battery containing WEEE aren't currently allowed in kerbside collection waste bins. But compliance of this is low, with the majority of people not aware of this limitation. This is worrying - there are, on average, 250 fires at recycling facilities caused by batteries each year in the UK.

The changing landscape of WEEE waste is also of concern, most recently with the surge in disposable vape pens in waste streams and landfill. Collection and proper processing of these is essential to recover the valuable metals used and to avoid environmental pollution. Kerbside collection and large scale recovery is possible and should be encouraged.

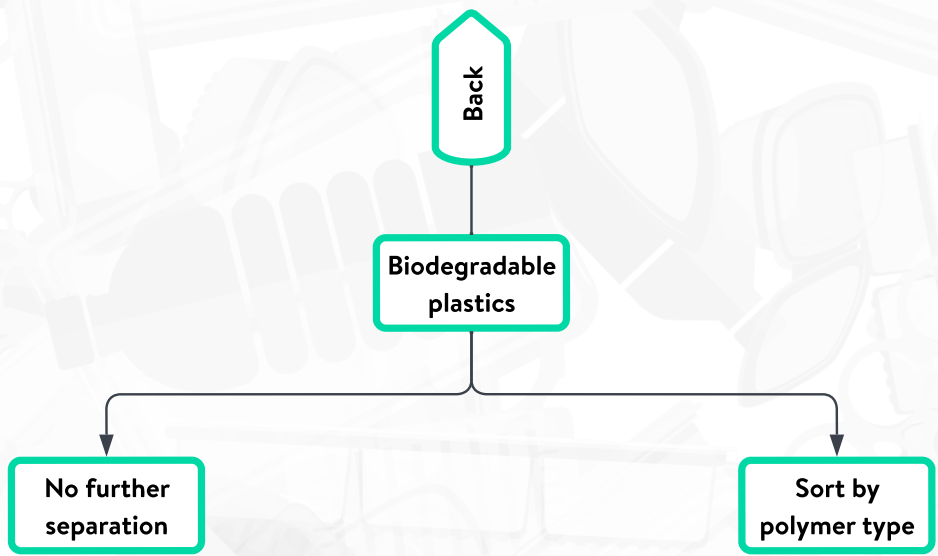

Back → Biodegradable plastics → No further separation | Sort by polymer type

The definitions of biodegradable plastics, bioplastics, and compostable plastics are vague and currently open to abuse for greenwashing purposes.[1] Their sustainability is questionable, because degradability and recyclability varies immensely by polymer type.

Without further sorting, the end-of-life fate of biodegradable plastics will deliver mixed results, as different plastic items will perform very differently in composting or recycling. On the other hand, sorting by polymer type is challenging due to the wide range of biodegradable plastics in low volumes. The cost of advanced sorting is generally not outweighed by the sustainability gains, making this a choice between two relatively negative fates.

Additionally, biodegradable plastics are viewed as contaminants in general plastic waste streams and present a clear hazard to reliable recycling of commodity plastics (especially PET).[2] The use of biodegradable plastics should be restricted to cases where they will assuredly enter the environment (marine, agriculture, etc.) or where segregated waste management systems are in place and enforced to avoid biodegradable plastics in household waste.[3]

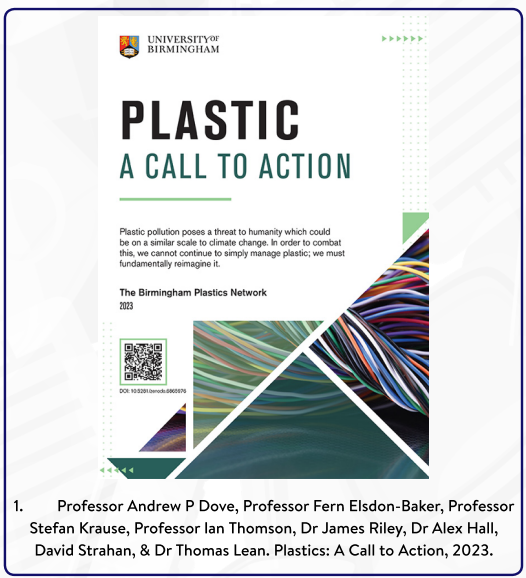

1. Professor Andrew P Dove, Professor Fern Elsdon-Baker, Professor Stefan Krause, Professor Ian Thomson, Dr James Riley, Dr Alex Hall, David Strahan, & Dr Thomas Lean. Plastics: A Call to Action, 2023.

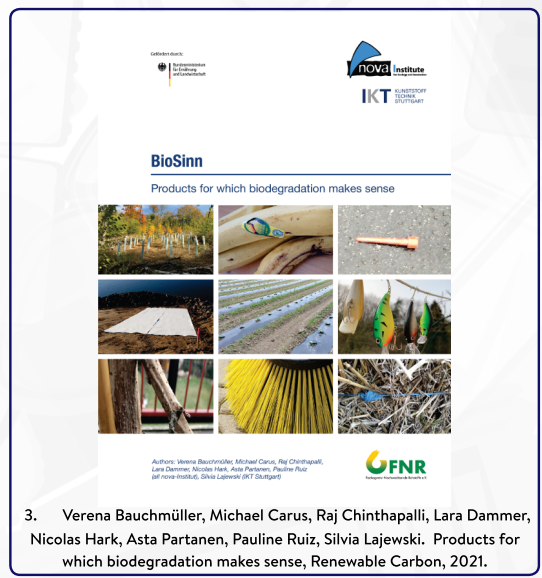

3. Verena Bauchmüller, Michael Carus, Raj Chinthapalli, Lara Dammer, Nicolas Hark, Asta Partanen, Pauline Ruiz, Silvia Lajewski. Products for which biodegradation makes sense, Renewable Carbon, 2021.

2. McLauchlin, A. R.; Ghita, O. R., Studies on the thermal and mechanical behavior of PLA-PET blends. Journal of Applied Polymer Science 2016.

# Biodegradable plastics

```
Back
 └─ Biodegradable plastics
     ├─ No further separation
     │   ├─ Industrial composting
     │   └─ Anaerobic digestion
     └─ Sort by polymer type
         ├─ Mechanical recycling
         ├─ Chemical recycling
         └─ EfW or landfill
```

Some plastic items claim to be compostable, often even home compostable, but 'real world' assessments show this is often not true.[1] Most polymers that are classed as 'compostable', such as the widely used PLA, will at minimum require an industrial composting system with higher temperature and controlled environment to degrade.[2,3] Other 'biodegradable' polymers meanwhile, will not degrade significantly within composting, further complicating this end-of-life option.[4]

While composting is more sustainable than landfilling, LCAs of industrial composting systems have shown that it can be a poor end-of-life fate for compostable polymers, with anaerobic digestion or mechanical recycling forming better choices.[4,5] All end-of-life decisions do however have significant drawbacks, highlighting the problematic nature of sustainable recycling of biodegradable polymers.

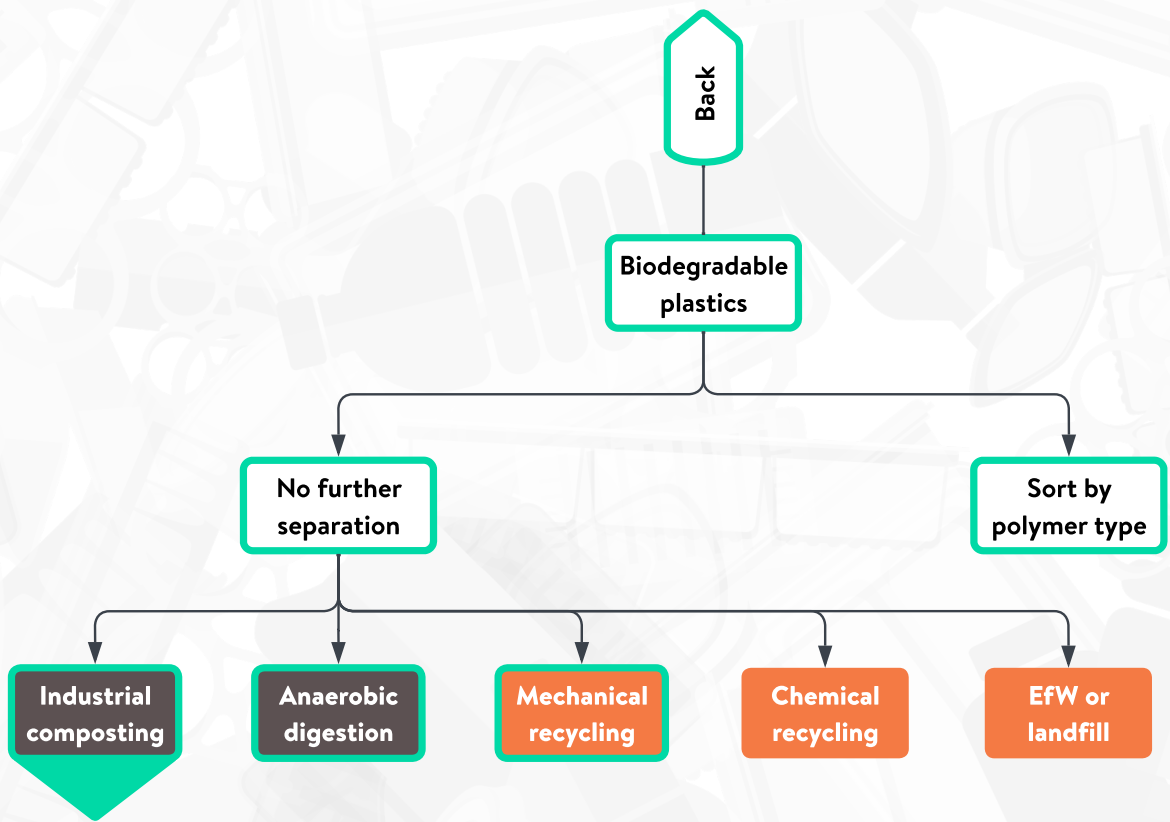

1. Purkiss, D.; Allison, A. L.; Lorencatto, F.; Michie, S.; Miodownik, M., The Big Compost Experiment: Using citizen science to assess the impact and effectiveness of biodegradable and compostable plastics in UK home composting. Frontiers in Sustainability, 2022.

2. Cafiero, L. M.; Canditelli, M.; Musmeci, F.; Sagnotti, G.; Tuffi, R. Assessment of Disintegration of Compostable Bioplastic Bags by Management of Electromechanical and Static Home Composters, Sustainability, 2021.

3. End-of-Life Options for Plastics (Ch. 6). In Sustainable Plastics: Environmental Assessments of Biobased, Biodegradable, and Recycled Plastics, Second Edition, 2022.

4. Rossi, V.; Cleeve-Edwards, N.; Lundquist, L.; Schenker, U.; Dubois, C.; Humbert, S.; Jolliet, O., Life cycle assessment of end-of-life options for two biodegradable packaging materials: sound application of the European waste hierarchy, Journal of Cleaner Production, 2015.

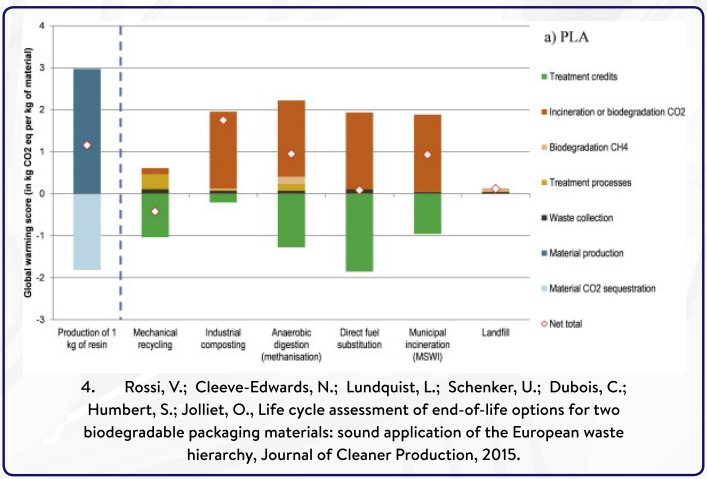

5. Narancic, T.; Verstichel, S.; Reddy Chaganti, S.; Morales-Gamez, L.; Kenny, S. T.; De Wilde, B.; Babu Padamati, R.; O'Connor, K. E., Biodegradable Plastic Blends Create New Possibilities for End-of-Life Management of Plastics but They Are Not a Panacea for Plastic Pollution, Environmental Science & Technology, 2018.

Jargon Buster

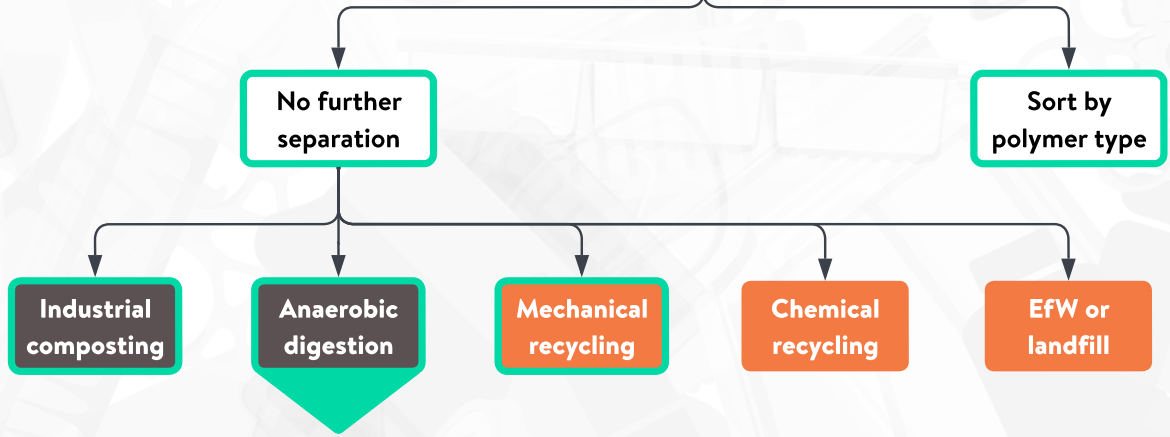

Anaerobic digestion, unlike aerobic composting, allows for the recovery of energy in the form of methane (alongside $CO_2$) from the complete biodegradation of susceptible polymers.[1] Therefore, anaerobic digestion has the potential to be a preferred end-of-life choice for many compostable polymers, with research showing better resource use in most cases.[2] Sustainability analysis and optimisation is required to guide the use of anaerobic digestion and to confirm which materials would be preferentially recycled through this method.[3]

But as with composting, the process results in the conversion of plastics into $CO_2$, making it a net contributor to global warming. End-of-life fates that result in the reuse of plastic resources, such as mechanical recycling will always be, more sustainable and should be favoured in the pursuit of a plastics circular economy.[4]

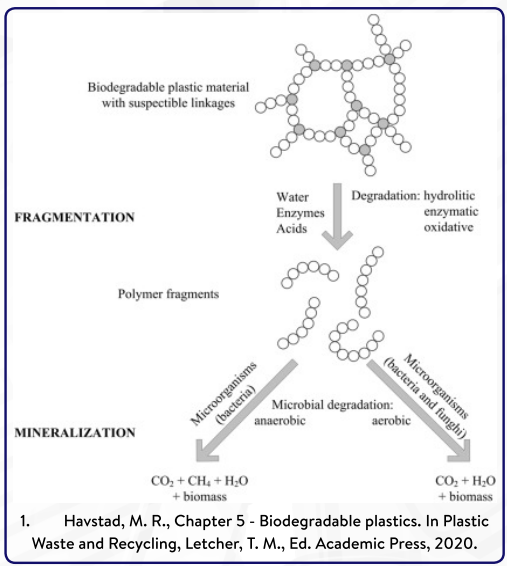

1. Havstad, M. R., Chapter 5 - Biodegradable plastics. In Plastic Waste and Recycling, Letcher, T. M., Ed. Academic Press, 2020.

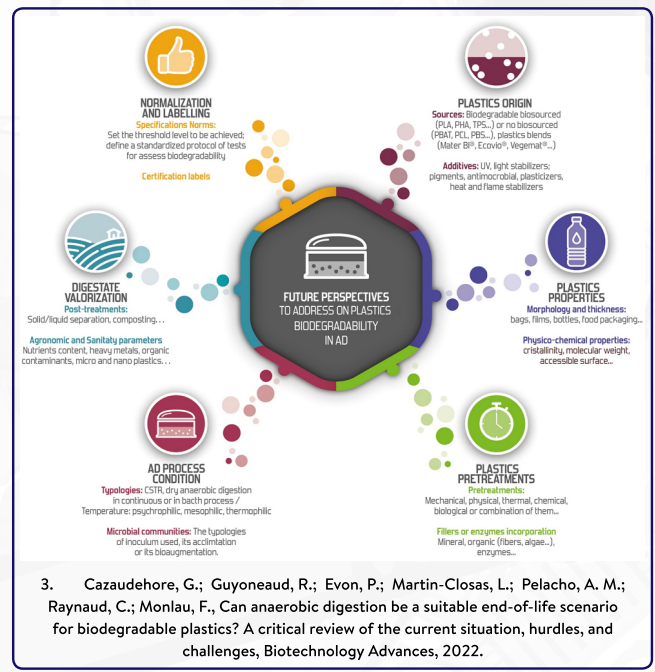

3. Cazaudehore, G.; Guyoneaud, R.; Evon, P.; Martin-Closas, L.; Pelacho, A. M.; Raynaud, C.; Monlau, F., Can anaerobic digestion be a suitable end-of-life scenario for biodegradable plastics? A critical review of the current situation, hurdles, and challenges, Biotechnology Advances, 2022.

4. Lamberti, F. M.; Roman-Ramirez, L. A.; Wood, J., Recycling of Bioplastics: Routes and Benefits. Journal of Polymers and the Environment 2020.

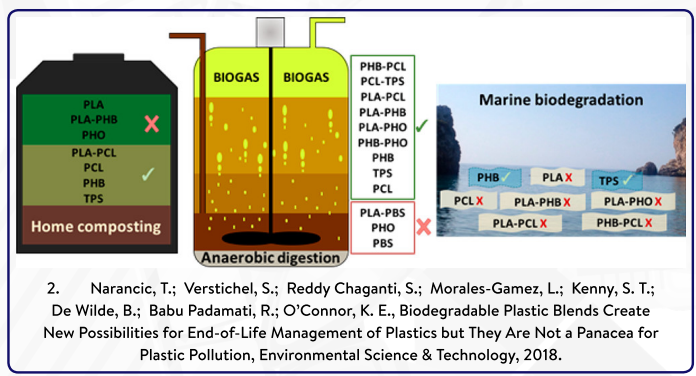

2. Narancic, T.; Verstichel, S.; Reddy Chaganti, S.; Morales-Gamez, L.; Kenny, S. T.; De Wilde, B.; Babu Padamati, R.; O'Connor, K. E., Biodegradable Plastic Blends Create New Possibilities for End-of-Life Management of Plastics but They Are Not a Panacea for Plastic Pollution, Environmental Science & Technology, 2018.

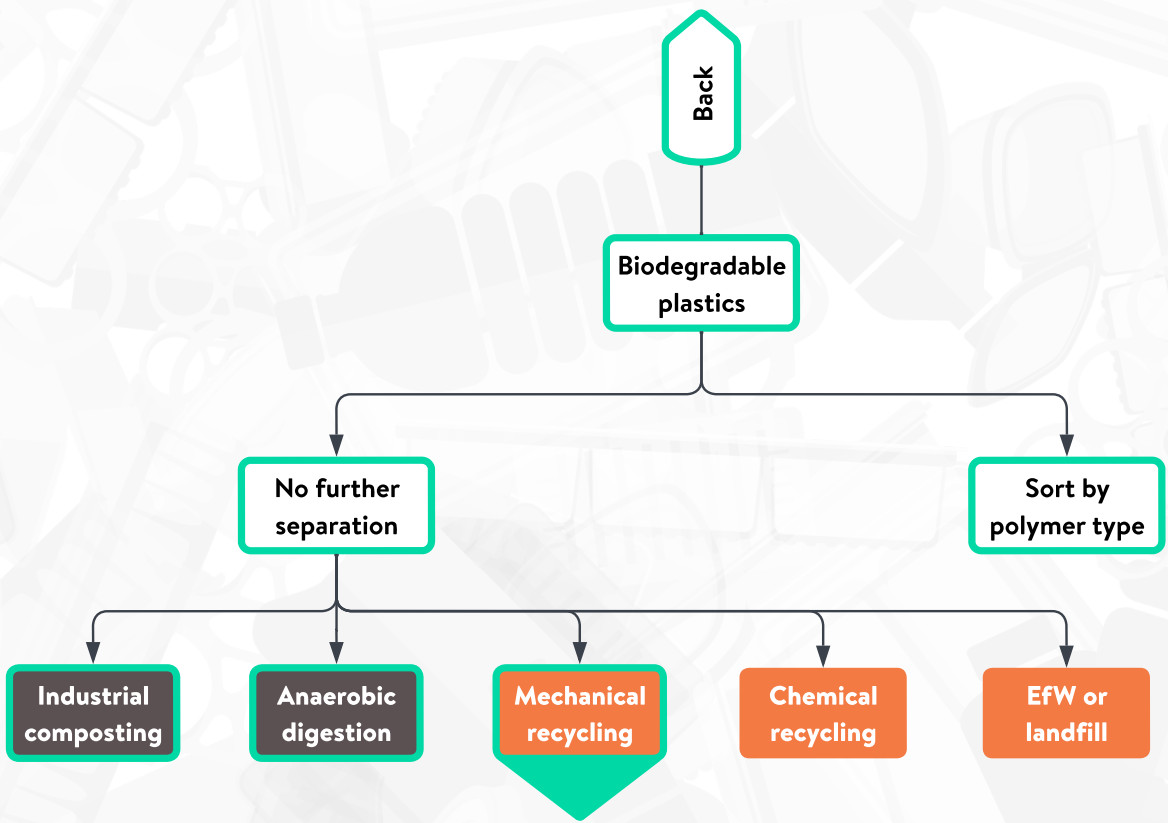

Mechanical recycling of mixed biodegradable plastics will not yield a product of any value, due to the mixture of polymer types with varying degradability. Reprocessing temperatures would inevitably lead to partial degradation and a highly inconsistent output.
Mechanical recycling is only viable after sorting by polymer type to extract the polymers that can provide value through this end-of-life fate.

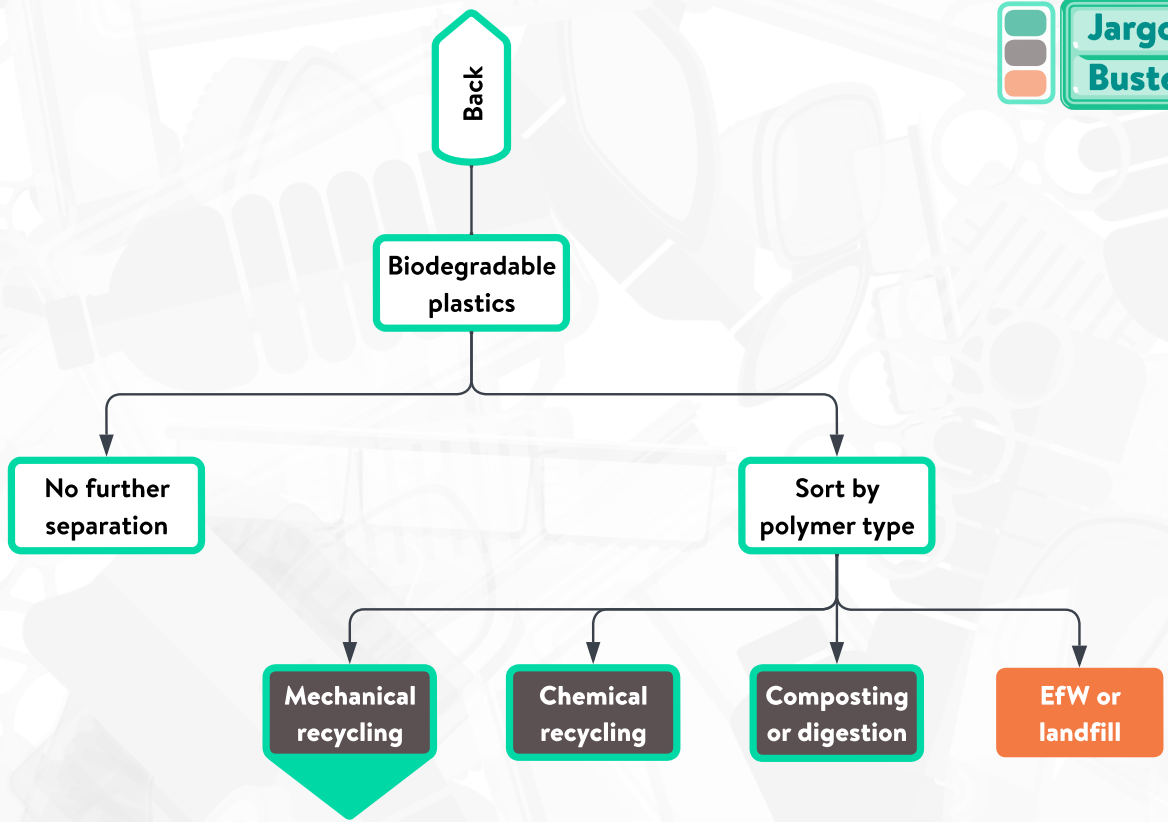

Many end-of-life biodegradable plastics can be mechanically recycled after sorting by polymer type, including the currently most abundant types such as PLA.[1] LCAs show that the environmental impact of mechanical recycling is far superior to composting, and is also preferable to chemical recycling.[2,3]

Biodegradable plastics degrade, and this means that mechanical recycling will result in loss of product properties faster than that seen for commodity plastics such as PET or PE, limiting the potential of continued mechanical recycling.[4] The degradation levels of recyclate will need to be assessed through, for example, melt flow rate, oxidation value, or bespoke recycled content analysis, which will guide choices between mechanical and chemical recycling.

Despite mechanical recycling being the best end-of-life fate for biodegradable plastics, the degradation that occurs from mechanical recycling and the need for enhanced sorting means that a better choice currently is to limit the presence of biodegradable plastics in household waste.

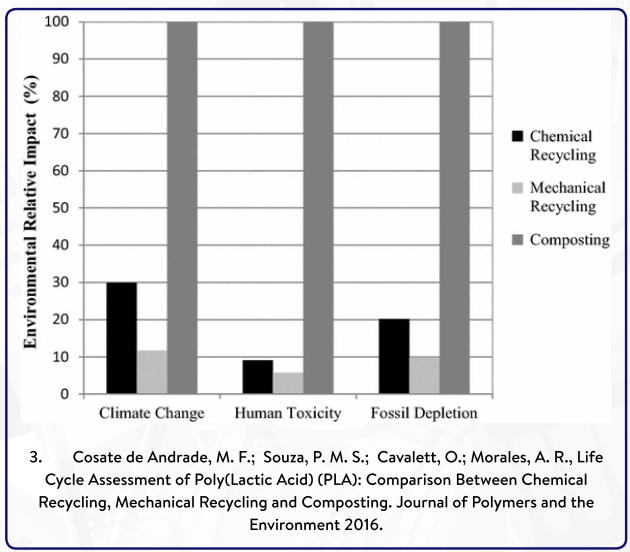
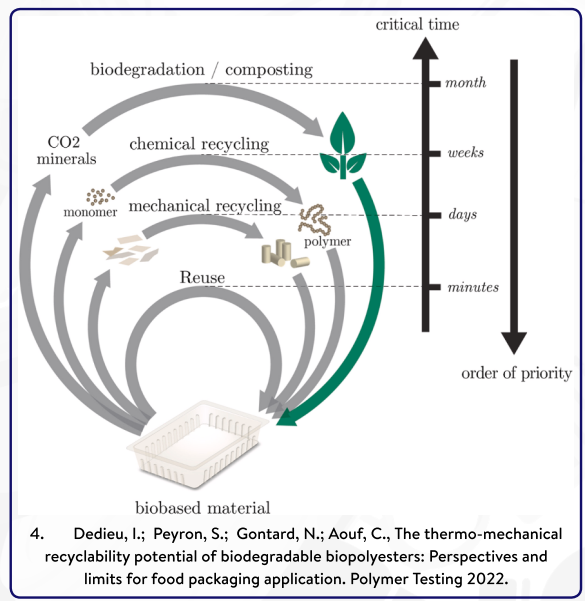

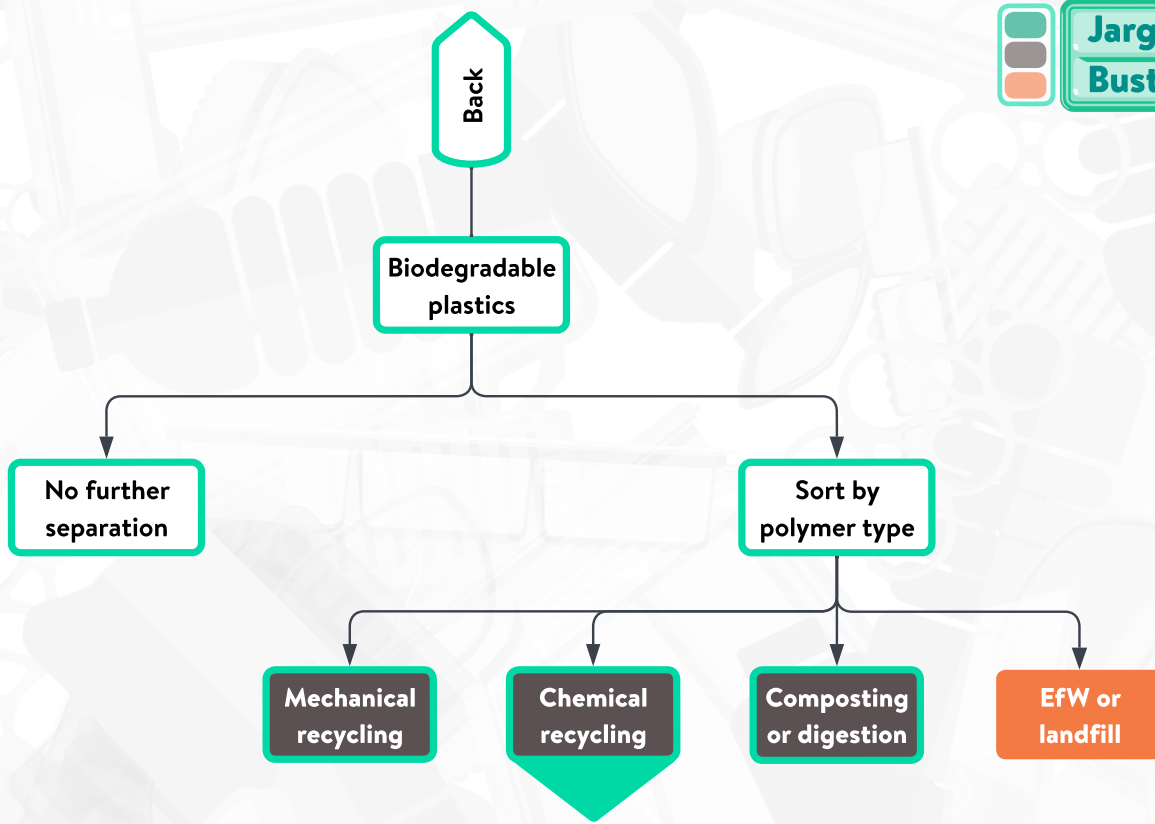

Biodegradable plastics that are no longer suitable for mechanical recycling due to excessive degradation and loss of desired properties, should be directed to chemical upcycling. LCA analysis shows that it is a far better end-of-life fate than composting, and is only behind mechanical recycling in terms of environmental impact.[1] There is ample literature showing the technical possibility to chemically recycle PLA, PCL, and PBAT, but nothing yet on a suitably large scale.[2,3,4]

Despite ample research and a range of patented technologies to enable chemical recycling of biopolymers, the waste streams of these materials remains small and overly complex, meaning that the creation of bespoke separation and recycling streams cannot currently be justified.[5] Efforts should be focussed on avoiding contamination of the large commodity plastic waste streams by limiting the prevalence of biodegradable plastics in household waste.

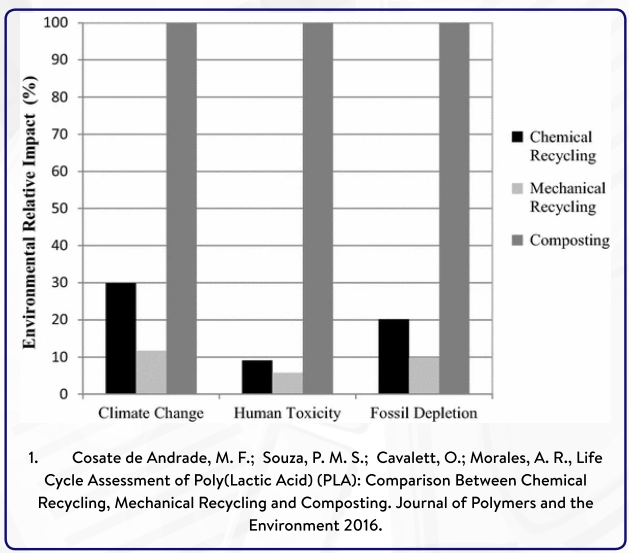

1. Cosate de Andrade, M. F.; Souza, P. M. S.; Cavalett, O.; Morales, A. R., Life Cycle Assessment of Poly(Lactic Acid) (PLA): Comparison Between Chemical Recycling, Mechanical Recycling and Composting. Journal of Polymers and the Environment 2016.

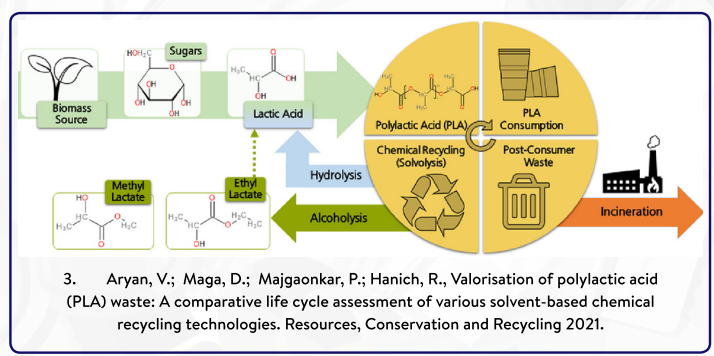

3. Aryan, V.; Maga, D.; Majgaonkar, P.; Hanich, R., Valorisation of polylactic acid (PLA) waste: A comparative life cycle assessment of various solvent-based chemical recycling technologies. Resources, Conservation and Recycling 2021.

4. Lamberti, F. M.; Roman-Ramirez, L. A.; Wood, J., Recycling of Bioplastics: Routes and Benefits. Journal of Polymers and the Environment 2020.

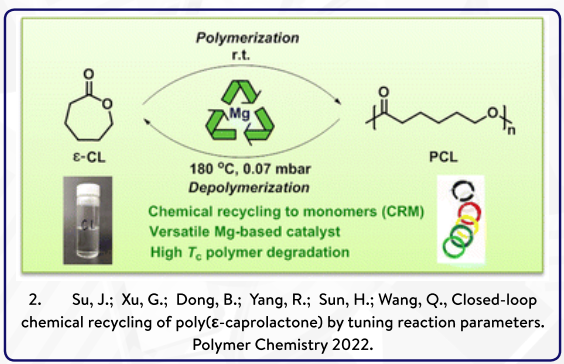

2. Su, J.; Xu, G.; Dong, B.; Yang, R.; Sun, H.; Wang, Q., Closed-loop chemical recycling of poly(ε-caprolactone) by tuning reaction parameters. Polymer Chemistry 2022.

5. Niaounakis, M., Recycling of biopolymers – The patent perspective. European Polymer Journal 2019.

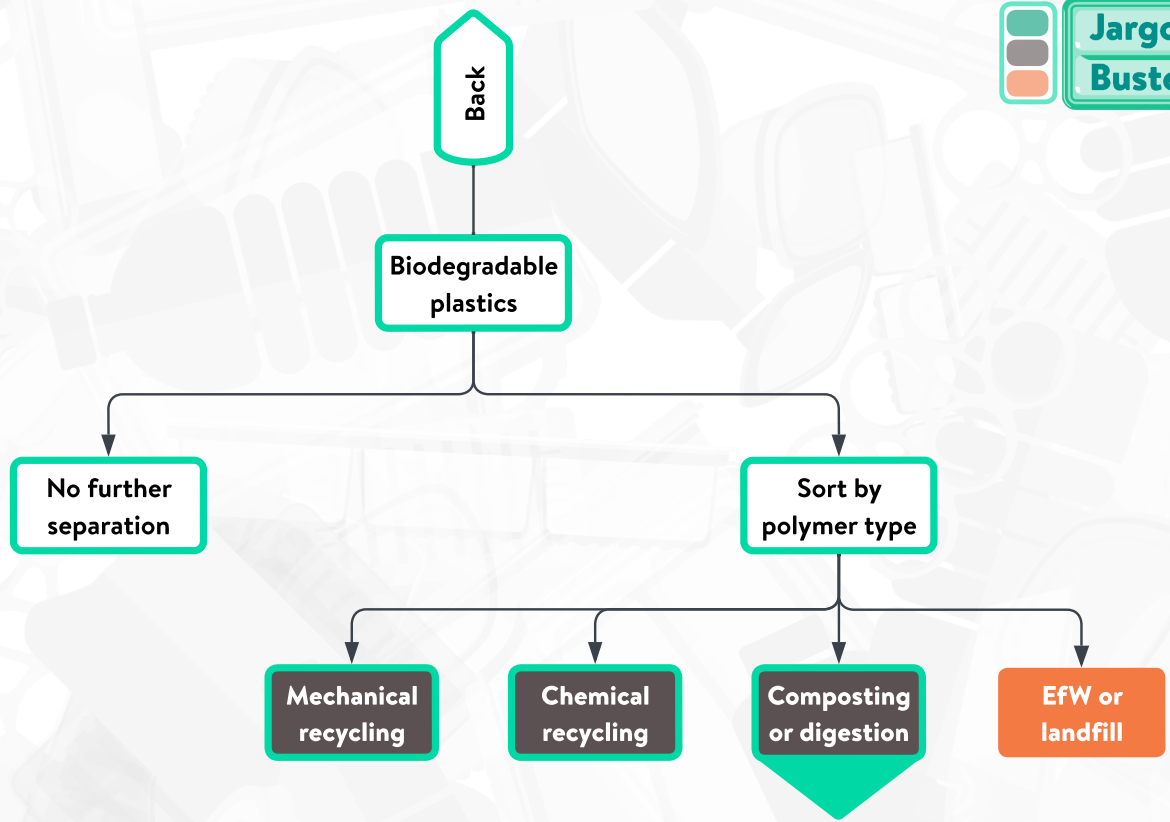

For plastic types that cannot be mechanically or chemically recycled, composting or anaerobic digestion must be considered. Sorting by polymer type can provide a more controlled composting or digestion process compared to processing a mixed biodegradable waste stream, but this end-of-life treatment should be developed for a mixed biodegradable plastics waste stream.

Sorting for composting is challenging due to the low volumes and high diversity within biodegradable plastics. Limiting the different types of polymers, addressing the ability to separate them from other plastics, and optimising their end-of-life treatment is essential if they are to be seen as recyclable or sustainable in household waste.

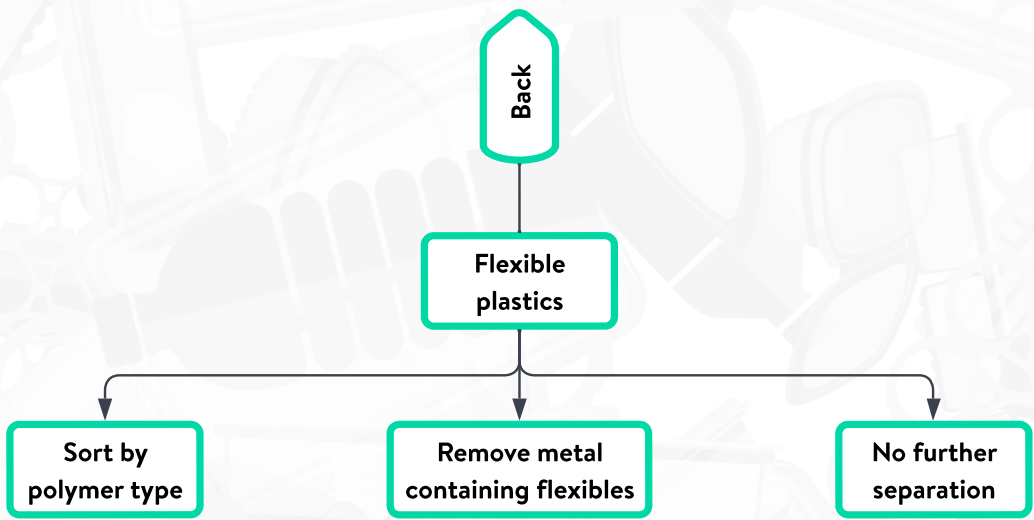

Flexibles are currently not recycled, and large sustainability gains are being lost. While optimal sorting and recycling may be hard to achieve, continuing to do nothing is unacceptable.

To maximise the long-term sustainability gains of flexible plastics, sorting by polymer type will be preferred. Mono-material PE and PP films can be separated by NIR to provide a high quality recyclate stream.[1] But this will only be valuable if the waste stream of flexible plastics is sufficiently clear of contaminants and of a large enough volume to justify required investments. This needs to be facilitated by consistent kerbside waste collection, but also requires all plastics following 'design to recycle' guidelines. Guidelines highlight that flexibles should: contain only one polymer type (PE or PP) and minimal amounts of additives; that all additives should be tested for their effect on recycling; and that the use of any non-compatible plastic types such as PVC should be avoided.[2,3,4]

Flexible plastics put onto market in the UK consist of around 80% PE and PP, but the lack of kerbside collections make post-consumer recycling nearly non-existent.[5] Recycling of non-consumer flexible plastics, typically films without colouring and containing fewer additives or polymer mixtures, is currently 95% of all plastic film recycling in the UK, mainly due to a lack of infrastructure and kerbside collection. Nevertheless, there is clearly recycling potential![5]

Short-term sustainability gains are possible by processing all flexible plastics without further separation.[1] The current diversity in the flexible plastics put on the UK market and the prevalence of multi-material packaging (containing more than one polymer type) make sorting difficult and reduce the quality of a mixed flexibles waste stream.[5] The current disposal route towards landfill must be avoided in favour of down-cycling to wood/concrete replacements, expanding chemical recycling, and mechanically recycling where possible.[1]

With consistent kerbside collection and larger waste volumes, the removal of metal containing flexibles for separate recycling may become favourable. This can improve the quality of the mixed flexibles waste stream and allow for recovery of valuable metals.

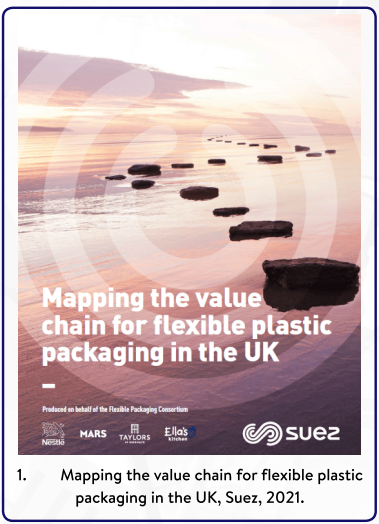

1. Mapping the value chain for flexible plastic packaging in the UK, Suez, 2021.

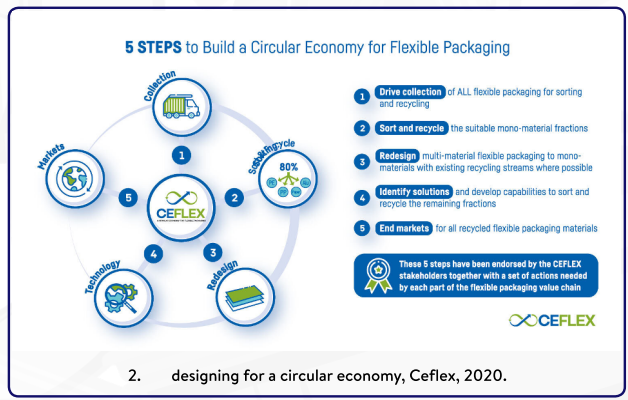

2. designing for a circular economy, Ceflex, 2020.

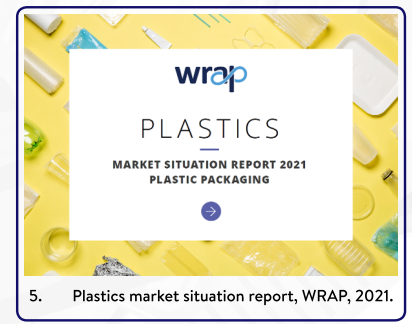

5. Plastics market situation report, WRAP, 2021.

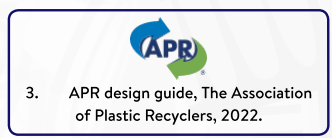

3. APR design guide, The Association of Plastic Recyclers, 2022.

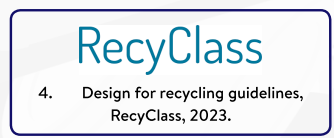

4. Design for recycling guidelines, RecyClass, 2023.

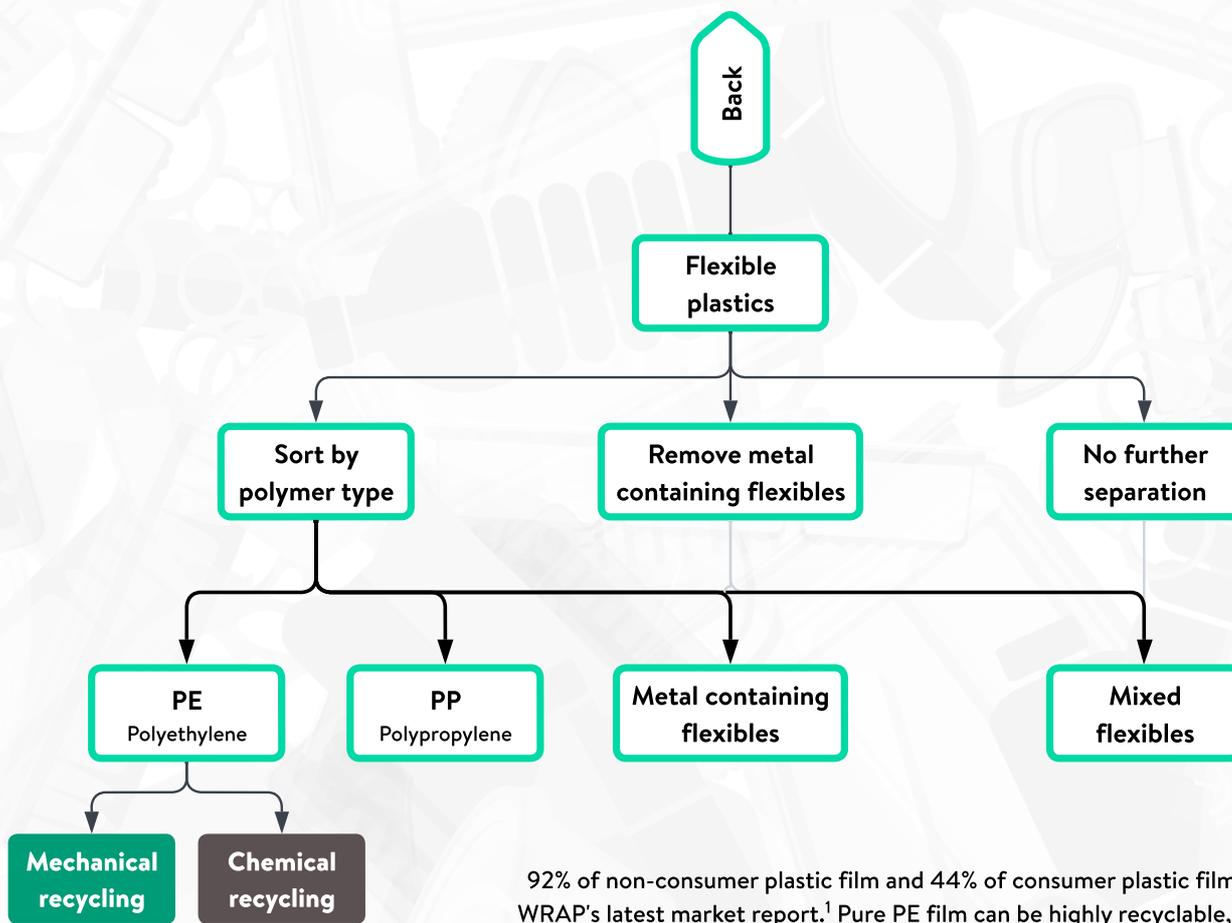

92% of non-consumer plastic film and 44% of consumer plastic film is PE, based on WRAP's latest market report.[1] Pure PE film can be highly recyclable, so its use should be preferred over other polymer types. 'Design for recycling' guidelines recommend using only mono-material PE (or PP) to improve consistency in flexibles waste, aid sorting and improve recyclate quality.[2]

Mechanical recycling of PE film is commercially viable, and already occurs for non-consumer waste streams where high purity waste is available. To expand this to consumer waste, significant capacity increases will be needed across the UK, as well as vast improvements in flexible waste composition.[3] The wide variety of plastic types in flexibles and the difficulty in sorting them, make polymer type specific processing highly unlikely. In the short-term, it is more important to target recycling of mixed flexibles, until 'design for recycling' becomes universally adopted.

A series of LCAs collated by WRAP confirm that the preferred end-of-life recycling choice for PE films is mechanical recycling.[4] As material degrades further from repeated recycling, the preferred end-of-life outcome should be chemical recycling, as suggested by the Birmingham Plastics Network and highlighted by Suez.[3,5]

1. Plastics market situation report, WRAP, 2021.

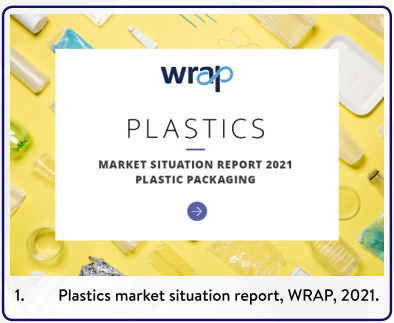

2. designing for a circular economy, Ceflex, 2020.

4. The Plastics Waste Hierarchy, WRAP, 2022.

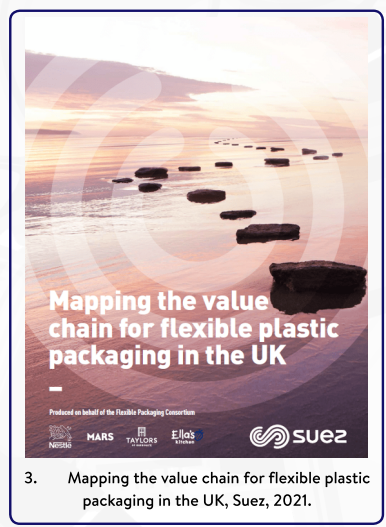

3. Mapping the value chain for flexible plastic packaging in the UK, Suez, 2021.

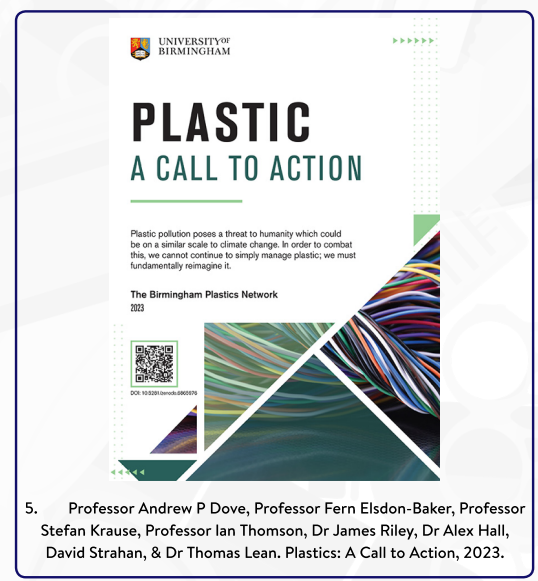

5. Professor Andrew P Dove, Professor Fern Elsdon-Baker, Professor Stefan Krause, Professor Ian Thomson, Dr James Riley, Dr Alex Hall, David Strahan, & Dr Thomas Lean. Plastics: A Call to Action, 2023.

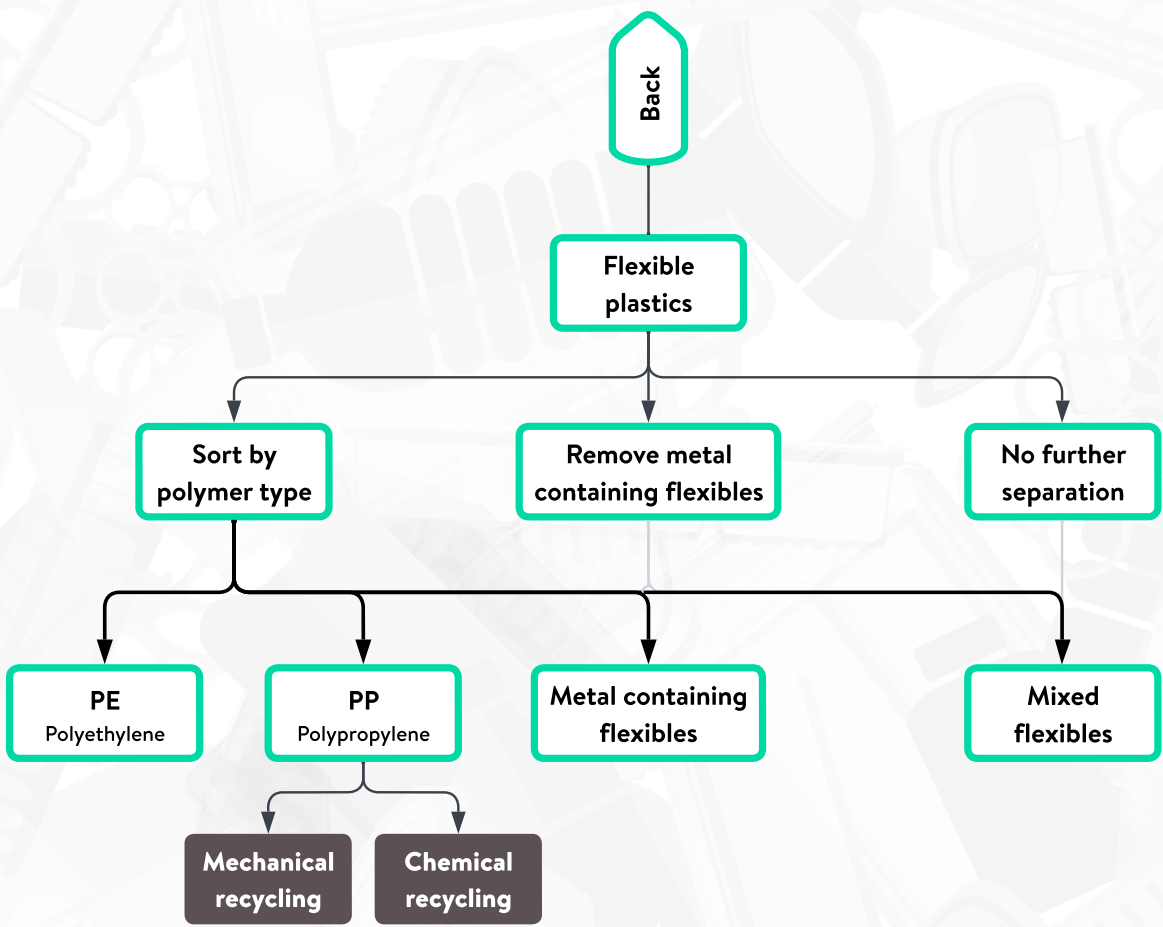

6% of non-consumer plastic film and 26% of consumer plastic film is PP, based on WRAP's latest market report.[1] Mechanical recycling of PP film is possible with existing infrastructure, though this is currently rare because segregated PP film waste is barely available.[3]

'Design for recycling' guidelines again suggest mono-material PP film with minimal to no additives and colourings, as with PE films.[2] Mechanically recycled PP film is often of lower quality than virgin PP films, similar to the output of unsorted flexibles recycling, so PE should be favoured where possible.[4] The bulkiness and tendency of flexible plastic waste to tangle makes sorting by polymer type difficult, meaning recycling of mixed flexible waste remains a more immediate priority and easier win.

LCAs collated by WRAP confirm that the most sustainable end-of-life recycling choice for PP films is still mechanical recycling.[5] As with PE film chemical recycling is the preferred fallback option when mechanical recycling is not possible, which is also reflected in mixed plastic recycling.[3]

1. Plastics market situation report, WRAP, 2021.

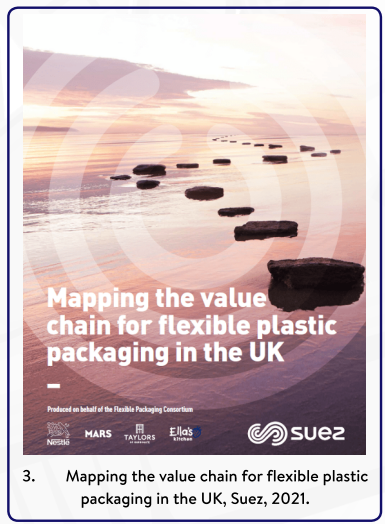

3. Mapping the value chain for flexible plastic packaging in the UK, Suez, 2021.

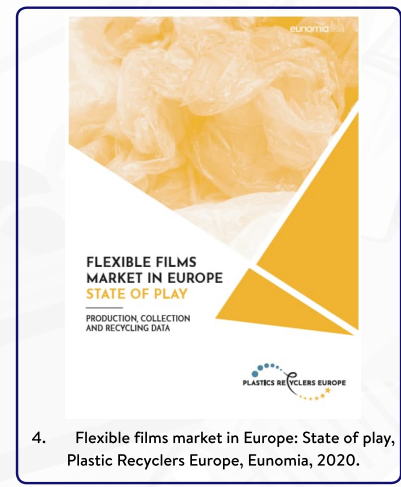

4. Flexible films market in Europe: State of play, Plastic Recyclers Europe, Eunomia, 2020.

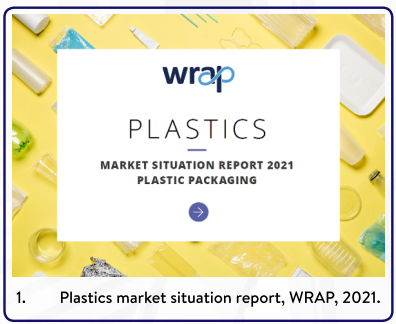

2. designing for a circular economy, Ceflex, 2020.

5. The Plastics Waste Hierarchy, WRAP, 2022.

```
V0.9 - public beta
Email any comments to Kris
```

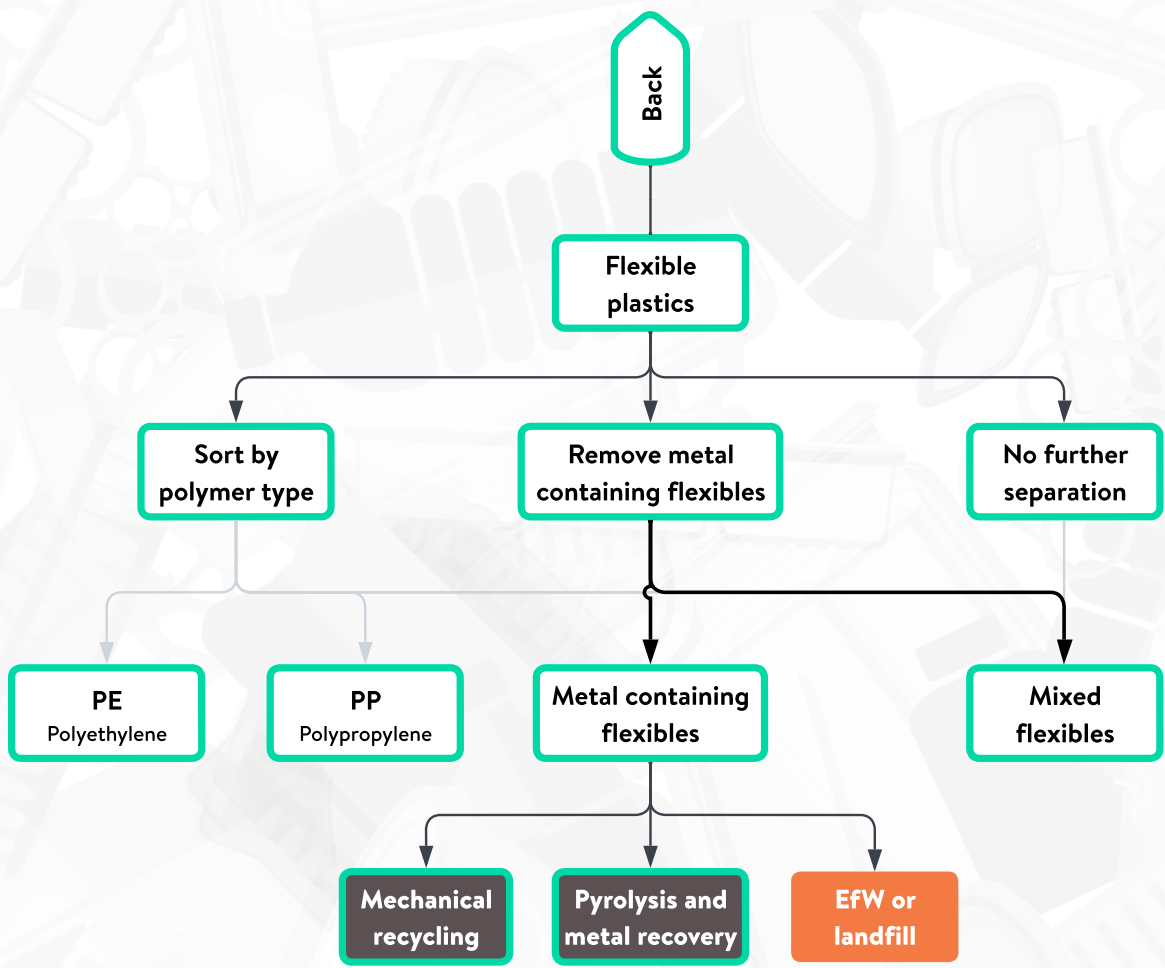

Metallised films and plastic/aluminium laminates are multi-material flexibles, which contain a metal layer that forms a problematic source of contamination in the recycling of flexible plastics.

They are commonly used for: food and drinks pouches; crisp packets; toothpaste tubes and candy wrappers. Reduction in the use of these materials should be highly encouraged, though their desired properties renders this difficult.

A feasibility study by WRAP has shown that the separation of aluminium containing flexibles can easily be achieved using existing sorting methods of eddy current separation, optical sorting, and air separation.[1] These materials are currently landfilled, but recycling of aluminium-polymer laminates is crucial for sustainability progress.

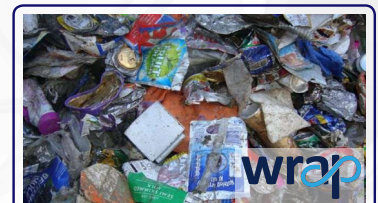

1. Recovery of laminated packaging from black bag waste, Wrap, 2012.

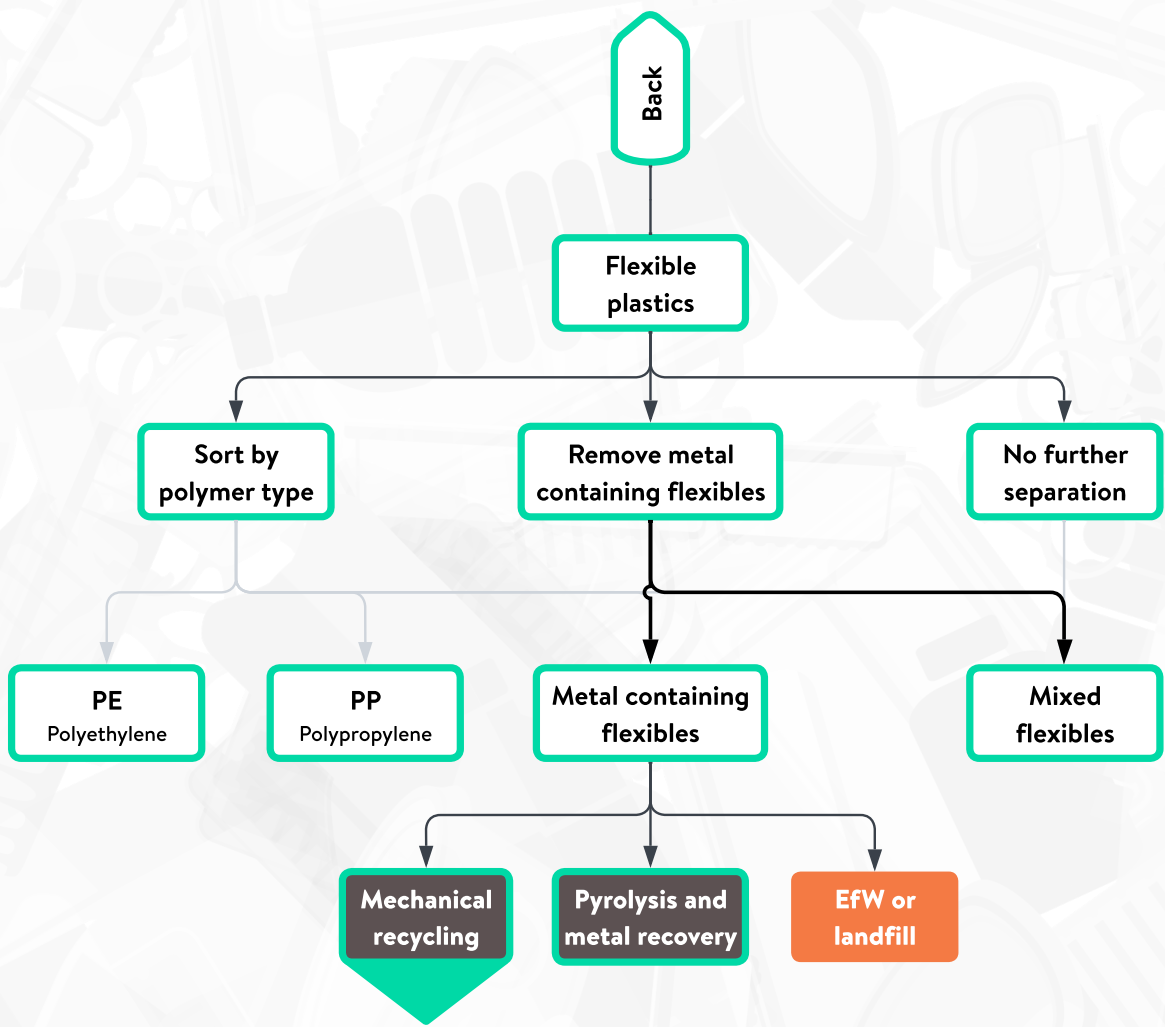

Mechanical recycling of the plastic in aluminium containing flexibles would be the most sustainable end-of-life option, though this is currently not technically feasible and research is limited. Industrial scale pyrolysis can be used to recycle the aluminium fraction, but the plastic fraction is not recycled in this process but rather converted to energy.

To mechanically recycle both the plastic and aluminium fractions, the layers must be separated, which is where the main challenge lies. Research into aseptic carton recycling, where a polymer aluminium mixture is also found, has shown the potential of solvent based separation followed by mechanical recycling, though only on lab scale currently.[2,3,4] Scale-up of potential solvent separation systems are still in their infancy, and hurdles still remain in the processing of aluminium containing flexibles.[5,6]

Guidelines for recyclable carton design by ACE UK, as well as the plastics design guide by the Association of Plastic Recyclers highlight potential design improvements, but currently don't describe aluminium containing laminates. These 'design for recycling' guidelines need to be expanded to include best practice advice for plastic/aluminium laminates, with the aim of enabling mechanical recycling while maintaining the pyrolysis end-of-life fate in the meantime.

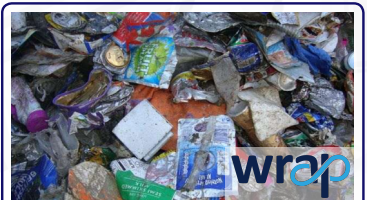

1. Recovery of laminated packaging from black bag waste, Wrap, 2012.

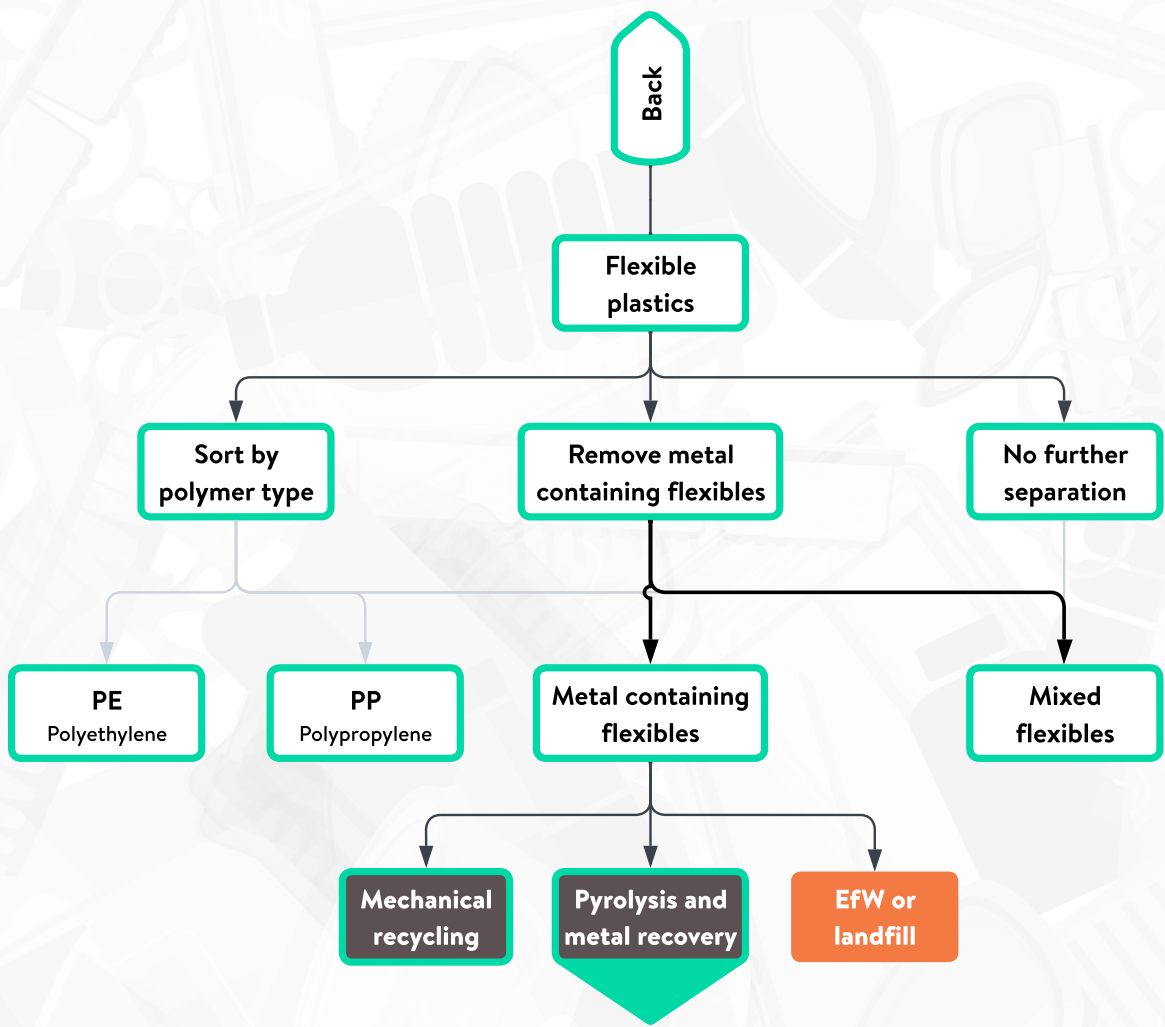

An end-of-life fate that is currently applicable on a large scale is microwave enabled pyrolysis through the Enval process (patent US20080099325A1). Plastic is converted to oils/gasses for the production of energy and the aluminium is recovered in flake form. A WRAP report[2] on this process showed it to be commercially viable if kerbside collection and sorting was rolled out and substantial sustainability gains were demonstrated compared to landfilling, while a recent Suez report[3] also highlights this method as a good end-of life outcome.

The total carbon emissions associated with this recycling system are approximately half of the emissions from the production of primary aluminium alone. Additionally, there are depletable resource savings by recycling aluminium, there is a net energy production through the oils produced in pyrolysis of the plastics, and landfilling of the material is avoided.[2] Further sustainability gains could be made by recycling the plastic fraction of the material, though this will require further research.

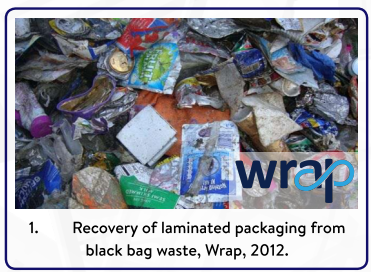

1. Recovery of laminated packaging from black bag waste, Wrap, 2012.

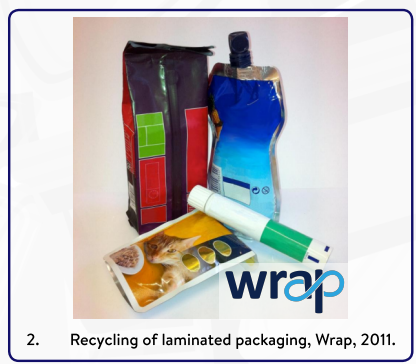

2. Recycling of laminated packaging, Wrap, 2011.

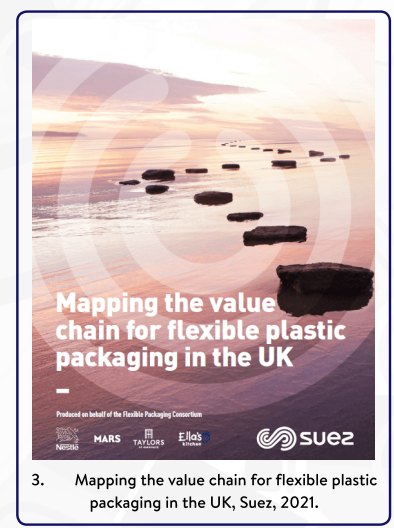

3. Mapping the value chain for flexible plastic packaging in the UK, Suez, 2021.

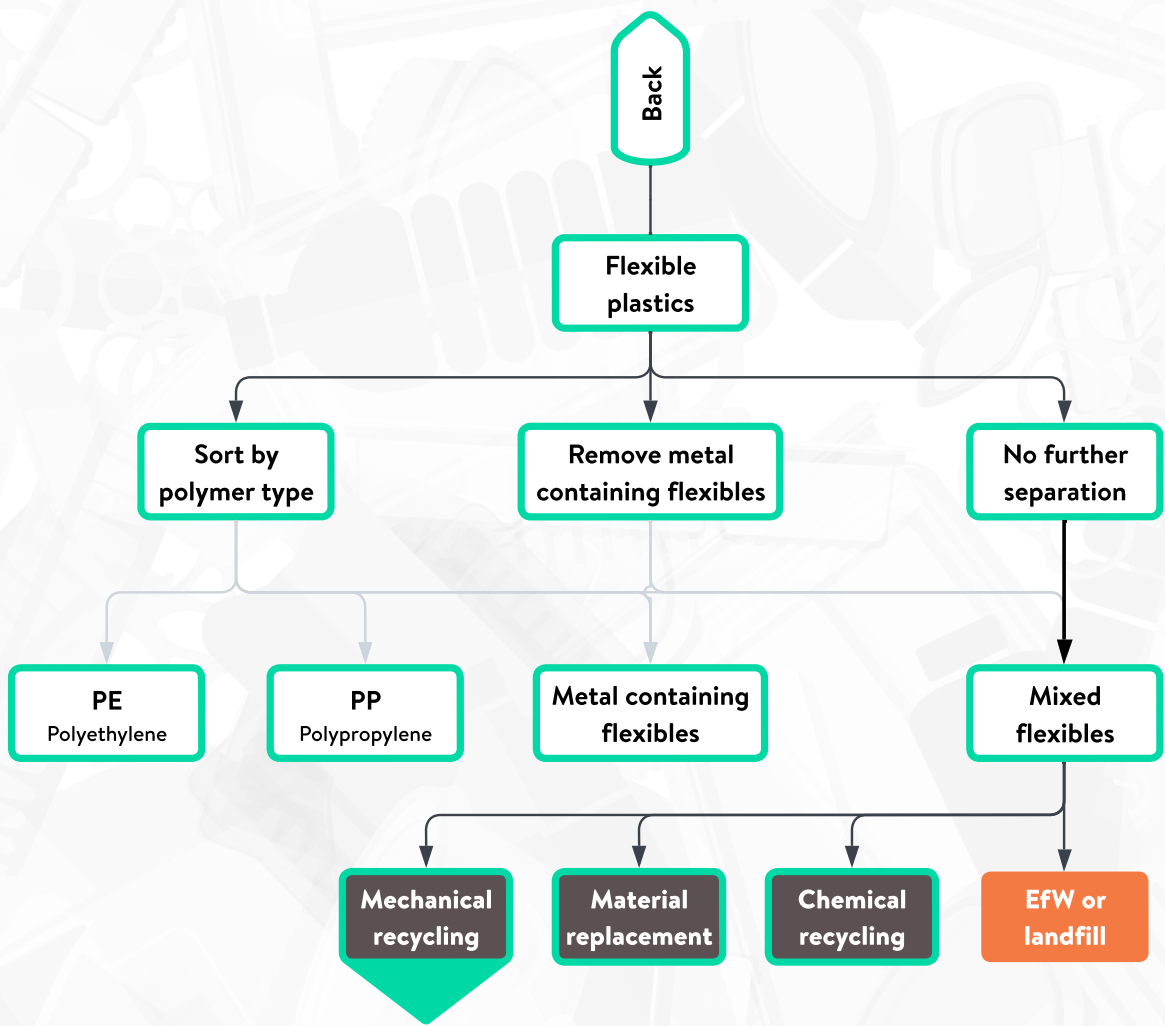

The prohibitive investment cost of further flexible plastic sorting means recycling as mixed flexibles should be targeted to provide significant short-term sustainability gains compared to the current landfilling and incineration.[1]

LCAs have shown that mechanical recycling is the preferred end-of-life outcome for all plastic products when feasible, including mixed plastic streams.[2] The output from mechanical recycling of a mixed flexibles stream will partially be able to replace virgin flexibles feedstock as well as provide 'down-cycled' plastic products with residual value.[1] The desired outcome of down-cycling should be less valuable flexible applications, such as refuse bags and non-food packaging. The next best outcome is then rigid plastic products like wood, concrete, and other material replacements.[1,3,4]

The quality of mixed flexibles mechanical recycling is severely affected by the composition of incoming waste, enforcing 'design to recycle' guidelines is essential in improving recycling quality and maximising flexible-to-flexible closed loop recycling.[5]

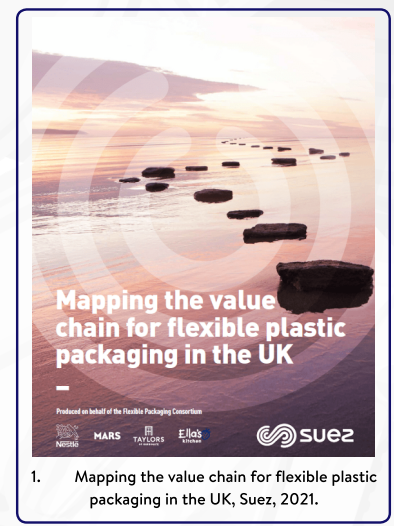

1. Mapping the value chain for flexible plastic packaging in the UK, Suez, 2021.

2. The Plastics Waste Hierarchy, WRAP, 2022.

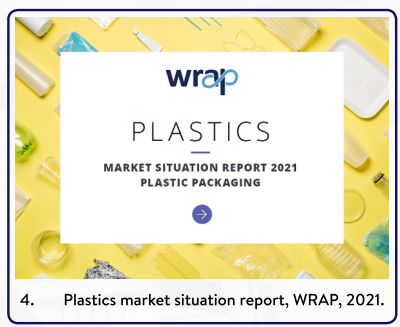

4. Plastics market situation report, WRAP, 2021.

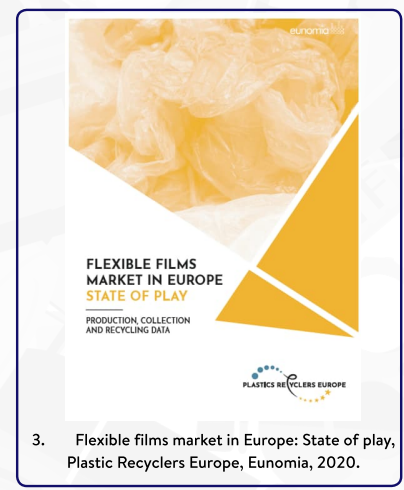

3. Flexible films market in Europe: State of play, Plastic Recyclers Europe, Eunomia, 2020.

5. designing for a circular economy, Ceflex, 2020.

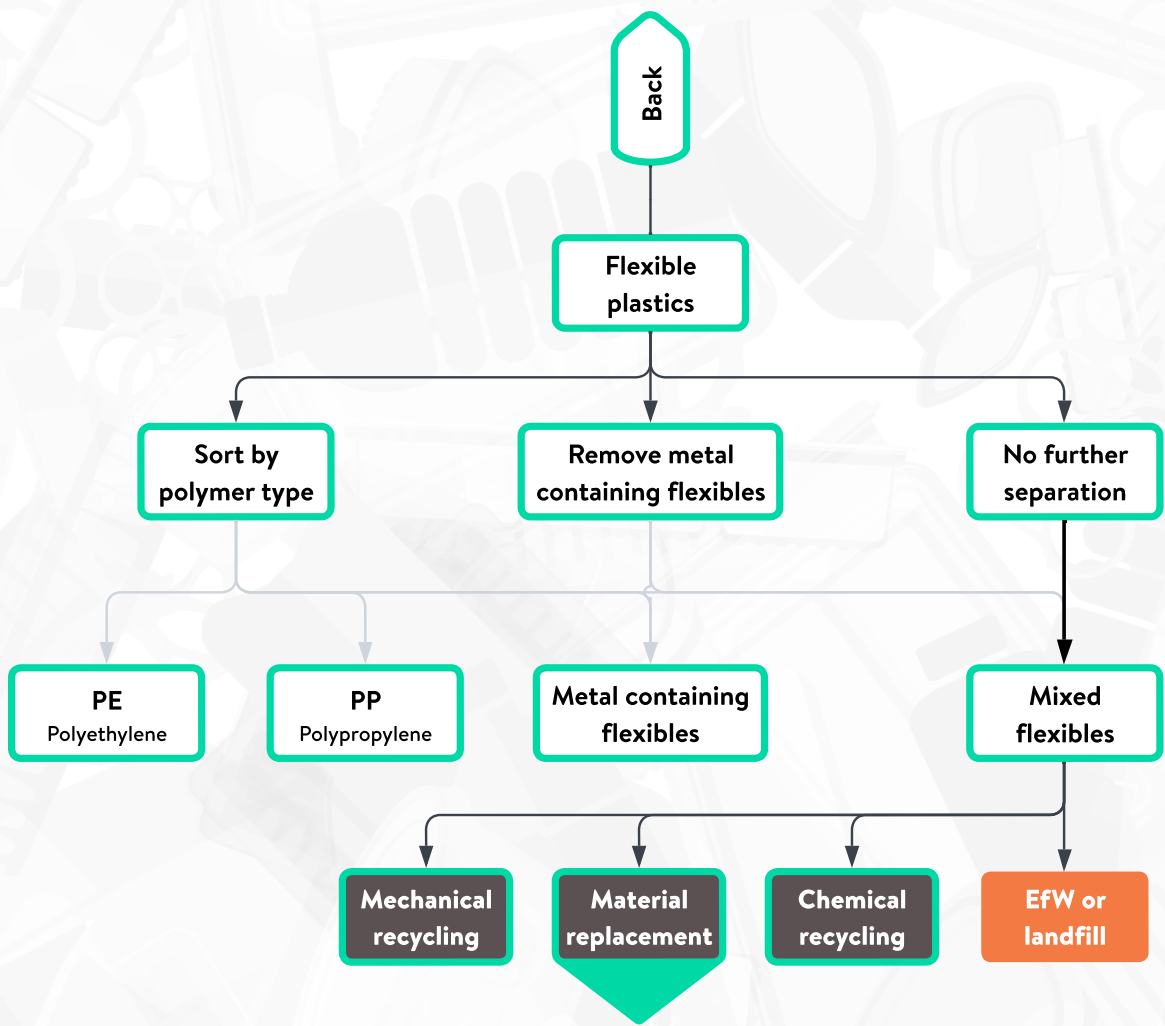

Down-cycling of flexible plastic waste into material replacements, to substitute wood or concrete for example, is a complementary end-of-life option behind mechanical recycling back into flexibles.[1] Plastic waste can be converted into a range of products such as pallets, containers, planters, fencing, furniture, playground equipment, railroad sleepers, boarding.

There is a large potential market for these products and the sustainability gains compared to landfill/incineration are significant. However, long-term focus should be on improving the waste stream quality and promote mechanical and chemical recycling aiming for a film-to-film lifecycle.

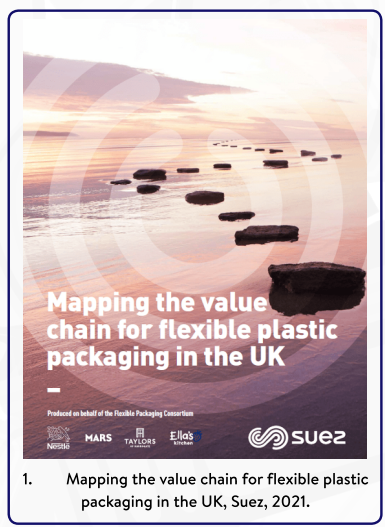

1. Mapping the value chain for flexible plastic packaging in the UK, Suez, 2021.

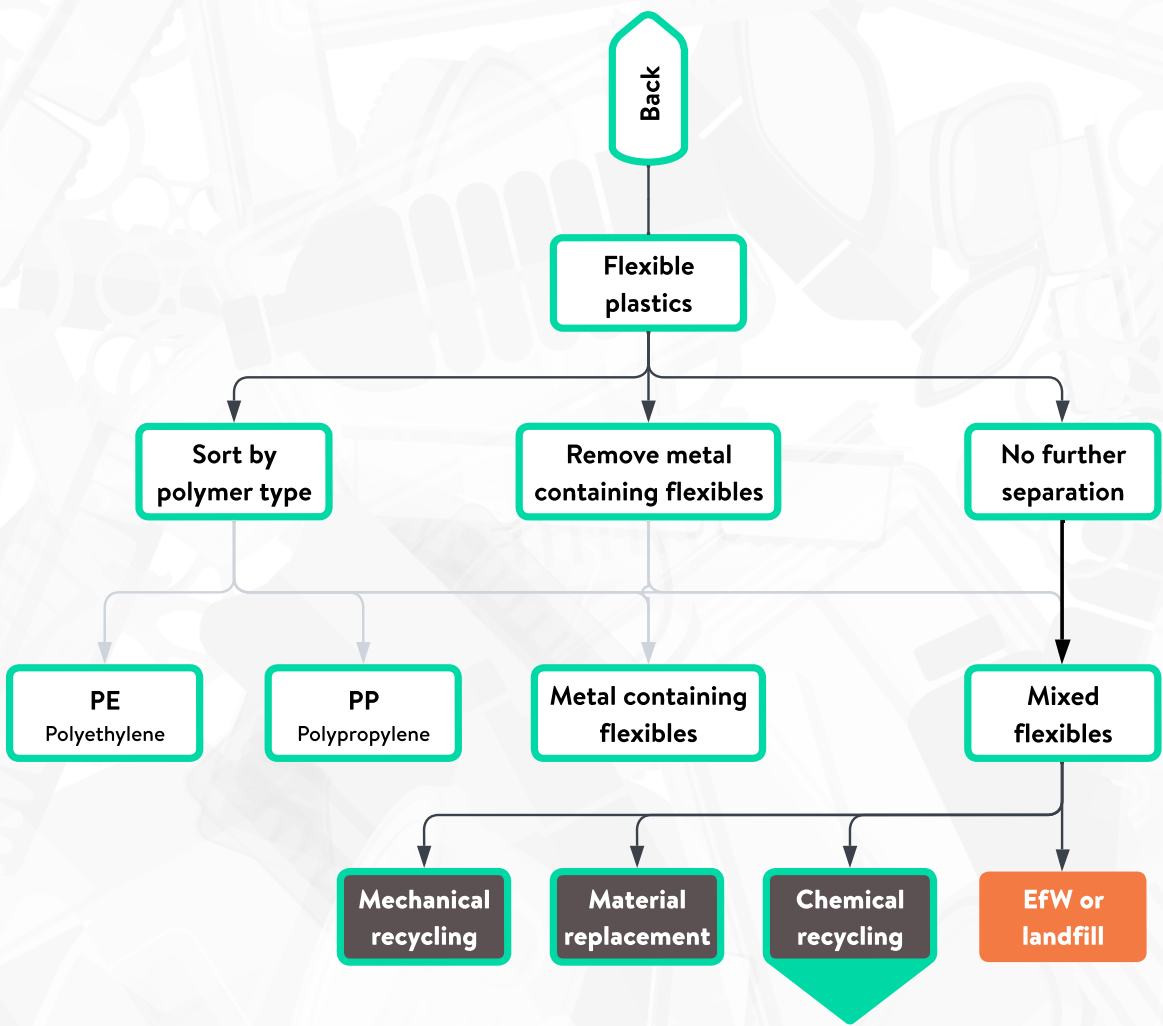

Chemical recycling of mixed flexibles can be considered as an alternative to mechanical recycling, especially in cases of difficult to recycle flexibles such as multi-material films. While chemical recycling capacity at scale develops, down-cycling to material replacements is the better fallback option to mechanical recycling.

Sustainability analysis is needed to guide the decision between non-selective chemical recycling, material replacement, and further sorting. Enforcing 'design to recycle' guidelines can improve the consistency of flexibles waste, which will influence which of these recycling fates should be preferred.[1,2]

Polyolefins (PE and PP) are the favoured plastic film types in design to recycle guidelines, but are challenging to chemically recycle due to their chemical structure. Pyrolysis or gasification, currently widely employed for EfW, can be used to recover plastic feedstock in polyolefin chemical recycling, and could become the preferred backup option after mechanical recycling.[3] The environmental impact of these techniques as a plastic-to-plastic recycling system must be scrutinised on a case-by-case basis as they develop further, because efficiency is often still low.[3,4,5] The conflation of pyrolysis in EfW and pyrolysis to feedstock must be avoided, as they are very different in terms of environmental impact and sustainability of plastics.

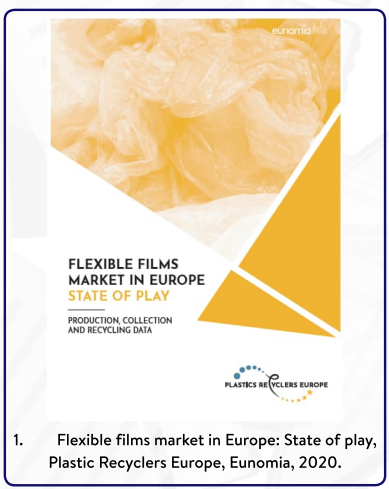

1. Flexible films market in Europe: State of play, Plastic Recyclers Europe, Eunomia, 2020.

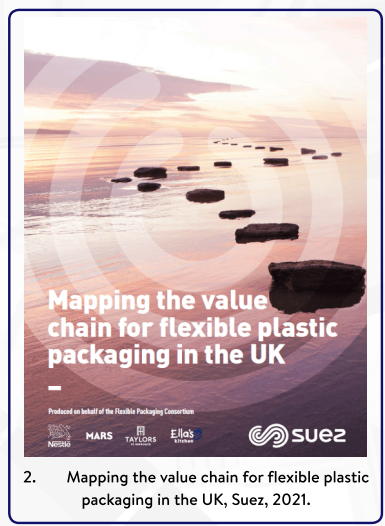

2. Mapping the value chain for flexible plastic packaging in the UK, Suez, 2021.

3. Non-Mechanical Recycling of Plastics, WRAP, 2019.

4. Redesigning the plastics system - the role of non-mechanical recycling, WRAP, 2022.

5. 7 Steps To Effectively Legislate On Chemical Recycling, Zero Waste Europe (ZWE) and the Rethink Plastic Alliance (RPa), 2020.

# Rigid plastics

- Back
  - Rigid plastics
    - Mono-material
    - **Multi-material**

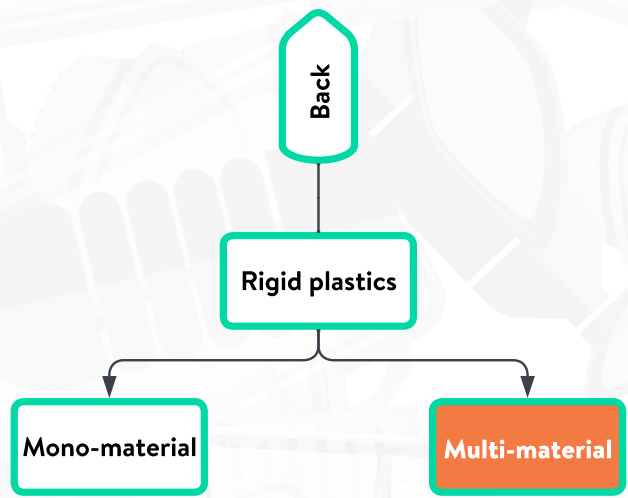

1.5 million tonnes of plastic packaging is put onto the UK market for household consumption annually, and over 78% of this is rigid plastic packaging.[1] While all local authorities offer kerbside collection of plastic bottles, 13% of councils do not collect pots, tubs, and trays (PTTs), meaning almost half of rigid plastics in household waste are sent to landfill.[1]

Both bottles and PTTs are recyclable with the sorting and processing infrastructure in place in the UK, but improvements are possible.[2] Consistent collection and sorting is needed on a national level, which will improve and stabilise recycled plastics quality.[3] Further to this, elimination of problem plastics and hard to recycle materials is needed.[4]

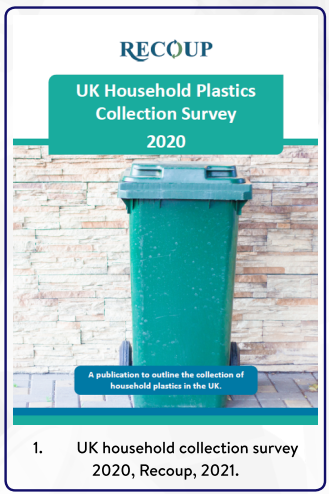

1. UK household collection survey 2020, Recoup, 2021.

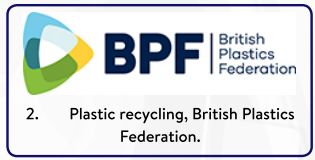

2. Plastic recycling, British Plastics Federation.

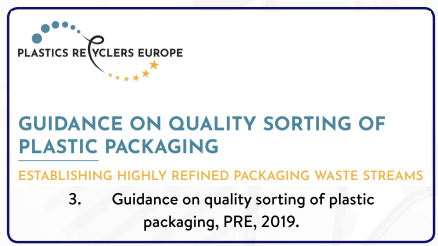

3. Guidance on quality sorting of plastic packaging, PRE, 2019.

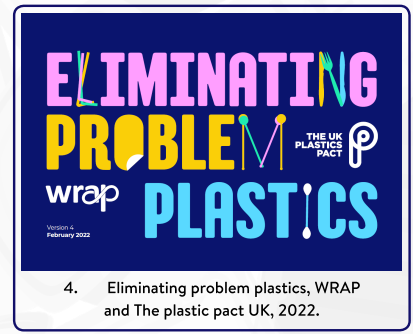

4. Eliminating problem plastics, WRAP and The plastic pact UK, 2022.

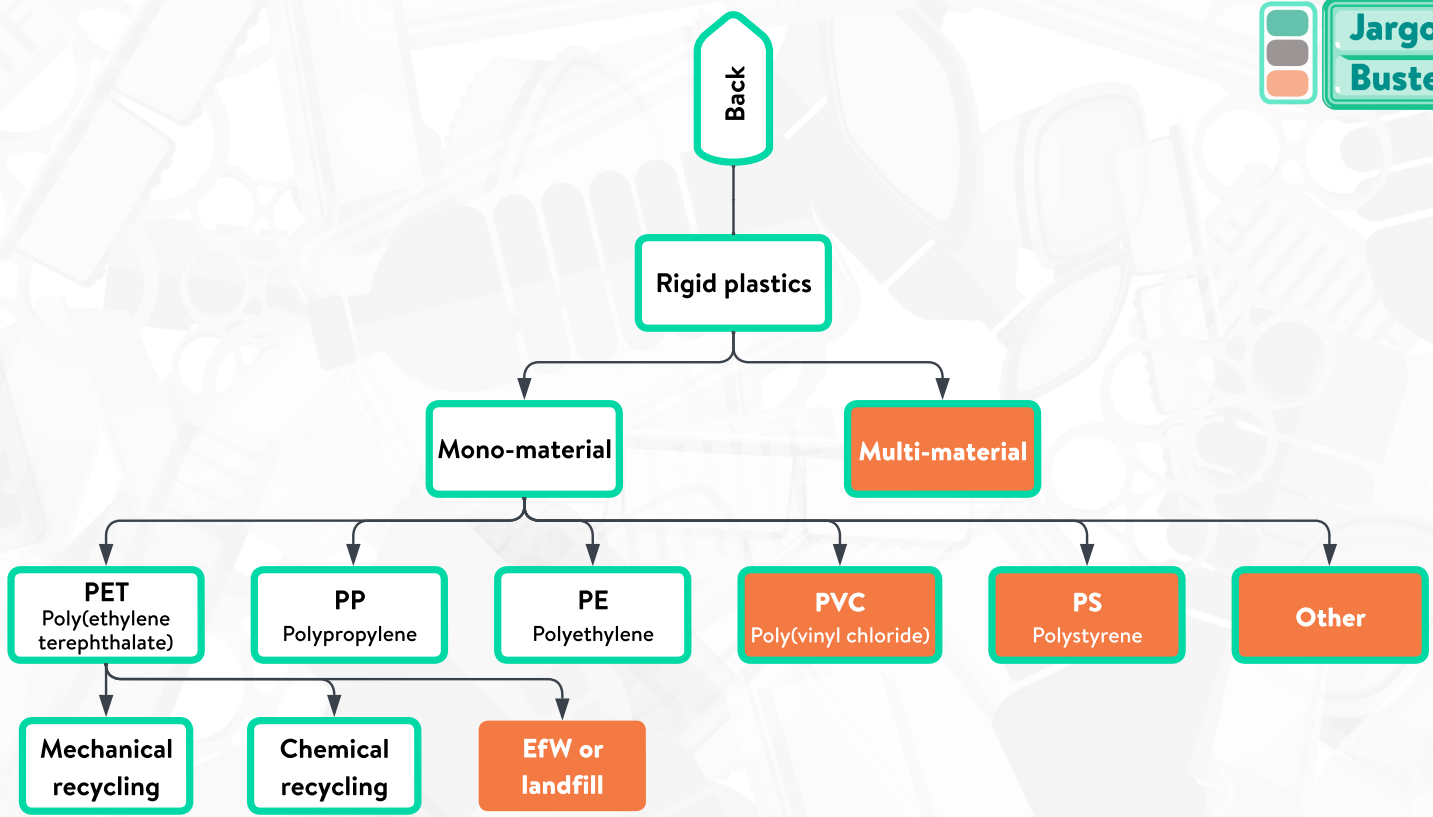

PET represents the largest fraction of rigid plastics in household waste, with both bottles and PTTs (pots, tubs & trays) often made from this plastic.[1] Due to its high density, separation of PET from PE and PP is easily achieved by float-sink separation.[2] PVC on the other hand cannot be separated out using this method and forms a severe contaminant and threat to quality PET recycling, which is why the elimination of PVC packaging is key.[3] New plastic materials can also form a risk to recycling. This must be addressed through clear legislation, assuring end-of-life fates are considered before items end up in household waste.

Further sorting of mono-material PET should be dependent on end-of-life fate. Detailed sorting is only needed when targeting the preferred mechanical recycling route. For chemical recycling routes, further sorting is not needed. While chemical recycling processes are showing promise, and can provide food-grade recyclate from all waste, they should currently be considered a fallback option for when mechanical recycling is not possible. Incineration or landfill should of course be avoided in all circumstances.

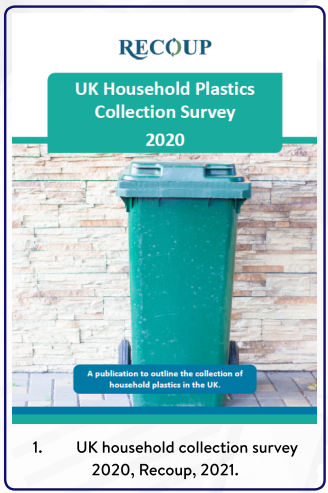

1. UK household collection survey 2020, Recoup, 2021.

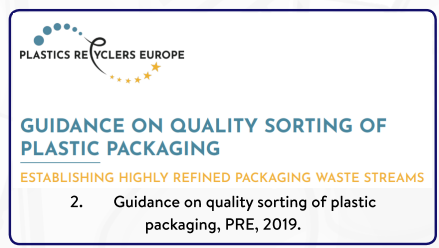

2. Guidance on quality sorting of plastic packaging, PRE, 2019.

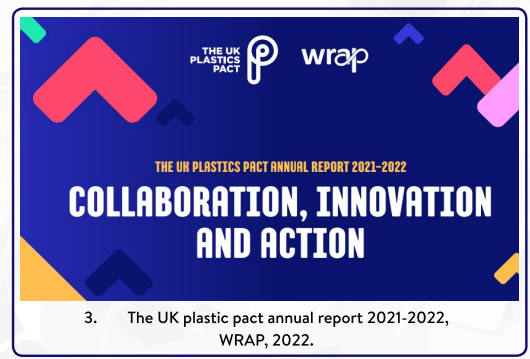

3. The UK plastic pact annual report 2021-2022, WRAP, 2022.

```
V0.9 - public beta
Email any comments to Kris
```

```
Back
 │
Rigid plastics
 ├── Mono-material
 │    ├── PET — Poly(ethylene terephthalate)
 │    │    ├── Mechanical recycling
 │    │    │    ├── Bottles
 │    │    │    └── Pots, Tubs, and Trays
 │    │    ├── Chemical recycling
 │    │    └── EfW or landfill
 │    ├── PP — Polypropylene
 │    ├── PE — Polyethylene
 │    ├── PVC — Poly(vinyl chloride)
 │    ├── PS — Polystyrene
 │    └── Other
 └── Multi-material
```

Jargon Buster

Over 70% of PET rigids are plastic bottles and a substantial mechanical recycling system is in place for them across the UK and Europe.[1] Sheet based PET is widely used for food packaging as PTTs, accounting for around 25% of PET rigids put on the market.[1]

Compared to PET bottles, the collection and recycling rates of PET PTTs are lower, despite their good recyclability. The main factor historically preventing the adoption of mechanical recycling of PET PTTs is that an estimated 60% were multi-material plastics, rendering them practically unrecyclable.[1] Thankfully, the prevalence of multi-materials is decreasing rapidly and PET PTT recycling should be encouraged.[2]

Before closed-loop mechanical recycling, PET PTTs and bottles must be separated due to their different viscosity.[1,3] Mixed recycling would result in a more brittle PET, unsuitable for high value PET bottles.[3]

Advanced sorting to separate food and non-food plastics could improve value and unlock closed loop recycling of food packaging, but innovation in sorting technology to improve efficiency and eliminate cross contamination is needed. Tagging of food safe plastics could enable this, with several pilot projects showing promise.[4,5,6,7] Legislation will be needed to ensure consistency across plastics waste and drive this emerging technology.

 ├── Food
 └── Non-food

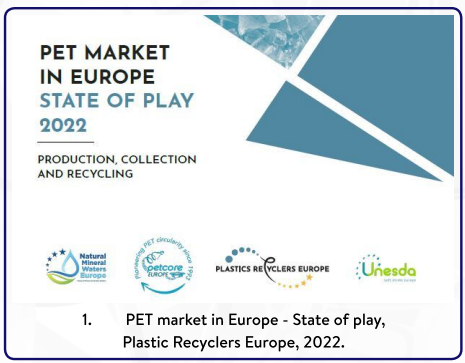
1. PET market in Europe - State of play, Plastic Recyclers Europe, 2022.

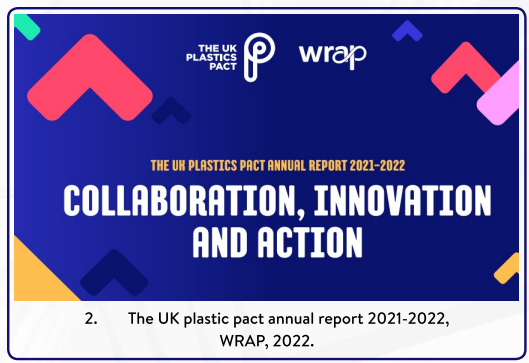
2. The UK plastic pact annual report 2021-2022, WRAP, 2022.

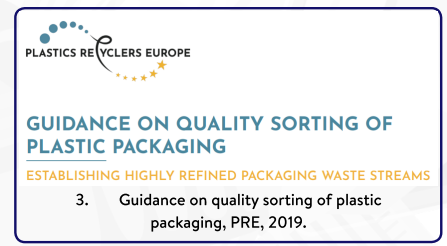
3. Guidance on quality sorting of plastic packaging, PRE, 2019.

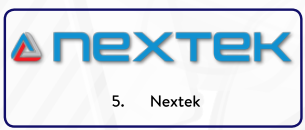
5. Nextek

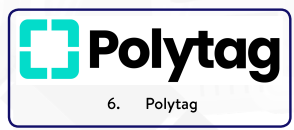
6. Polytag

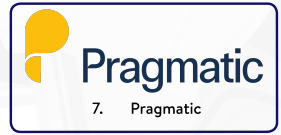
7. Pragmatic

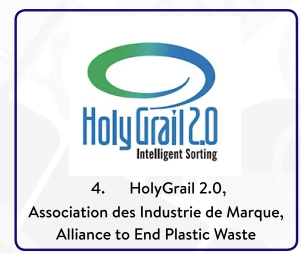
4. HolyGrail 2.0, Association des Industrie de Marque, Alliance to End Plastic Waste

V0.9 - public beta
Email any comments to Kris

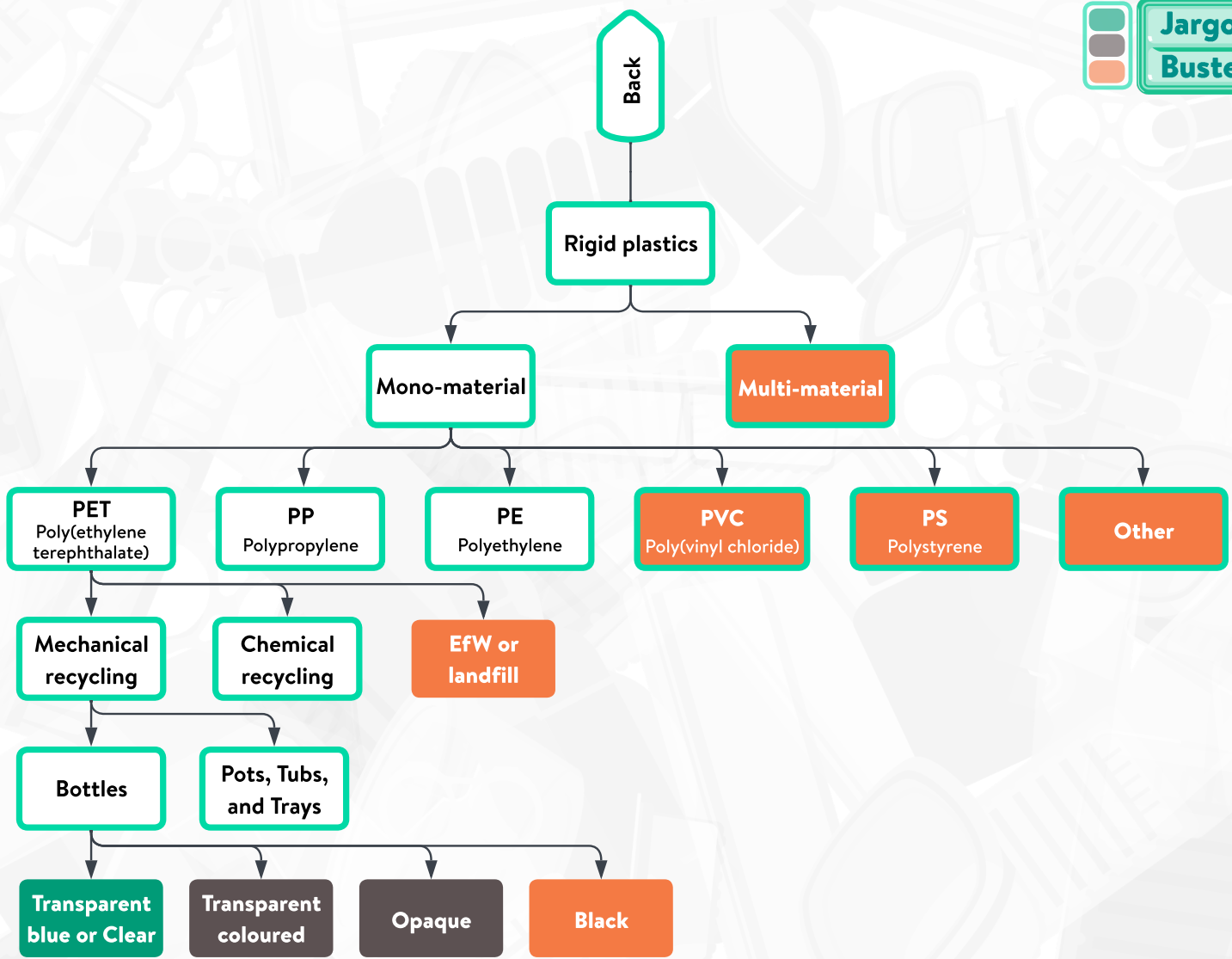

Colour sorting of PET bottles retains recyclate quality and value, avoiding mixed bottle recycling that results in inferior opaque PET (aka down-cycling).[1,2] Transparent blue and clear PET bottles are most desirable in mechanical recycling for food contact and have the highest recyclate value. Light-blue tinted and clear PET bottles are mixed, as the food-safe pigments can mask slight yellowing from mechanical recycling.[3] Other transparent light colours may be incorporated into clear PET bottle recycling, but pigments must first be assessed for their effect on recyclate quality.

Separate mechanical recycling of transparent coloured PET bottles will allow them to be reused in the same application, but low volumes of various colours limits commercial viability. Opaque PET bottles reduce the clarity of recycled PET and must be processed separately. Opaque PET can be mechanically recycled separately if enough of it is collected, but sorting costs are higher and NIR sorting systems struggle to separate it from clear PET.[3]

Improving adherence to design for recycling guidelines is essential to optimising the PET bottle waste stream further. Guidelines by EPBP, RecyClass, RECOUP, and APR, all highlight the need to minimise/avoid colour, limit non-PET features such as labels and closures when possible, and to test all additives for their compatibility with sorting and recycling systems.[3,4]

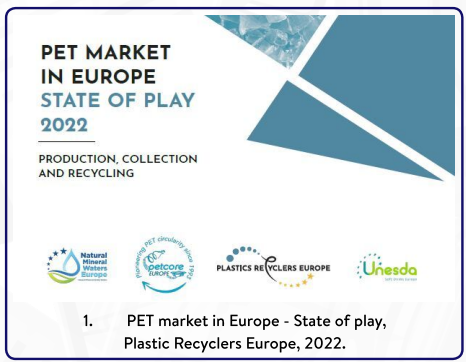

1. PET market in Europe - State of play, Plastic Recyclers Europe, 2022.

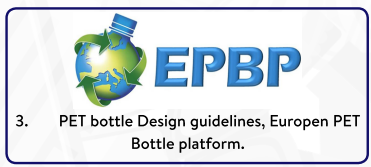

3. PET bottle Design guidelines, Europen PET Bottle platform.

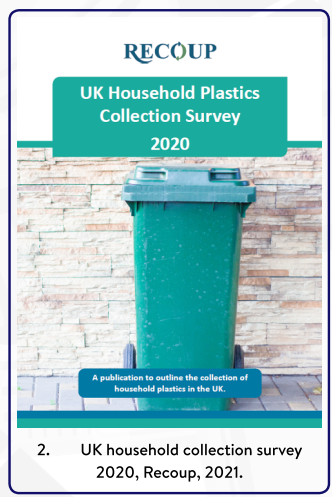

2. UK household collection survey 2020, Recoup, 2021.

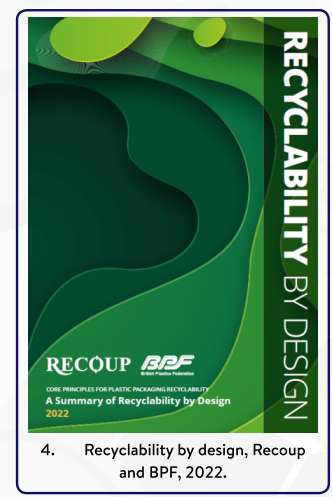

4. Recyclability by design, Recoup and BPF, 2022.

```
V0.9 - public beta
Email any comments to Kris
```

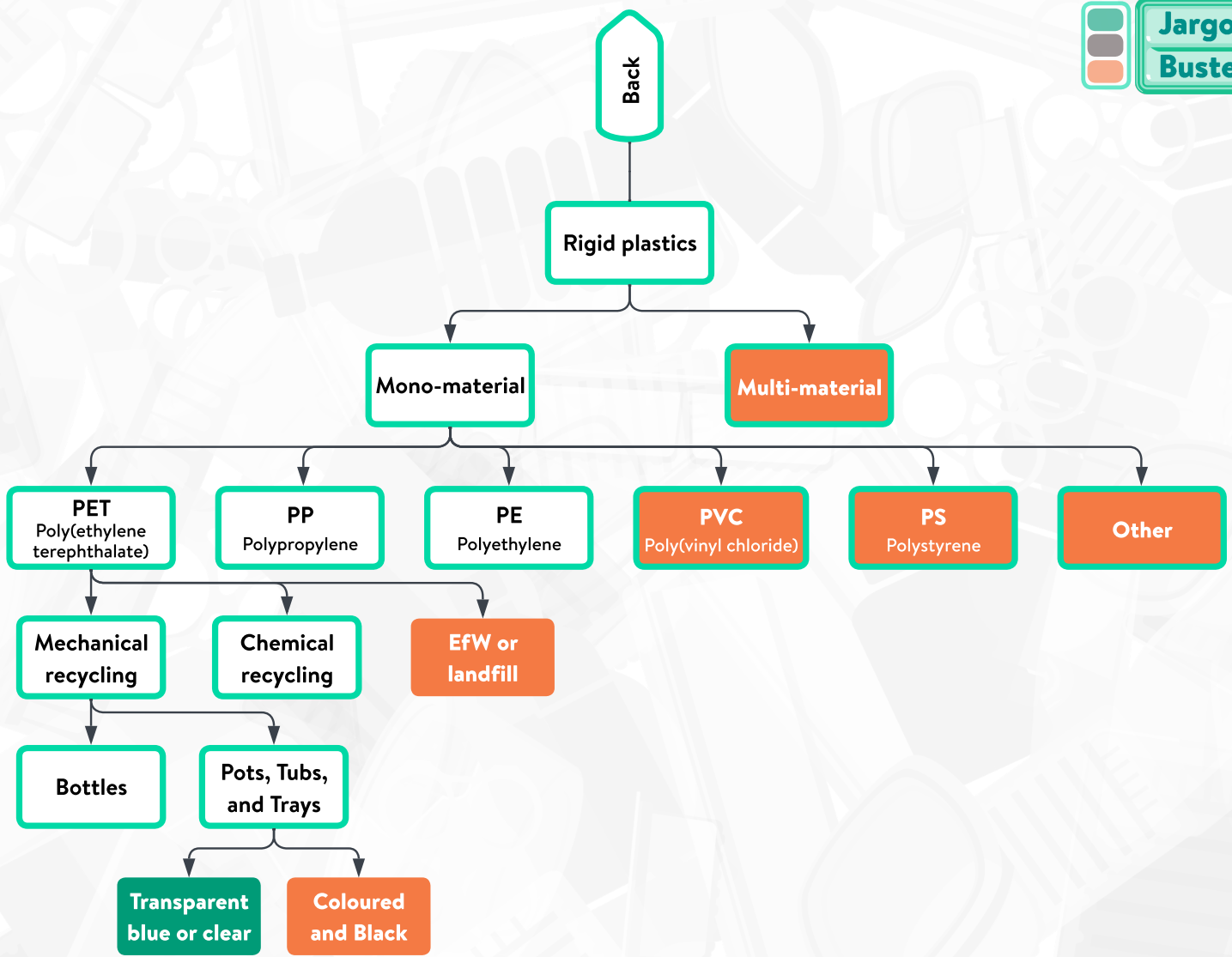

Lower collection rates and the historical prevalence of multi-material PTTs has hindered mechanical recycling of sheet based PET.[1,2] Black trays have also been very damaging to the recycling process, as they are invisible to common sorting systems and contaminate any mechanical recycling process and should be eliminated from production.[1,2,3] Coloured PET PTTs are rare, but they must be separated out to maintain high value transparent PET after recycling.

As with Bottles, design for recycling is essential in assuring high quality recyclate and avoiding detrimental contamination. Guidance advice involves avoiding colourants, using compatible (clear PET) or separable (low density PE or PP) lidding, and only applying adhesives that are fully removed during washing.[4,5,6]

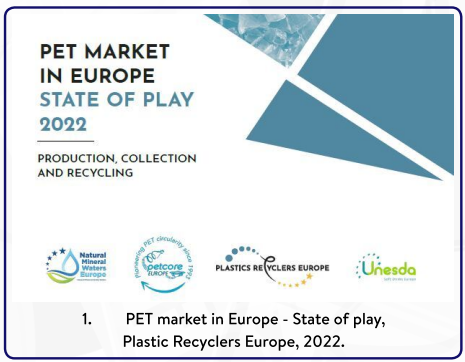

1. PET market in Europe - State of play, Plastic Recyclers Europe, 2022.

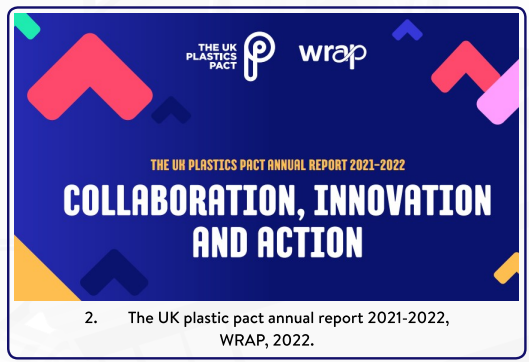

2. The UK plastic pact annual report 2021-2022, WRAP, 2022.

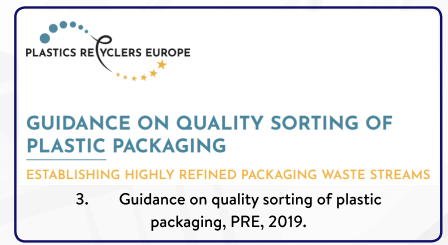

3. Guidance on quality sorting of plastic packaging, PRE, 2019.

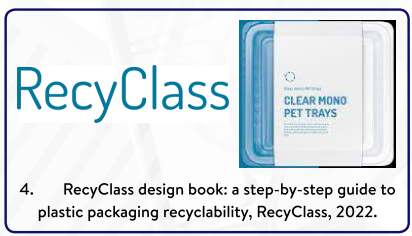

4. RecyClass design book: a step-by-step guide to plastic packaging recyclability, RecyClass, 2022.

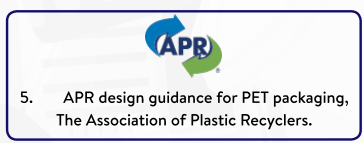

5. APR design guidance for PET packaging, The Association of Plastic Recyclers.

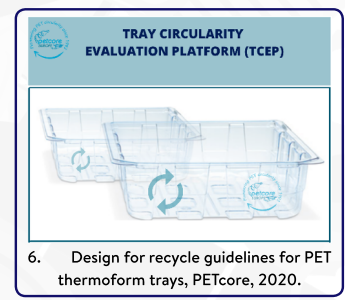

6. Design for recycle guidelines for PET thermoform trays, PETcore, 2020.

```
V0.9 - public beta
Email any comments to Kris
```

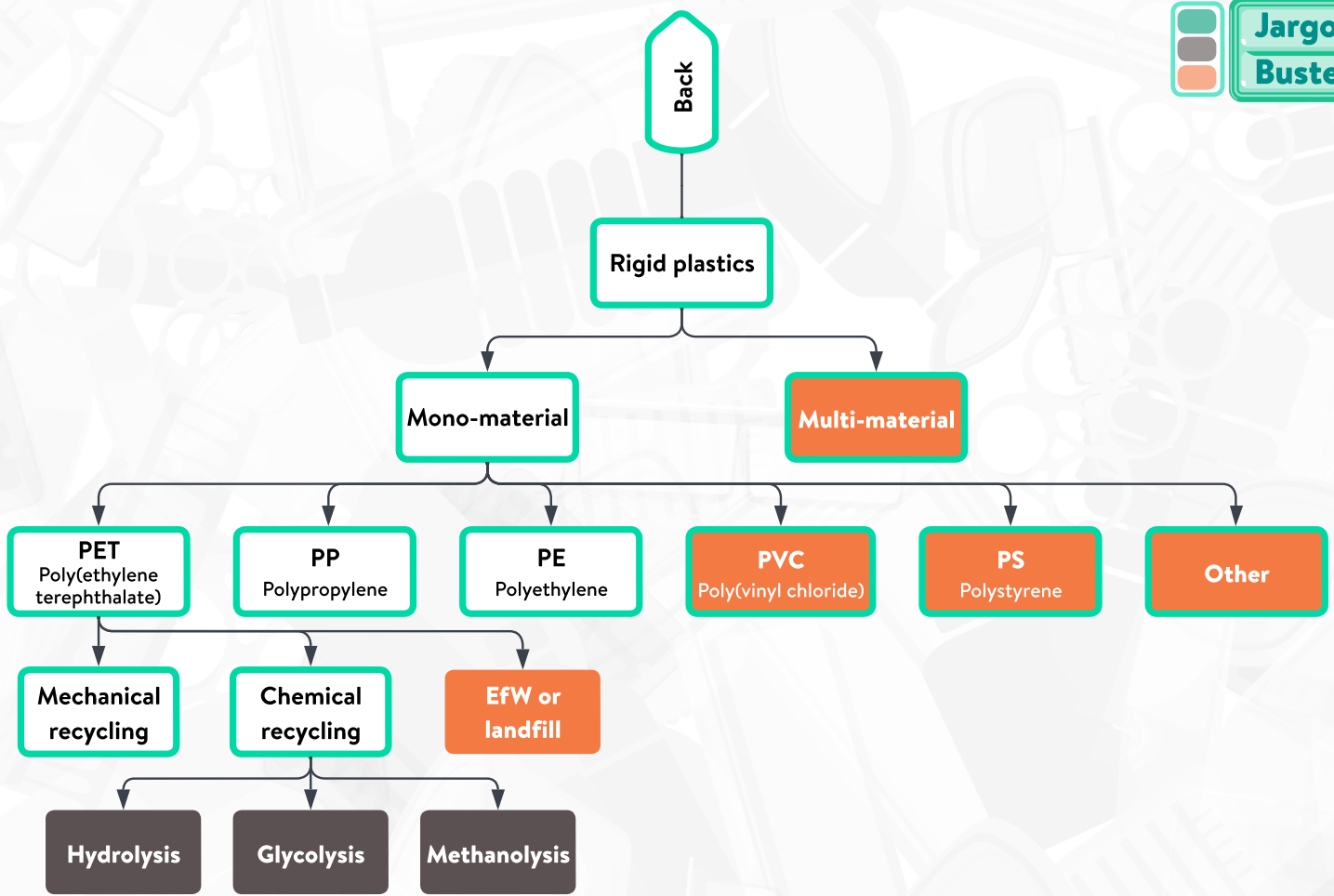

Chemical recycling of PET is a rapidly growing area of research, with several methods of depolymerisation emerging. It has the potential to provide a complementary end-of-life outcome for items that are difficult to mechanically recycle due to their composition or deterioration.[1] While it should be seen as the preferred fall-back option after mechanical recycling, the technology needs further development and critical analysis before it can be considered an 'at-large' end-of-life fate.[2]

Technical, economical, and environmental comparisons between hydrolysis, methanolysis, glycolysis, and mechanical recycling of PET have shown that chemical recycling may be a preferred end-of-life fate for highly contaminated PET waste, but the route of minimal environmental impact remains maximising the mechanical recycling potential of waste.[3] While greenhouse gas emissions of chemical recycling is higher than that of mechanical recycling, it is considerably lower than that of producing virgin PET.[1]

There is currently a need for recycled PET in food-grade applications, but no realistic method of making this from mixed PET waste streams.[2]  Regulations need to be updated in this field, to ensure that recycled PET can be safely used in food contact PET items and by extension avoid the use of multi-layers to hide recycled plastic inside items.

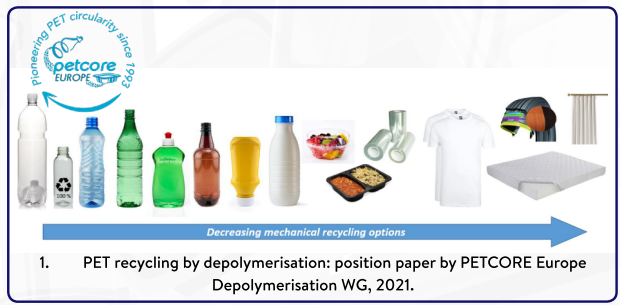

1. PET recycling by depolymerisation: position paper by PETCORE Europe Depolymerisation WG, 2021.

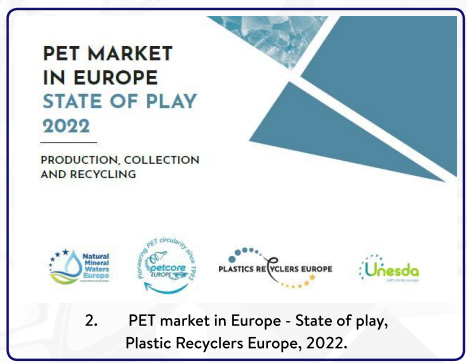

2. PET market in Europe - State of play, Plastic Recyclers Europe, 2022.

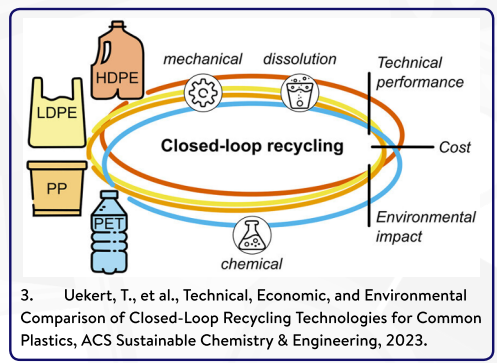

3. Uekert, T., et al., Technical, Economic, and Environmental Comparison of Closed-Loop Recycling Technologies for Common Plastics, ACS Sustainable Chemistry & Engineering, 2023.

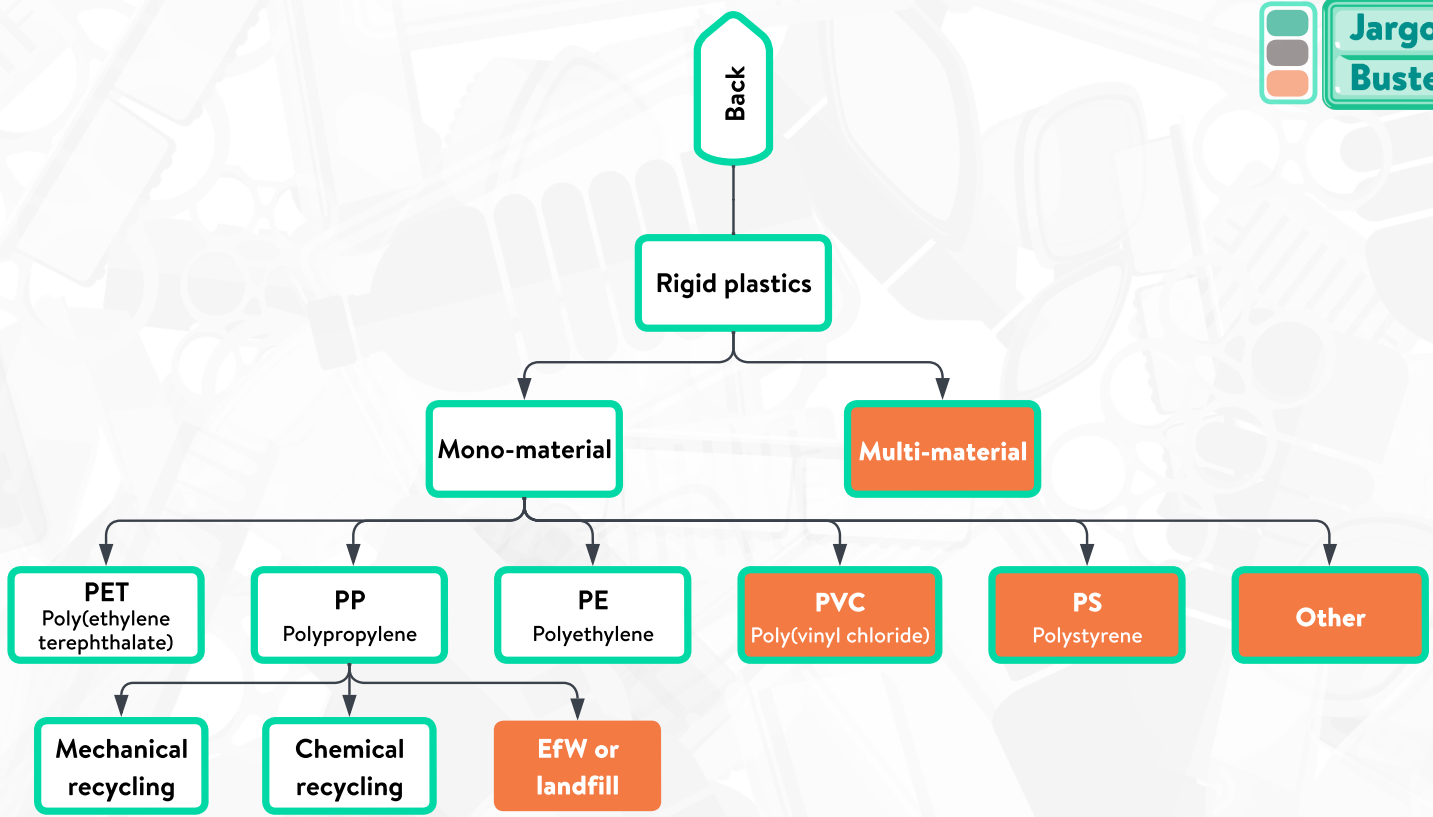

Around 20% of rigid plastics in household waste are made of PP. Over 90% of rigid PP put on the consumer market in the UK are in the from PTTs and other non-bottle applications.[1] The low density of PP (and PE) means that separation from PET and other dense plastics is often done through float-sink separation. Sorting of rigid PP from other low-density plastics, like PE, is possible with Near-Infrared technology that is widely used in the recycling industry.[2]

Further sorting of PP by colour or shape should be dependent on end-of-life fate, though larger waste volumes are needed as advanced sorting is currently not viable due to low kerbside collection rates, insufficient processing, and thus infrastructural capacity.[1,2,3]

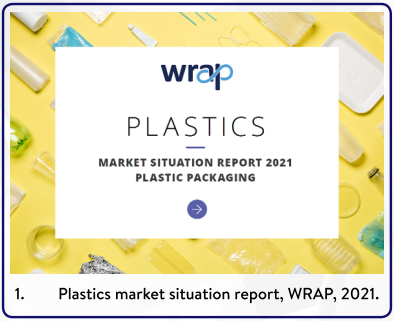

1. Plastics market situation report, WRAP, 2021.

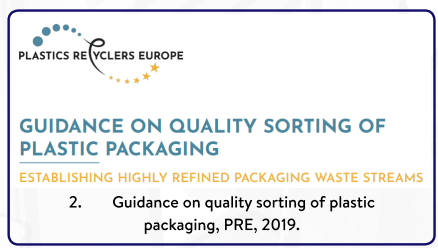

2. Guidance on quality sorting of plastic packaging, PRE, 2019.

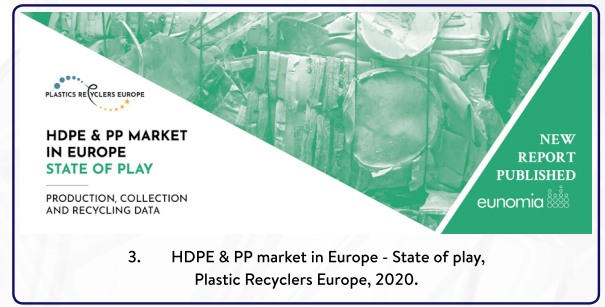

3. HDPE & PP market in Europe - State of play, Plastic Recyclers Europe, 2020.

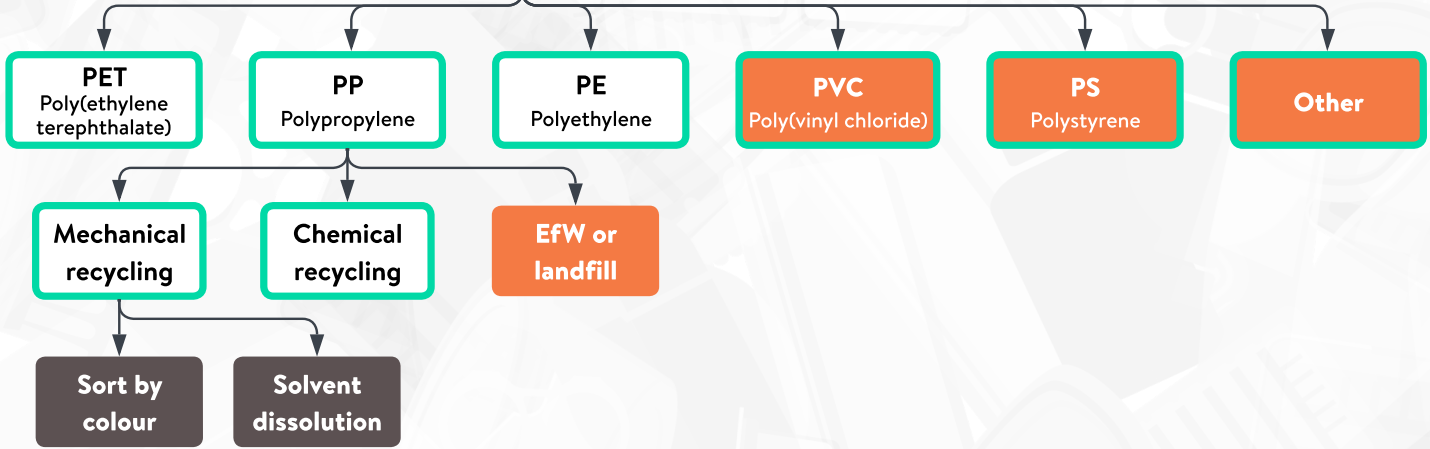

Type sorting of PP rigids (bottles, trays, etc.) is not yet valuable as they are mostly PPTs of similar viscosity, but should become part of future sorting. Separating transparent, white, and coloured items is currently not done due to low kerbside collection rates and low processing capacity.[1,2] PP is suited for mechanical recycling, and this must be the primary end-of-life fate, but degradation from continued recycling is of concern and improvements are needed to optimise this end-of-life outcome.[1,3]

Solvent dissolution, a process that removes contaminants and colours prior to mechanical recycling, is a promising way to improve the process, but the technology is in its infancy. Economic and environmental analysis has shown that solvent dissolution prior to mechanical recycling of PP can be profitable and environmentally sound, if the technology is developed sufficiently.[3] Pilot plants for solvent dissolution of PP are being developed, and showing potential.[4,5,6]

'Design for recycling' is also an important method to improve recyclability of this waste stream. It stresses the importance of minimising colourants and contaminating additives, as well as using PP in all closures and attachments to maximise recyclability and ease of sorting.[7,8]

Advanced sorting to separate food and non-food plastics could unlock closed loop recycling of high-value food packaging. Tagging of food safe plastics could enable this, with several pilot projects showing promise. Legislation will be needed to ensure consistency in this emerging technology.

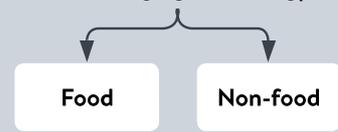

1. HDPE & PP market in Europe - State of play, Plastic Recyclers Europe, 2020.
2. Guidance on quality sorting of plastic packaging, PRE, 2019.
3. Uekert, T., et al., Technical, Economic, and Environmental Comparison of Closed-Loop Recycling Technologies for Common Plastics, ACS Sustainable Chemistry & Engineering, 2023.
4. ISOPREP
5. NEXTLOOPP
6. PureCycle
7. APR design guidance for PP, The Association of Plastic Recyclers.
8. RecyClass design book: a step-by-step guide to plastic packaging recyclability, RecyClass, 2022.

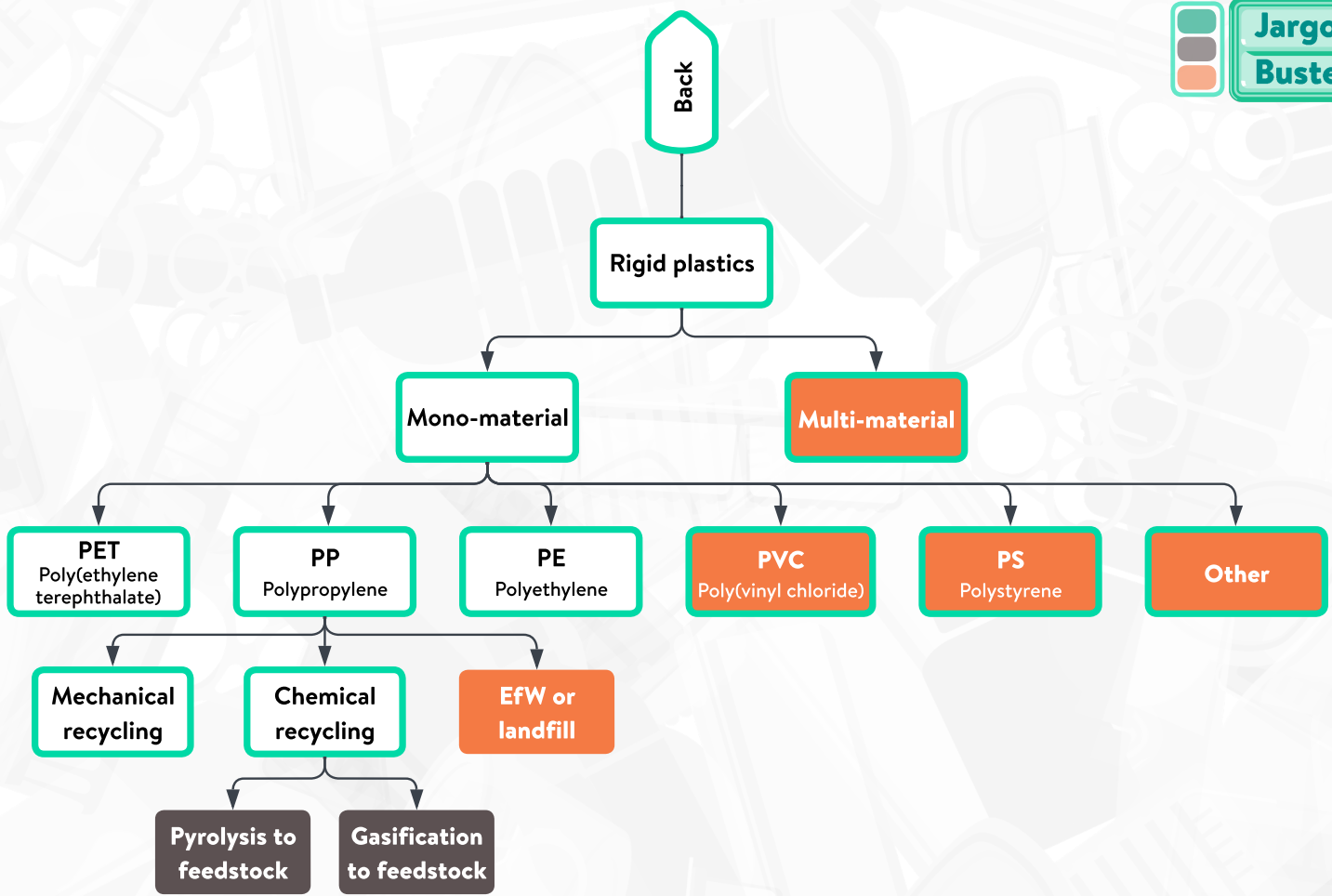

Chemical recycling of PP would not require further sorting, and many emerging methods target a mixed waste stream of polyolefins (PP and PE in both flexible and rigid form), potentially reducing the sorting burden.[1] Depolymerisation of polyolefins is not possible due to the chemical structure, so chemical recycling is limited to feedstock recycling through pyrolysis or gasification techniques.[2] These techniques are currently used for Energy from Waste (EfW), converting plastics into fuel instead of converting into feedstock to make new plastics.[1,2,3] The environmental impact of these techniques for chemical recycling must be scrutinised on a case-by-case basis as they develop further, with a focus on plastic to plastic recycling only, as EfW should be avoided if possible.[4,5]

Chemical recycling should not be viewed as an alternative to better design for recycling and large scale mechanical recycling of plastics. Rather, it should be limited to being a backup of mechanical recycling options for degraded and contaminated plastic waste.[4,5] While chemical recycling is essential in achieving sustainable closed loop recycling of plastics, environmental impact assessments clearly show it must be secondary to mechanical recycling.[6]

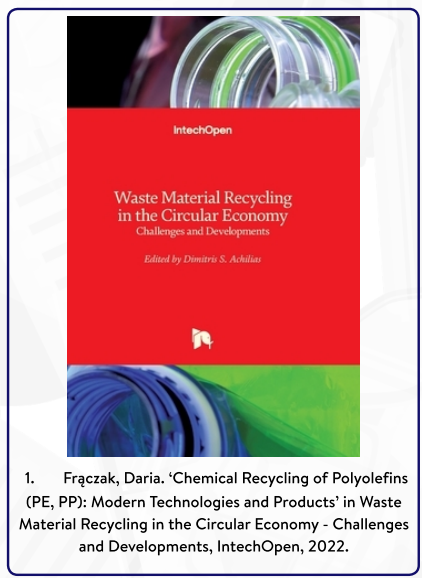

1. Frączak, Daria. 'Chemical Recycling of Polyolefins (PE, PP): Modern Technologies and Products' in Waste Material Recycling in the Circular Economy - Challenges and Developments, IntechOpen, 2022.

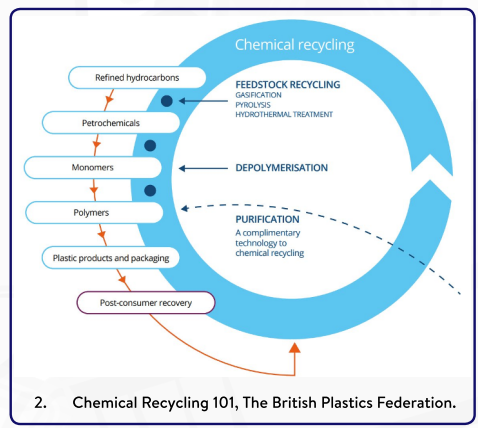

2. Chemical Recycling 101, The British Plastics Federation.

5. 7 Steps To Effectively Legislate On Chemical Recycling, Zero Waste Europe (ZWE) and the Rethink Plastic Alliance (RPa), 2020.

3. Non-Mechanical Recycling of Plastics, WRAP, 2019.

4. Redesigning the plastics system - the role of non-mechanical recycling, WRAP, 2022.

6. The Plastics Waste Hierarchy, WRAP, 2022.

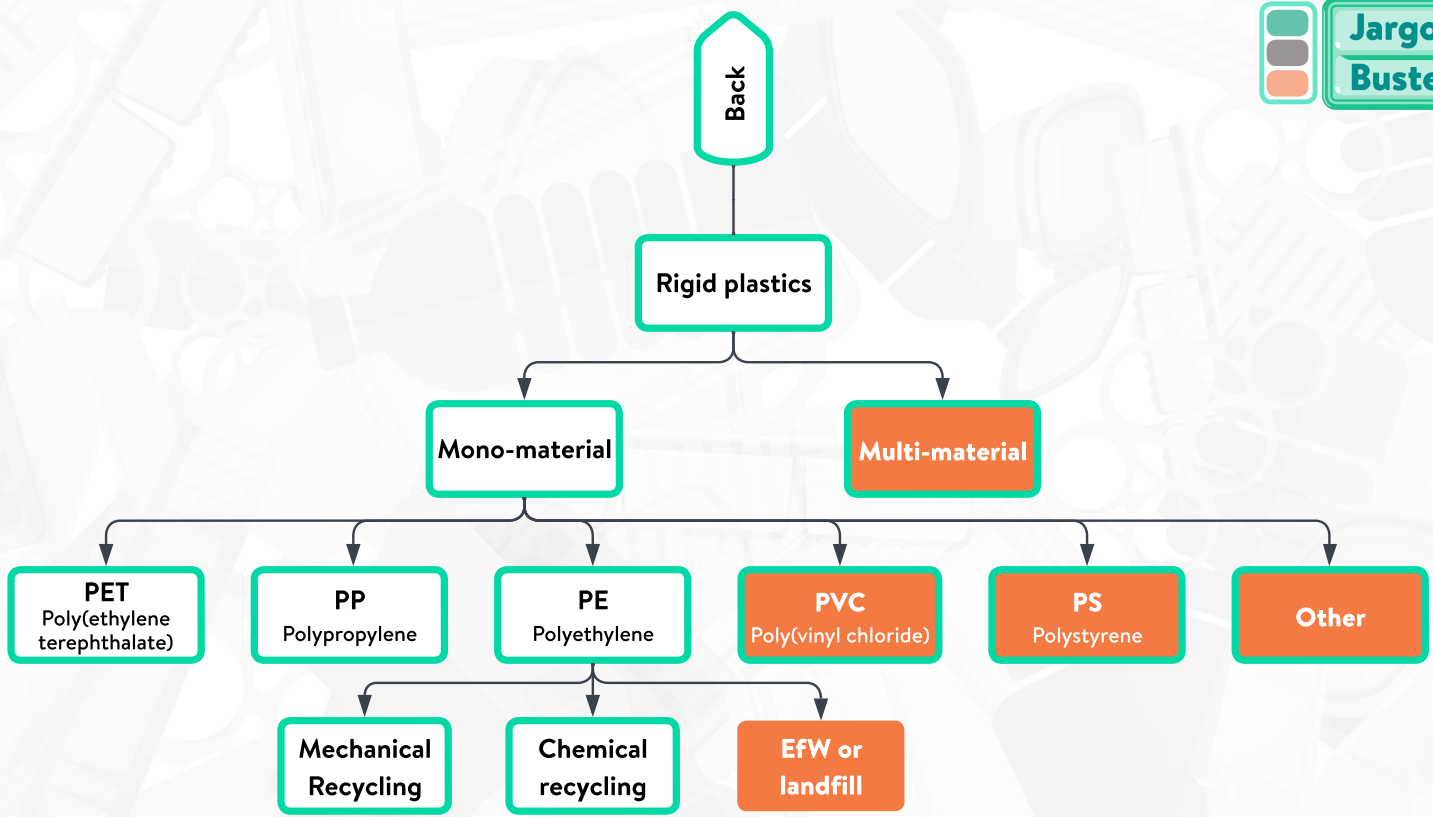

Approximately 28% of rigid plastics in household waste are made of PE, and over 85% of these PE plastics are in the form of bottles.[1] These bottles are often colourless (translucent milk bottles), or white (cleaning sprays), while coloured items are rare. The low density of PE (and PP) means that separation from PET and other dense plastics is often done through float-sink separation. Sorting of PE from other low density plastics, like PP, is possible with Near-Infrared technology that is widely used in the recycling industry.[2]

The need for further sorting within the rigid PE waste fraction will depend on end-of-life outcome and the types of PE items in the waste stream.[3]

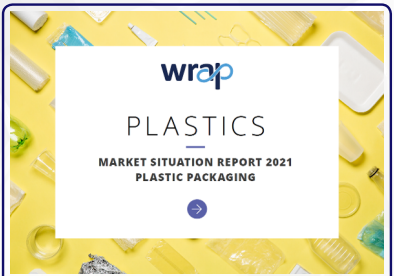

1. Plastics market situation report, WRAP, 2021.

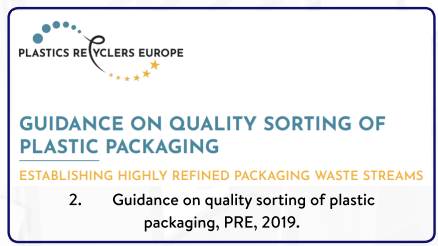

2. Guidance on quality sorting of plastic packaging, PRE, 2019.

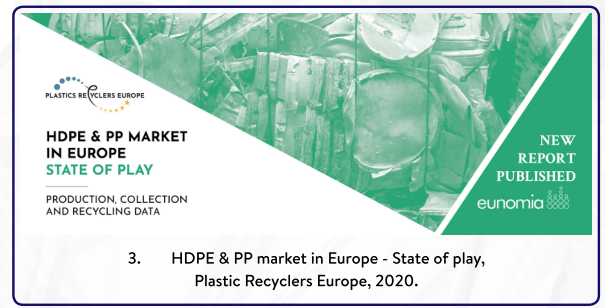

3. HDPE & PP market in Europe - State of play, Plastic Recyclers Europe, 2020.

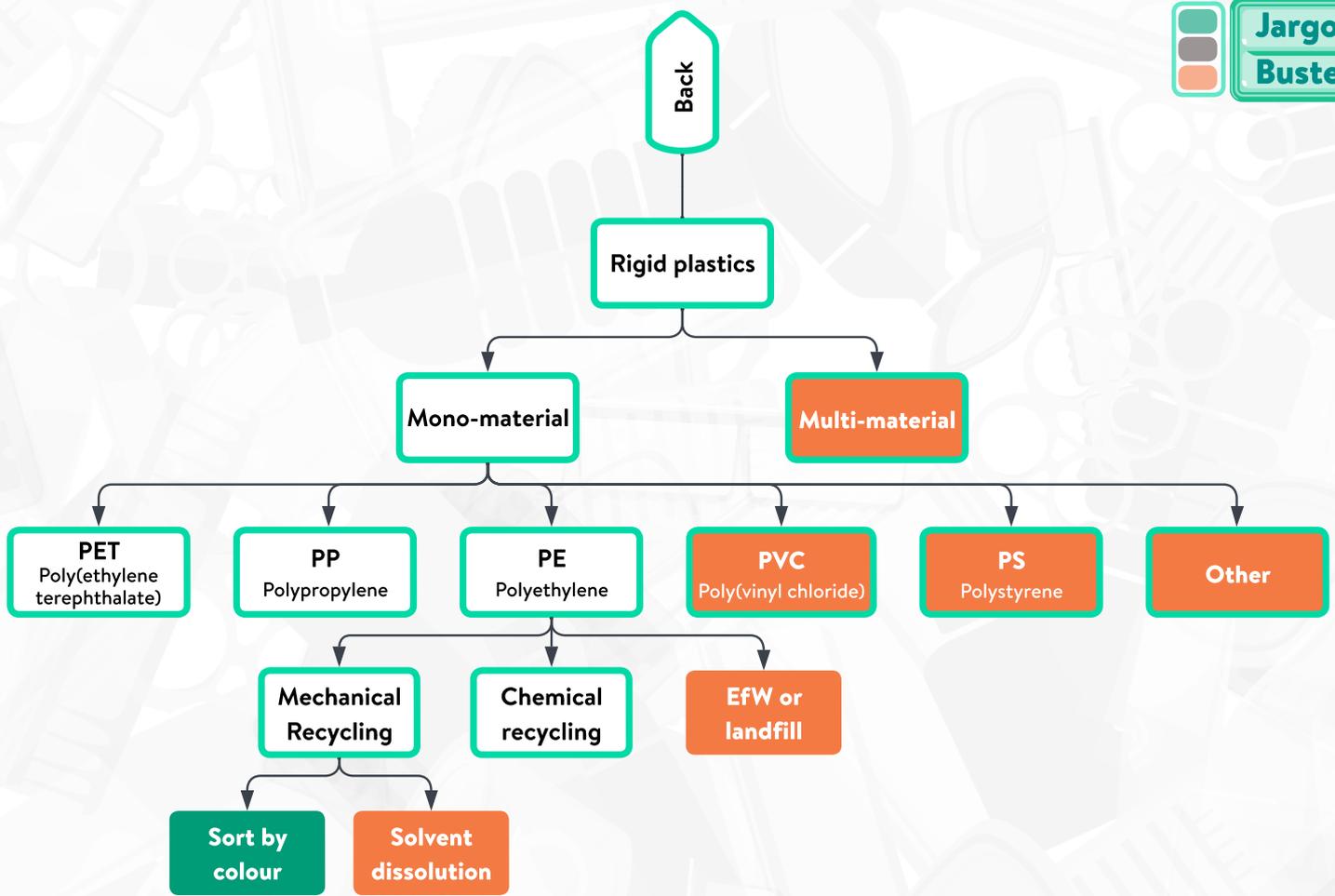

Rigid PE (HDPE) is extremely well suited for mechanical recycling, and this must be the primary end-of-life fate.[1] Best sorting practice for HDPE could involve the separation of bottles from PTTs. While the large majority of HDPE waste is bottles, the viscosity of the mixed waste stream may not always be acceptable for HDPE bottle production.[2,3] Coloured, white, and natural HDPE items should be sorted and processed separately, because colour severely impacts the value of recyclate. This is already done at some existing facilities, material volumes permitting.[3,4]

To maximise recyclability and ease of sorting, 'design for recycling' guidelines set out the importance of minimising colourants and contaminating additives, as well as using HDPE in all closures and attachments.[5,6]

Solvent dissolution, a method of dissolving plastics to purify plastics prior to mechanical recycling is an attractive way to optimise mechanical recycling, with potential for PP waste. However, economic and environmental analysis show that it will not be a good choice for HDPE.[7]

Advanced sorting to separate food and non-food plastics could unlock closed loop recycling of high-value food packaging. Tagging of food safe plastics could enable this, with several pilot projects showing promise. Legislation will be needed to ensure consistency in this emerging technology.

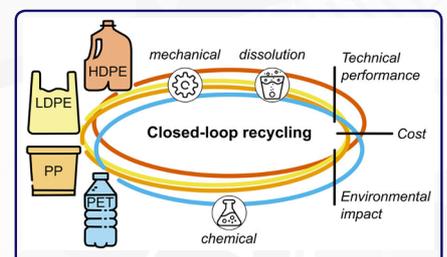

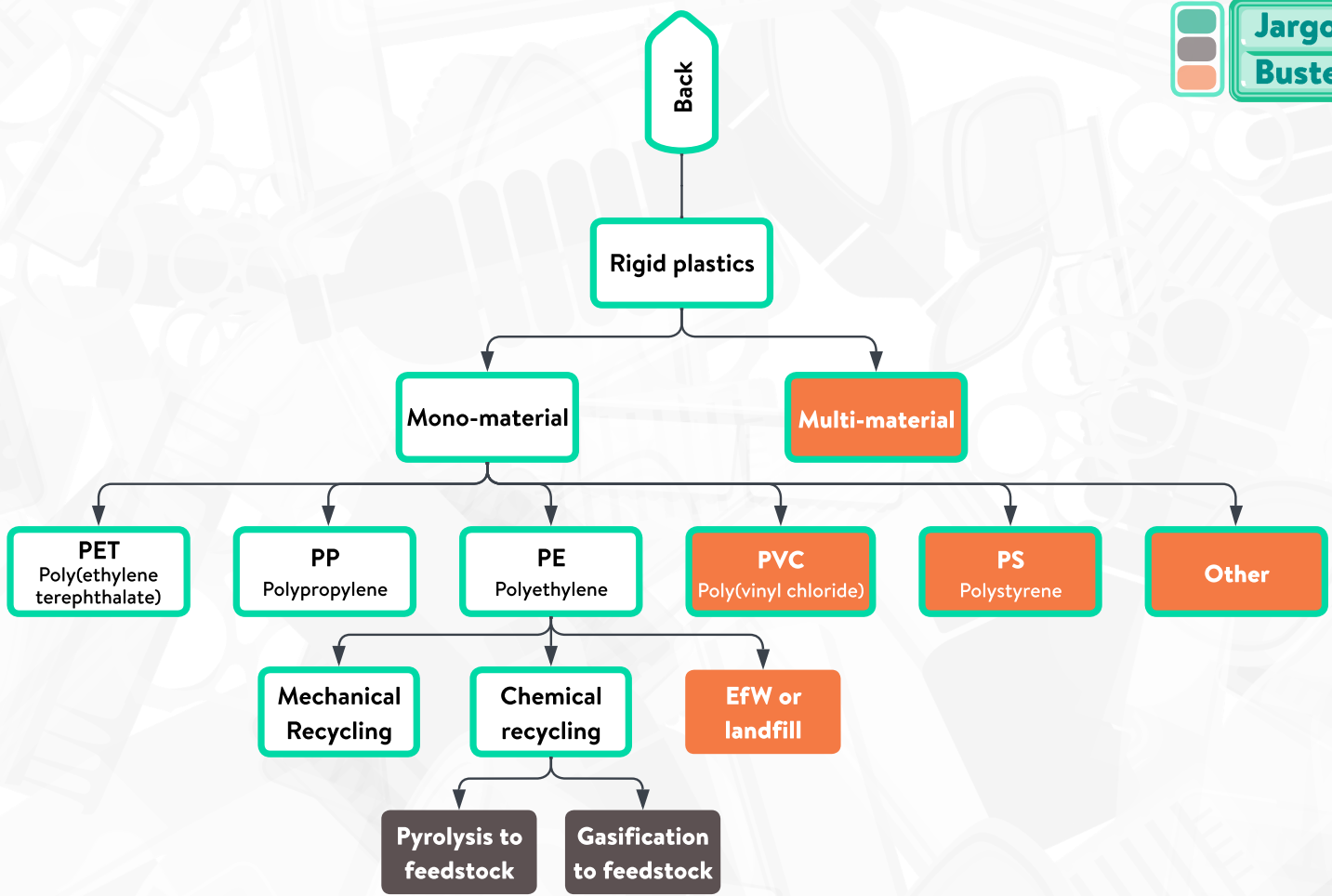

PE chemical recycling would not require further sorting and could potentially be processed as a mixed waste stream of polyolefins (PP and PE in both flexible and rigid form).[1] Chemical recycling of polyolefins is possible through pyrolysis or gasification techniques, converting plastics to feedstock molecules that can be used to make new plastics.[2] Pyrolysis and gasification are currently used for EfW, converting plastics into fuel instead of transforming back into plastics.[1,2,3] The environmental impact of these techniques for chemical recycling will need to be assessed as they develop further to assure that they are the most sustainable option for deteriorated plastic waste that cannot be mechanically recycled.[4,5]

Chemical recycling should not be viewed as an alternative to better design for recycling and large scale mechanical recycling of plastics, but rather it should be limited to being a backup of mechanical recycling options for degraded and contaminated plastic waste.[4,5] While chemical recycling is essential in achieving sustainable closed loop recycling of plastics, environmental impact assessments clearly show it should be secondary to mechanical recycling.[6]

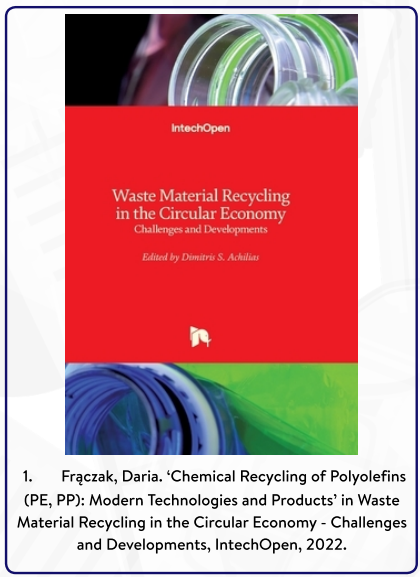

1. Frączak, Daria. 'Chemical Recycling of Polyolefins (PE, PP): Modern Technologies and Products' in Waste Material Recycling in the Circular Economy - Challenges and Developments, IntechOpen, 2022.

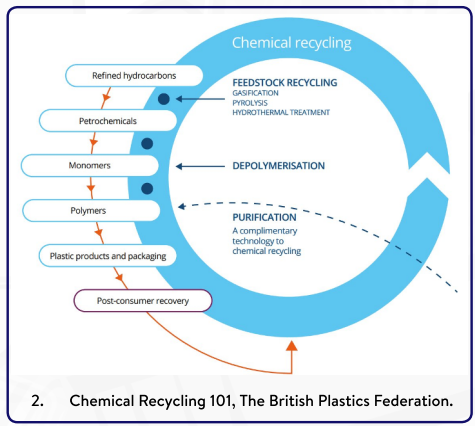

2. Chemical Recycling 101, The British Plastics Federation.

3. Non-Mechanical Recycling of Plastics, WRAP, 2019.

4. Redesigning the plastics system - the role of non-mechanical recycling, WRAP, 2022.

5. 7 Steps To Effectively Legislate On Chemical Recycling, Zero Waste Europe (ZWE) and the Rethink Plastic Alliance (RPa), 2020.

6. The Plastics Waste Hierarchy, WRAP, 2022.

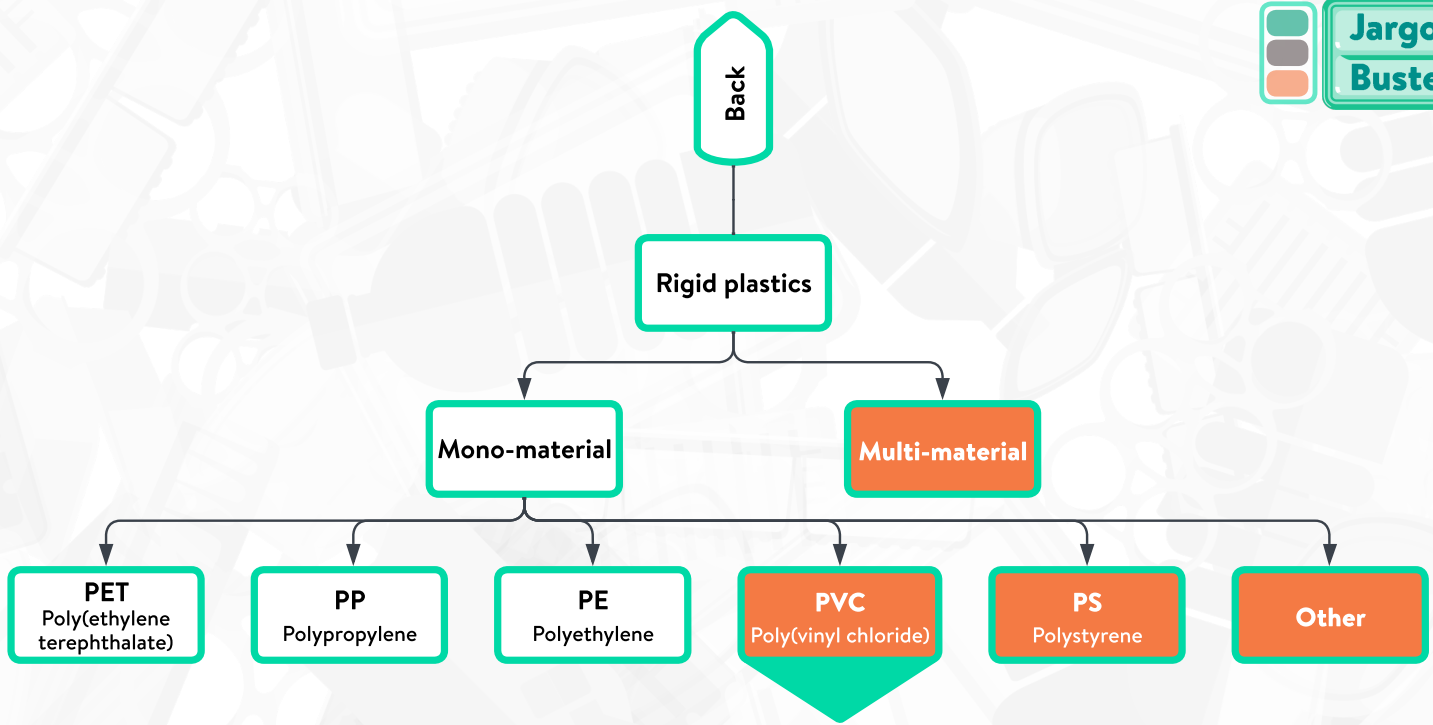

While PVC is used for a range of rigid packaging applications, the recycling of PVC waste is potentially harmful due to the chemicals that are released upon degradation.[1] WRAP included PVC packaging on their list problem plastics that should be eliminated,[2] because it can't be recycled and forms a severe contaminant in PET recycling.[3]

In the latest UK Plastics Pact annual report it was highlighted that PVC packaging had fallen by 90%, showing the positive effect of their call for elimination.[4] An area where the use of PVC persists is in pharmaceutical packaging, mainly pill blisters. While this sector has legitimate reasons to keep PVC packaging for now, innovation in this area must be encouraged.[2]

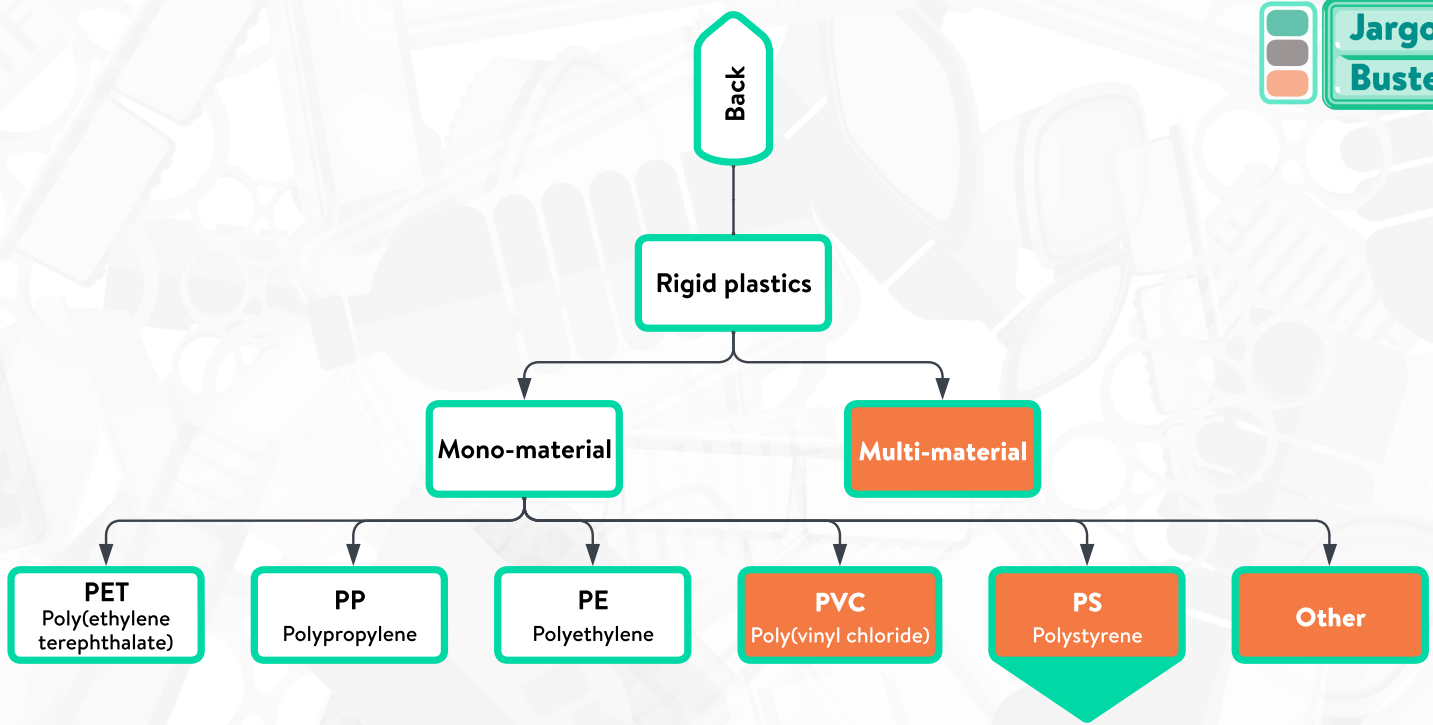

2% of all consumer plastic packaging is made of PS, often used for yoghurt pots or as expanded Polystyrene (ePS) for food takeaway containers.[1]

Recycling of this plastic is not considered commercially viable because it is only present in small amounts, and its low density makes collection and processing challenging.[1,2] This is why it was included on the UK Plastic Pact's list of problem plastics that should be eliminated from household waste, but progress on its elimination is poor so far and more needs to be done to replace it with more recyclable alternatives (PET, PE, PP).[3]

PS can be mechanically recycled under certain conditions, which is why business-to-business PS waste can be acceptable where effective processing and recycling is demonstrated, though the cost remains prohibitive.[1,2,4] ePS additionally requires solvent dissolution prior to any recycling attempt, further adding to processing costs. This technique is more suitable for bespoke recycling streams such as ePS home insulation at end-of-life, but cost remains a barrier with little supporting legislation.[5]

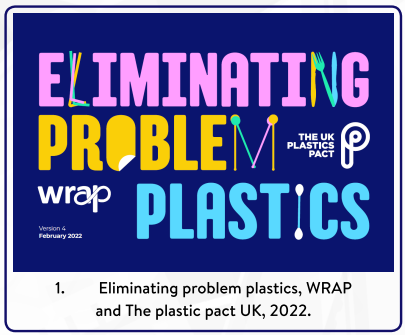

1. Eliminating problem plastics, WRAP and The plastic pact UK, 2022.

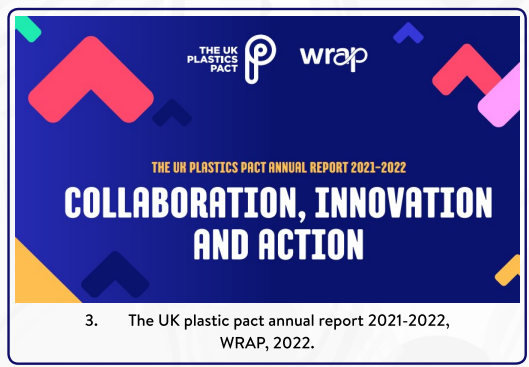

3. The UK plastic pact annual report 2021-2022, WRAP, 2022.

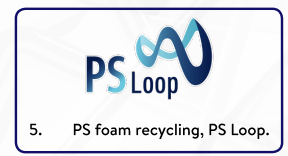

5. PS foam recycling, PS Loop.

2. Schyns, Z. O. G.; Shaver, M. P., Mechanical Recycling of Packaging Plastics: A Review, Macromolecular Rapid Communications, 2021.

4. Maharana, T.; Negi, Y. S.; Mohanty, B., Review Article: Recycling of Polystyrene, Polymer-Plastics Technology and Engineering, 2007.

6. Gil-Jasso, N. D.; Segura-Gonzalez, M. A.; Soriano-Giles, G.; Neri-Hipolito, J.; Lopez, N.; Mas-Hernandez, E.; Barrera-Diaz, C. E.; Varela-Guerrero, V.; Ballesteros-Rivas, M. F., Dissolution and recovery of waste expanded polystyrene using alternative essential oils, Fuel, 2019.

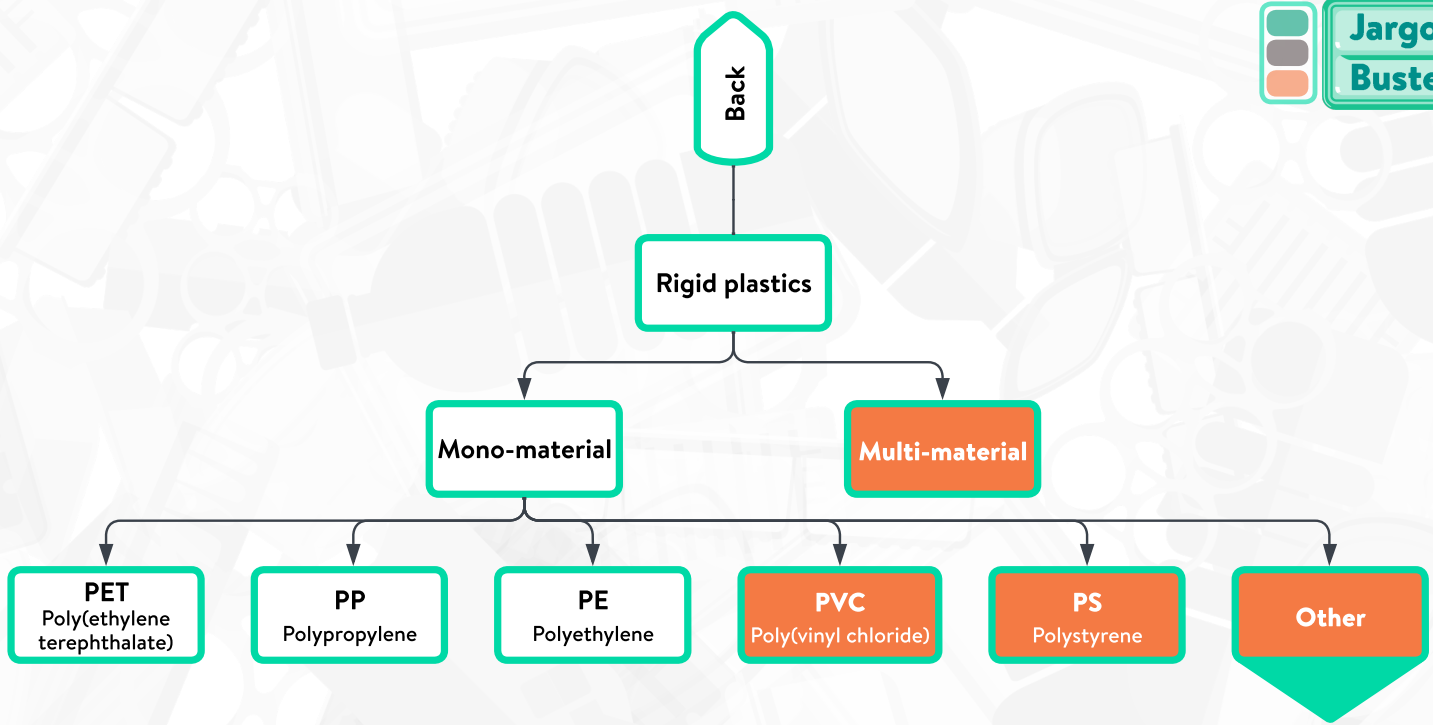

Over 5% of plastic waste currently falls beyond the sorting categories shown in this hierarchy and used throughout recycling industry.[1] These materials are diverse and generally unknown, meaning they can complicate and contaminate the overall recycling process.

Adding other plastic types to the market should only be allowed after end-of-life recycling can be guaranteed and any negative effect on other recycling systems can be avoided.

Newly developed plastics should always be assessed for their impact on recycling before being put into the market. This means the following questions must be addressed:

- What infrastructure is needed to separate the plastic waste from other items?
- How will the material perform within the existing sorting infrastructure?
- What effect will the material have on other plastic streams if it enters as a contaminant?
- Which sustainable recycling pathways are available and how will they be enforced?

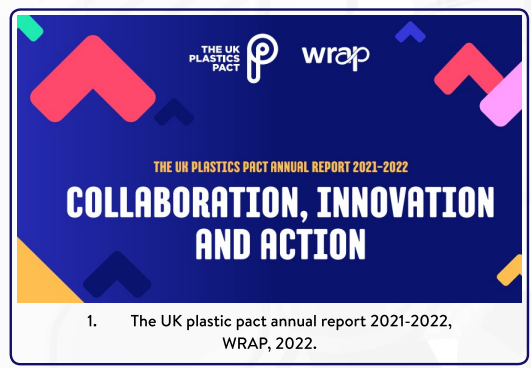

1. The UK plastic pact annual report 2021-2022, WRAP, 2022.

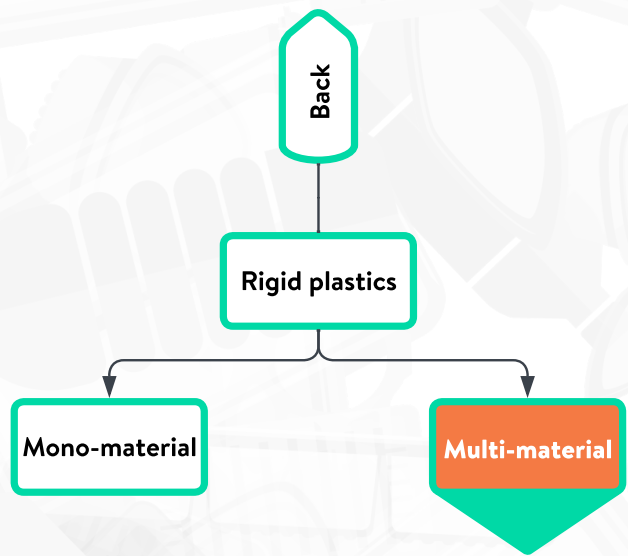

Multi-layered rigid plastic items containing more than one polymer resin type are practically impossible to sort, making them unrecyclable. During the sorting process, optical, density or spectroscopic sorting systems (e.g. NIR) will wrongly place multi-material rigids into mono-material waste streams. Contamination of different resin types will reduce the recyclate quality in the mono-material streams and diminish the value of recycling.

In 2021, 4000 tonnes of multi-material rigid plastic was used in primary packaging. This represents just 0.4% of the total plastic packaging market and is a reduction of over 70% compared to 2020.[1] The UK plastics pact identified multi-material rigid plastics as non-recyclable and categorised it as problematic/unnecessary, and this has thankfully led to a significant reduction in prevalence.[1]

Further elimination of multi-material rigids should be strived for, to safeguard high quality sorting and recycling of other rigid plastic packaging. However, complete elimination may be impossible because multi-material plastics can have unique properties. While there may be a place for them on the market, they are a contaminant to recycling and must be removed.

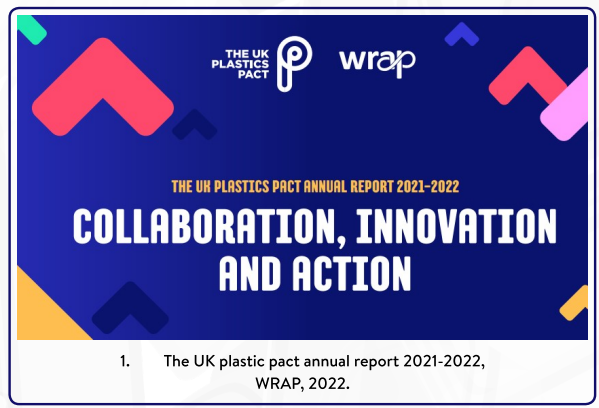

1. The UK plastic pact annual report 2021-2022, WRAP, 2022.

# Jargon Buster

| Term | Definition | Context |
|---|---|---|
| Biodegradable | (of a substance or object) capable of being decomposed by bacteria or other living organisms and thereby avoiding pollution. | This does not mean the material should be released into the environment, as rate of biodegradation varies. The term is often misused or misleading. Quite often, if a plastic item is only marked as biodegradable, it is actually a plastic that has an additive in it that speeds up fragmentation into microplastics. This is not good. |
| Chemical recycling | A process that changes the chemical structure of a polymer to produce raw materials for the manufacture of products (e.g. monomers). | This purposefully excludes the conversion of plastics to fuels or energy. Chemical recycling within the hierarchy only targets a plastic-to-plastic path, not linear plastic-to-oil/gas paths, such as energy from waste. Extracting energy from waste is not recycling. |
| Dry mixed recycling (DMR) | A broad term used to describe all types of dry recyclable waste. | Items accepted in DMR waste collections are dependent on the collection and recycling systems employed in local areas. The aim of this hierarchy is to include all items that could be considered suitable, including contaminants, and maximise recycling. |
| Energy from Waste (EfW) | A process used to generate energy in the form of electricity, heat, or combustible fuel commodities from the treatment of waste. | EfW does not recycle materials, but is often wrongly categorised as chemical recycling. It is part of a linear economy for plastics, replacing landfill pollution with air pollution. True recycling should always be targeted and EfW should only be a last resort end-of-life fate for unrecyclable waste. |
| Environmentally friendly | not harmful to the environment | A very broad term that without further detail could mean anything, including being used to greenwash a product or service. |
| Flexible plastics | Items made from primarily unmoulded plastic (e.g. bags, films) | An overarching term used to describe malleable plastic items. More on the varying terminology of flexibles can be found here. |
| Greenwashing | The practice of making an unsubstantiated or misleading claim about the environmental benefits of a product, technology or service. | The practice of companies launching adverts, campaigns, products etc. under the pretence that they are environmentally beneficial/benign, often in contradiction to a product or organisations environmental and sustainability record in general. |
| Mechanical recycling | Processing of plastics into secondary raw material without significantly changing the chemical structure. | A recycling method where the polymer chains are not chemically disrupted during processing, ideally resulting in like-for-like items being produced. |
| Recyclable | Able to be recycled. | Materials with evidence of a recycling process at pilot plant scale or larger are considered recyclable in the hierarchy. |
| Resin identification codes | A set of symbols that may appear on plastic products to identify the main plastic resin out of which the product is made. 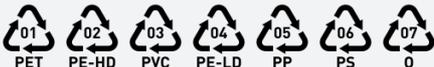 | The chasing arrows in resin identification codes often imply recyclability to the public. However, there is no link between the symbols and recyclability. The codes are not used in waste sorting and do not match with optimal recycling streams. For example PET bottles, trays, and films with the  symbol are not compatible in mechanical recycling and are processed separately. |
| Sustainable | Able to be maintained at a certain rate or level. | This is very broad term that without further detail could mean anything, including being used to greenwash a product or service. |

# Conclusion

With plastic production continuously increasing and recycling rates stubbornly low, change is urgently needed. Information on plastic waste sorting, processing, and recycling is often scattered and presented for a very technical audience, failing to reach decision makers across the plastics value chain. Opportunities to achieve real progress in plastics recycling are being missed, hampered by confusing and potentially conflicting narratives or evidence.

To encourage sustainable progress through informed decision making, we developed a 'Plastics Hierarchy of Fates' tool that summarises technical information in a more accessible manner. We have set out the merits and perils of potential decisions in plastics waste processing and recycling in our tool, with a forward-looking perspective on maximising circularity. By scrutinising plastic waste recycling as a whole system, we show that decisions are interlinked and the order in which decisions are made is important. As different recycling methods will have different sensitivity towards contamination, needing different sorting levels, and different decisions can unlock material value, these interrelationships are important. Good recycling practices are only useful with suitable sorting processes to create valuable products. Many factors will play a role in a more sustainable circular plastics economy, and through the 'Plastics Hierarchy of Fates' we believe a path can be found to achieve meaningful sustainability progress.

# References

Literature references are embedded within the various segments of the 'Plastics Hierarchy of Fates' tool. All references are clickable hyperlinks to the original source.